\documentclass[preprint,12pt]{elsarticle}

\usepackage{subfig}
\usepackage[colorlinks,allcolors=blue]{hyperref}
\usepackage{graphicx}
\usepackage{caption}
\usepackage{amssymb}
\usepackage{amsmath}
\usepackage{listings}
\newcommand{\vphi}{\mbox{\boldmath{$\phi$}}}
\newcommand{\vc}{\mbox{\boldmath{$c$}}}
\newcommand{\vmu}{\mbox{\boldmath{$\mu$}}}

\journal{Computer Physics Communications}

\begin{document}

\begin{frontmatter}



\title{MICROSIM: A high performance phase-field solver based on CPU and GPU implementations}


\author[inst1]{Tanmay Dutta}

\affiliation[inst1]{organization={Department of Materials Engineering},
            addressline={Indian Institute of Science}, 
            city={Bangalore},
            postcode={560012}, 
            state={Karnataka},
            country={India}}

\affiliation[inst2]{organization={Department of Metallurgical and Materials Engineering},
            addressline={IIT Madras}, 
            city={Chennai},
            postcode={600036}, 
            state={Tamil Nadu},
            country={India}}

\affiliation[inst3]{organization={Department of Materials Science and Metallurgical Engineering},
            addressline={IIT Hyderabad}, 
            city={Hyderabad},
            postcode={502284}, 
            state={Telangana},
            country={India}}
\affiliation[inst4]{organization={Center for development of Advanced Computing},
            addressline={Pune University Campus}, 
            city={Pune},
            postcode={411007}, 
            state={Maharashtra},
            country={India}}     
\affiliation[inst5]{organization={Department of Metallurgical Engineering and Materials Science},
            addressline={IIT Bombay, Powai}, 
            city={Mumbai},
            postcode={400076}, 
            state={Maharashtra},
            country={India}}          
\affiliation[inst6]{organization={Department of Scientific Computing, Modeling and Simulation
},
            addressline={Savitribai Phule Pune University}, 
            city={Pune},
            postcode={411007}, 
            state={Maharashtra},
            country={India}}  
\author[inst2,inst5]{Dasari Mohan}
\author[inst3]{Saurav Shenoy}
\author[inst4]{Nasir Attar}
\author[inst4]{Abhikshek Kalokhe}
\author[inst1]{Ajay Sagar}
\author[inst1]{Swapnil Bhure}
\author[inst2]{Swaroop S. Pradhan}
\author[inst1]{Jitendriya Praharaj}
\author[inst1]{Subham Mridha}
\author[inst1]{Anshika Kushwaha}
\author[inst6]{Vaishali Shah}
\author[inst5]{M. P. Gururajan}
\author[inst4]{V. Venkatesh Shenoi}
\author[inst2]{Gandham Phanikumar}
\author[inst3]{Saswata Bhattacharyya}
\author[inst1]{Abhik Choudhury}

\begin{abstract}
The phase-field method has become a useful tool for the simulation of classical metallurgical phase transformations as well as other phenomena related to materials science. The thermodynamic consistency that forms the basis of these formulations lends to its strong predictive capabilities and utility. However, a strong impediment to the usage of the method for typical applied problems of industrial and academic relevance is the significant overhead with regard to the code development and know-how required for quantitative model formulations. In this paper, we report the development of an open-source phase-field software stack that contains generic formulations for the simulation of multi-phase and multi-component phase transformations. The solvers incorporate thermodynamic coupling that allows the realization of simulations with real alloys in scenarios directly relevant to the materials industry. Further, the solvers utilize parallelization strategies using either multiple CPUs or GPUs to provide cross-platform portability and usability on available supercomputing machines. Finally, the solver stack also contains a graphical user interface to gradually introduce the usage of the software. The user interface also provides a collection of post-processing tools that allow the estimation of useful metrics related to microstructural evolution. 
\end{abstract}




\end{frontmatter}


\section{Introduction}
\label{sec:introduction}
Materials modeling is being widely utilized by the community both for the fundamental understanding of materials phenomena and processes as well as in applied problems related to materials discovery, process optimization, and control. Typically, modeling in materials science involves methods that span several decades in lengths and time scales. For example, at the electronic scale, there are density functional-based methods for electronic structure and energetic predictions, whereas there are also methods related to property predictions based on the grain structure, such as crystal plasticity and further finite element-based homogenization methods for the estimation of properties at the macroscopic scale. Given this large breadth in scales, a comprehensive understanding of the materials phenomena typically requires multiscaling approaches involving information transfer across scales. This is broadly the approach in integrated computational materials engineering (ICME). This endeavor requires efficient models and software at the respective scales (electronic, atomistic, microscopic, mesoscopic, and macroscopic) for this framework to be useful. While fairly standard methods for software development are available for electronic, atomistic, and macroscopic structure and property estimation (for e.g., LAMMPS, SPPARKS, Quantum Espresso, and MOOSE)the microscopic and mesoscopic scales have limited possibilities, particularly where the focus is on questions related to microstructure evolution. In this context, while methods like
cellular automata (CA) and 
kinetic Monte Carlo (KMC) exist for quick predictions for certain phase transformations, the phase-field method has emerged as the method of choice, given its strong predictive capabilities and generic applicability. 
\par
The phase-field method is widely used for simulating microstructure evolution during phase transformations. The method has progressed quite rapidly in the past four decades since its inception, where quantitative model formulations are available to simulate most phase transformations relevant to metallurgy. In particular, there are approaches to couple phase-field models with thermodynamic databases based on the CALPHAD formalism that have increased the scope of this method's usage from 
classical problems in condensed matter physics to engineering applications. Further efforts are underway towards developing generic software that promotes the diversification of applications. For this purpose, some notable commercial packages exist, such as MICRESS \citep{Micress}, PACE3D \citep{PACE3D}, and OpenPhase \citep{OpenPhase}. While MICRESS \citep{Micress} provides a direct connection with thermodynamic databases, PACE3D \citep{PACE3D} and OpenPhase \citep{OpenPhase} provide alternate strategies for the same. Similarly, there are efforts underway for the development of specific phase-field modules in larger multi-physics packages such as MOOSE \citep{lindsay2022moose} and PRISMS-PF \citep{prisms}, providing a more generic framework for the implementation of arbitrary phase-field models. Similarly, modules based on Python, such as FEniCS \citep{FEniCS, FEniCS1, FEniCS2, FEniCS3}, allow quick implementation of phase-field models. 

\par While these software and modules provide viable options, there are certain deficiencies. Firstly, with respect to the phase-field software, each commercial package is engineered to utilize a particular type of phase-field model formulation that limits the scope for modifications and extensions. Secondly, while PACE3D \citep{PACE3D} and OpenPhase \citep{OpenPhase} provide scope for parallelization on several thousands of CPUs, the same is not possible for MICRESS \citep{Micress}, and this limits its utility on available supercomputing machines. Further, both PACE3D \citep{PACE3D} and OpenPhase \citep{OpenPhase} provide parallelization possibilities only using CPUs but not graphical processing units (GPUs), which are getting widely integrated into most state-of-the-art supercomputers. Finally, while the modules in MOOSE \citep{lindsay2022moose} and PRISMS-PF \citep{prisms} are open-source, they are limited by the type of discretization (finite element method or FEM) and the type of parallelization (to the best of knowledge of the authors, the available literature does not contain data with respect to simulations performed using both CPUs and GPUs).  Therefore, in summary, while software and specific modules are available for the scientific community, there is certainly scope for providing greater flexibility with respect to available model formulations, possible hardware utilization, and discretization. This forms the motivation for the development of our open-source phase-field software MICROSIM (MICROstructure SIMulator). These HPC solvers will also address the bottleneck in current ICME implementations, namely, microstructure evolution.
\par
The present paper describes the capabilities of MICROSIM, which is a software stack that is a combination of stand-alone generic phase-field solvers as well as modules utilizing multiphysics frameworks such as OpenFOAM \citep{openfoam} and AMReX \citep{AMReX,AMReX_short}. Similarly, the stack provides different discretization possibilities, such as finite volume method (FVM), finite difference (FD), and fast Fourier transform (FFT), that provide the user greater flexibility. The stack also encodes different models, such as the grand-potential (GP) \citep{Abhik2012} and the Kim-Kim-Suzuki (KKS) \citep{Kim1999} formulations, that are combined either with the multi-phase field method or the orientation-field-based methods \citep{henry2012orientation, granasy2005el, granasy2005mse, granasy2008} for multi-grain simulations. These methods are coupled to thermodynamic databases to simulate phase transformations involving real alloys. Further, the stack has parallelized solvers using multiple CPUs and GPUs that provide greater access to available hardware on supercomputing machines. Finally, a user-friendly graphical user interface (GUI) allows for easy interaction with the software stack. This interface is also coupled with a host of post-processing tools routinely utilized in the community to extract microstructural features during simulations. Thus, the stack is a unique combination of models, discretization, and implementation that provides flexibility beyond anything that is presently available in the community.
\par
We begin with brief details about the formulation of the GP \citep{Abhik2012} and the KKS \citep{Kim1999} phase-field models that are the bases of the solver modules. Next, we elaborate on the coupling of the solver modules with the thermodynamic databases based on the CALPHAD method, followed by the implementation of these phase-field models using different discretization (FD, FVM and FFT) and parallelization (CPU, and GPU) schemes. Subsequently, we compare and benchmark the simulation results using the different solver modules and comment on their scalability. Additionally, simulation test cases are depicted that highlight the overall capabilities of the software stack.
Finally, the GUI for generating input files to run the simulations, as well as analyzing and visualizing the results, is discussed.
 
\section{Formulation}
\label{sec:formulation}
MICROSIM broadly contains two different types of models: the GP and the KKS formulations. In the following, we briefly describe the models in detail.
\subsection{Grand-potential formulation}
\label{subsec:gp}
In the GP framework, phase evolution is determined by the phenomenological minimization of a functional formulated as the grand-potential functional \citep{Abhik2012}. Order parameter fields $\phi_\alpha$ are utilized for describing the distribution of the $N$ phases in the system. They follow the constraints:
$$
\phi_\alpha \in [0, 1]  \quad \text{and} \quad \sum_{\alpha}^{N} \phi_\alpha = 1.
$$

The grand-potential functional reads:
\begin{flalign}
\Omega (T, \mu, \vphi) = \int_{V}^{} \left[ \psi (T, \mu, \vphi) + \epsilon a(\vphi, \nabla \vphi) + \dfrac{w(\vphi)}{\epsilon} + 
f_{el}\left(\mathbf{u},\vphi\right) \right] dV.
\label{eq:grand_eqn}
\end{flalign}

Here, $\epsilon$ is the width of the diffuse interface, which is chosen to resolve the smallest feature in the resulting morphology accurately. The term $w(\vphi)$ can be either a multi-obstacle or a multi-well potential. The formulation for a double obstacle potential is:
\begin{flalign}
w(\vphi) = 
\begin{cases}
 \sum_{\substack{\alpha < \beta \\ \delta \neq \alpha \neq \beta}}^{N, N} \dfrac{16}{\pi^2} \gamma_{\alpha \beta} \phi_\alpha \phi_\beta + \gamma_{\alpha \beta \delta} \phi_\alpha \phi_\beta \phi_\delta & \text{if $ \phi \in [0, 1] $,} \\
\infty & \text{otherwise,}
\end{cases}
\label{Obstacle}
\end{flalign}

while the multi-well potential is of the form:
\begin{flalign}
w(\vphi) = 
\begin{cases}
 \sum_{\substack{\alpha < \beta \\ \delta \neq \alpha \neq \beta}}^{N, N} 9\gamma_{\alpha \beta} \phi_\alpha^{2} \phi_\beta^{2} + \gamma_{\alpha \beta \delta} \phi_\alpha \phi_\beta \phi_\delta,
\end{cases}
\label{Well}
\end{flalign}

$\gamma_{\alpha\beta}$ is the isotropic surface energy of the 
the $\alpha$-$\beta$ interface. $\gamma_{\alpha\beta\delta}$ are third-order terms added in order to suppress the adsorption of a third phase at a binary interface. $a(\vphi, \nabla \vphi)$ is the gradient energy term and $ \psi (T, \mu, \vphi) $ is the grand-potential. The anisotropy in the interface energy is incorporated in the gradient energy term as:
\begin{flalign}
a(\phi, \nabla \phi) = \sum_{\alpha < \beta}^{N, N} \gamma_{\alpha \beta} \left[ a_c(\vec{q}_{\, \alpha \beta}) \right]^2 |\vec{q}_{\, \alpha \beta}|^2,
\label{Eqn_a}
\end{flalign}

where $a_c$ is the anisotropy function of the $\alpha$ - $\beta$ interface normal vector, $\vec{q}_{\, \alpha \beta} = \phi_\alpha \nabla \phi_\beta - \phi_\beta \nabla \phi_\alpha$. For four-fold anisotropy, the function $a_c\left(\vec{q}_{\,\alpha\beta}\right)$ in Eq.~(\ref{Eqn_a}) becomes:
\begin{align}
a_c\left(\vec{q}_{\,\alpha\beta}\right) &= 1-\delta_{\alpha\beta}\left(3-4\dfrac{q_x^{4} + q_y^{4} + q_z^{4}}{\left|\vec{q}_{\,\alpha\beta}\right|^{4}}\right), 
\end{align}

where $\delta_{\alpha\beta}$ is the strength of the anisotropy, while $\vec{q}_{\,\alpha\beta}={q_x,q_y,q_z}$, where $q_{x},q_{y},q_{z}$ are the components of the vector $\vec{q}_{\,\alpha\beta}$. For imparting a particular rotation through an arbitrary angle; we have the following form for the anisotropy function:
\begin{flalign}
a(\phi, \nabla \phi) = \sum_{\alpha < \beta}^{N, N} \gamma_{\alpha \beta} \left[ a_c(\vec{q'}_{\, \alpha \beta}) \right]^2 |\vec{q}_{\, \alpha \beta}|^2,
\label{eq:aniso_eqn}
\end{flalign}

where $\vec{q'}_{\,\alpha\beta}={q_x',q_y',q_z'}$ corresponds to the rotation of the $\vec{q}_{\,\alpha\beta}$ by an arbitrary angle 
given by the rotation matrix $R$.
\par
The evolution equation follows from (Eq.~(\ref{eq:grand_eqn})), through the classical Allen-Cahn type formulation, which reads,
\begin{flalign}
\tau \epsilon \frac{\partial \phi_{\alpha}}{\partial t}=& \epsilon\left(\nabla \cdot \frac{\partial a(\phi, \nabla \phi)}{\partial \nabla \phi_{\alpha}}
-\frac{\partial a(\phi, \nabla \phi)}{\partial \phi_{\alpha}}\right) 
-\frac{1}{\epsilon} \frac{\partial w(\phi)}{\partial \phi_{\alpha}}-\frac{\partial \psi(T, \mu, \phi)}{\partial \phi_{\alpha}} -\nonumber \\
&\frac{\partial f_{el}\left(\mathbf{u},\vphi\right)}{\partial \phi_\alpha} -\lambda,
\label{phi_eqn}
\end{flalign}

where $\tau$ is the relaxation constant that controls the kinetics of the phase, and $\lambda$ is the Lagrange multiplier utilized for imposing the constraint $\sum_{\alpha}^{N} \phi_\alpha = 1$, i.e., the sum of phase-fields is one (here the phase-fields may be associated with volume fractions). $\psi$ is the grand-potential density in Eq.~(\ref{eq:grand_eqn}), which is obtained at any point as a weighted sum of the grand-potential densities of each phase present at that point: 
\begin{flalign}
\psi = \sum_{\alpha}^{N} h_\alpha(\vphi)\psi^\alpha,
\end{flalign}

where $h_\alpha(\vphi)$ can be an interpolation function of third order:
\begin{flalign}
h_\alpha(\vphi) = 3{\phi_\alpha}^2 - 2{\phi_\alpha}^3 + 2\phi_\alpha \sum_{\substack{ \beta, \gamma \neq \alpha \\ \beta < \gamma }}^{N, N} \phi_\beta \phi_\gamma,
\label{eqn:hphi}
\end{flalign}


$\psi^\alpha = \psi^\alpha(\mu, T)$ is the grand-potential density of phase $\alpha$ and can be written as:
\begin{flalign}
\psi^\alpha(T, \mu) = f^\alpha(c^\alpha(T, \mu)) - \mu c^\alpha(T, \mu),
\end{flalign}

where $f^\alpha$ is the Helmholtz free energy density per unit volume. Under constant pressure and volume, $f^\alpha$, the Helmholtz free energy density differs from the Gibbs free energy density by a constant.
\par
The conservation equation for each of the components follows:
\begin{align}
\dfrac{\partial c_i}{\partial t} &= \nabla \cdot \left(M_{ij}\nabla\mu_j\right),
\label{Mass_conservation}
\end{align}
 
where $M_{ij}=\sum_{\alpha=1}^{N}\left[D_{ij}^{\alpha}\right]\left[\dfrac{\partial c_i^{\alpha}}{\partial \mu_{ij}}\right]h_\alpha\left(\vphi\right)$. This can be converted into a diffusion potential update equation by linearizing based on the dependencies on the individual variables as: 
\begin{align}
&\left\lbrace\dfrac{\partial \mu_i}{\partial t}\right\rbrace = \left[\sum_{\alpha=1}^N 
h_\alpha\left(\vphi\right)\dfrac{\partial c_i^\alpha\left(\vmu,
T\right)}{\partial\mu_j}\right]^{-1}_{ij}\Big\lbrace\nabla\cdot \left(\sum_{j=1}^{K}M_{ij}
\left(\vphi\right)\nabla\mu_j - j^{at}_{i}\right)
- \nonumber\\
&\sum_{\alpha}^N c^\alpha_{i}\left(\vmu,T\right)\dfrac{\partial
h_\alpha\left(\vphi\right)}{\partial t} - \dfrac{\partial T}{\partial t}\sum_{\alpha}^N \left(\dfrac{\partial c^\alpha_{i}\left(\vmu,T\right)}{\partial T}\right)_{\vmu}
h_\alpha\left(\vphi\right) \Big\rbrace,
\label{Mu_explicit_temperature}
\end{align}

using the composition constraint $c_i = \sum_{\alpha=1}^{N} c_i^\alpha\left(\vmu\right)h_\alpha\left(\vphi\right)$. A correction referred to as the anti-trapping current, $j^{at}_{i}$ \citep{Karma2001, Abhik2012} has been incorporated to Eq.~(\ref{Mu_explicit_temperature}) and the explicit formulation is detailed in \citep{Abhik2012}
for one-sided solidification problems where the diffusivity in the solid is negligible compared to the liquid. The anti-trapping flux is given by:
\begin{equation}
j^{at}_{i} = \sum_{\alpha=1}^{N-1} j^{at,\alpha}_{i} \left(- \frac{\nabla \phi_{\alpha}}{|\nabla \phi_{\alpha}|} \cdot \frac{\nabla \phi_{liq}}{|\nabla \phi_{liq}|}\right),
\label{jatr}
\end{equation}
where $j^{at,\alpha}_{i}$ is written as:
\begin{equation}
j^{at,\alpha}_{i} = \frac{\pi \epsilon \phi_\alpha^0 [1 - h_\alpha(\phi_\alpha^0)]}{4\sqrt{\phi_\alpha^0(1 - \phi_\alpha^0)}} (c_{i}^\alpha-c_{i}^{liq}) \frac{\partial \phi_{\alpha}}{\partial t} \frac{\nabla \phi_{\alpha}}{|\nabla \phi_{\alpha}|},
\end{equation}

for the case where the multi-obstacle function is used as the potential. Conversely, in the case where the multi-well potential is used, the anti-trapping current reads,

\begin{equation}
j^{at,\alpha}_{i} = \frac{\epsilon}{2\sqrt{2}} (c_{i}^\alpha-c_{i}^{liq}) \frac{\partial \phi_{\alpha}}{\partial t} \frac{\nabla \phi_{\alpha}}{|\nabla \phi_{\alpha}|},
\end{equation}

where $\phi_\alpha^0$ is the equilibrium phase-field profile in 1D, which is part of a sinus curve for the case of the obstacle potential and a hyperbolic tangent profile for the case of a multi-well potential.

\subsection{Kim-Kim-Suzuki formulation}
\label{subsec:kks}
In the KKS model, the total free energy function of the system is given by,

\begin{flalign}
	F = \int_V^{} \left[ \frac{\varepsilon^2}{2} a(\vphi, \nabla \vphi) + g(\vphi) + f(c, T, \vphi) \right] dV,
	\label{fe}
\end{flalign}
where $\varepsilon$ is the gradient energy coefficient, the anisotropy function, $a(\phi, \nabla \phi)$ is given in Eq.~(\ref{eq:aniso_eqn}). g(\vphi), the multi-well potential is of the form given in Eq.~(\ref{Well}) for the implementation in OpenCL. Using the multi-well potential typically leads to quite strong third-phase adsorption for multi-phase simulations, where a small amount of a third phase percolates from the multi-phase regions of the simulations into binary interfaces. This phenomenon is undesirable as it modifies the effective interfacial energy. For a better control of the interfacial properties, the following form of the potential is also coded in the CUDA module of the solver, giving provision or flexibility to the user in the case where the third-phase adsorption is strong. This formulation will also be encoded in the OpenCL module in future versions.

\begin{flalign}
g(\vphi) = 
 \sum_{\alpha=1}^{N} \Theta_\alpha \phi_\alpha^{2} (1-\phi_\alpha)^{2} + \sum_{\alpha=1}^{N}\sum_{\alpha < \beta}^{N} \Theta_{\alpha \beta} \phi_\alpha^{2} \phi_\beta^{2} + \sum_{\alpha=1}^{N}\sum_{\alpha < \beta}^{N}\sum_{\beta < \delta}^{N} \Theta_{\alpha \beta \delta} \phi_\alpha^{2} \phi_\beta^{2} \phi_\delta^{2},
\label{Well1}
\end{flalign}

where $\Theta_\alpha$, $\Theta_{\alpha \beta}$, and $\Theta_{\alpha \beta \delta}$ are anti-phase domain boundary energy.


$f$ is the total Helmholtz free energy density of all phases:
\begin{flalign}
f\left(c, T, \vphi\right) = \sum_{\alpha=1}^{N} h_\alpha(\vphi)f^\alpha\left(c^\alpha, T\right),
\end{flalign}
where $h_\alpha(\vphi)$ is given in Eq.~(\ref{eqn:hphi}). In the KKS formulation, the free energy of the system is an interpolation of the 
individual free energy densities of the phases at the respective phase compositions. While the mass conservation equations are written in terms of the global composition field for each component, the deconvolution to the individual phase compositions for utilization in the equation for the phase evolution as well as the diffusion equation is performed by imposing the following conditions \citep{Kim1999}:
\begin{flalign}
	\frac{\partial f^\alpha}{\partial c_{i}^\alpha} = \frac{\partial f^{liq}}{\partial c_{i}^{liq}} \equiv {\mu}_i,
	\label{eqmu}
\end{flalign}
along with composition constraint is given by:
\begin{flalign}
	c_i = \displaystyle \sum_\alpha^N h_\alpha(\vphi)c_{i}^\alpha,
	\label{cint}
\end{flalign}

It can be shown that the above two conditions are equivalent to the treatment in the grand-potential formulation, where the equivalence in the diffusion potentials is implicit in the derivation, and the composition constraint follows from the interpolation of the grand-potential densities of the phases. 

The evolution equation for the phase-field variable is:
\begin{flalign}
&\frac{1}{M_{\phi}} \frac{\partial \phi_{\alpha}}{\partial t}=\frac{\varepsilon^2}{2} \left(\nabla \cdot \frac{\partial a(\phi, \nabla \phi)}{\partial \nabla \phi_{\alpha}}-\frac{\partial a(\phi, \nabla \phi)}{\partial \phi_{\alpha}}\right) - \frac{\partial g(\phi)}{\partial \phi_{\alpha}} \nonumber\\
&-\frac{\partial }{\partial \phi_{\alpha} } \left[ f(c, T, \phi) - \displaystyle\sum\limits_{i=1}^n  c_{i} {\mu}_i \right]-\lambda,
\label{aceq}
\end{flalign}
where ${M_{\phi}}$ is phase-field mobility \citep{Kim2007}.
The composition evolution is given by:
\begin{flalign}
\frac{\partial c_i}{\partial t} = \nabla \cdot (M_{ij} \nabla \mu_j - j^{at}_i),
\label{cheq}
\end{flalign}
and the linearized form of the above equation in terms of diffusion potentials is represented in Eq.~(\ref{Mu_explicit_temperature}). The form of the anti-trapping current is the same as in the preceding section.

\subsection{Elasticity implementation}
\label{subsec:elst}
For the simulation of precipitation reactions that are influenced by coherency stresses, the respective functionals in the GP and KKS formulations described in the preceding sections need to be modified by adding contributions from the elastic energy arising from the mismatch between the lattice parameters of the matrix and the precipitate. In the diffuse-interface framework, appropriate interpolation strategies need to be utilized for the stiffness and the eigenstrain matrices. For all the solver modules, we have utilized the following form of the elastic energy density that reads,
\begin{align}
f_{el}(\mathbf{u}, \phi) &= \dfrac{1}{2}C_{ijkl}(\vphi)(\epsilon_{ij} - \epsilon^*_{ij}(\vphi))(\epsilon_{kl} 
- \epsilon^*_{kl}(\vphi)),
\end{align} 
where the total strain field can be computed from the periodic displacement field $\mathbf{u}$ as:
\begin{align}
 \epsilon_{ij} &= \frac{1}{2}\left(\frac{\partial u_i}{\partial x_j} + \frac{\partial u_j}{\partial x_i}\right)
 \label{strain},
\end{align}
while the elastic constants $C_{ijkl}$ and eigenstrain $\epsilon^*_{ij}$ can be expressed as:
\begin{align}
 C_{ijkl}(\vphi) &= \sum_{\alpha=1}^{N}C^{\alpha}_{ijkl}\phi_\alpha\\ \nonumber
 \epsilon^*_{ij}(\vphi) &= \sum_{\alpha=1}^{N}\epsilon^{*\alpha}_{ij}\phi_\alpha. 
\end{align}
To simplify the equations, without any loss of generality, it is additionally imposed that the eigenstrain exists only in precipitate phases, which makes the eigenstrain in the matrix phases $\epsilon^{* \beta}_{ij}=0$.
\par

Finally, mechanical equilibrium is imposed for the solution of the displacement fields. Here, the solution methodologies differ among the different solver modules. For the GP formulation implemented using FD with MPI, the solution to the mechanical equilibrium state is achieved iteratively by solving the damped wave equation written as: 
\begin{align}
  \rho\frac{d^2\mathbf{u}}{dt^2} + b\frac{d\mathbf{u}}{dt} &= \nabla \cdot \boldsymbol{\sigma}, 
  \label{mech_equilibrium_1ppt}
\end{align}
where $\boldsymbol{\sigma}$ is the Cauchy stress tensor. Eq.~(\ref{mech_equilibrium_1ppt}) is solved until the equilibrium is reached, i.e $\nabla\cdot\boldsymbol{\sigma}=\mathbf{0}$. The density $\rho$ and damping coefficient $b$ are chosen to achieve convergence in the fastest possible time.

Similarly, for the OpenFOAM multiphysics module, the displacement fields are solved iteratively using the wave equation, wherein 
Eq.~(\ref{mech_equilibrium_1ppt}) becomes:
\begin{align}
  \rho\frac{d^2\mathbf{u}}{dt^2} &= \nabla \cdot \boldsymbol{\sigma}. 
  \label{mech_equilibrium_1ppt_of}
\end{align}

For the KKS formulation, the same mechanical equilibrium problem is solved in Fourier space implemented in CUDA. Presently, only the case of homogeneous stiffness tensor is implemented. The algorithm is detailed in the following.

\subsubsection{Elastic equilibrium in a multiphase setting} \label{sec:2}
    For a generic multi-phase setting, where the order parameter $\vphi = \{\phi_1, \phi_2, \hdots, \phi_N\}$ ($N$ being the number of phases in the system), the elastic stress can be expressed as:
    \begin{equation}
        \label{eqn:42}
        \sigma_{ij} = C_{ijkl}(\vphi)e_{kl}(\vphi),
    \end{equation}
    where $C_{ijkl}(\vphi)$ is given as
    \begin{equation}
        \label{eqn:43}
        C_{ijkl}(\vphi) = \bar{C}_{ijkl} + \Delta C_{ijkl}(\vphi),
    \end{equation}
    \begin{equation}
        \label{eqn:44}
        \Rightarrow C_{ijkl}(\vphi) = \bar{C}_{ijkl} + \sum_\alpha^N \Delta C_{ijkl}^\alpha h_\alpha(\vphi),
    \end{equation}    
    with $e_{kl}(\vphi)$ as,
    \begin{equation}
        \label{eqn:45}
        e_{kl}(\vphi) = \bar{\epsilon}_{kl} + \epsilon_{kl} - \sum_\alpha^N \epsilon_{kl}^{*\alpha} h_\alpha(\vphi).
    \end{equation}
    The term $h_\alpha(\vphi)$ in Eq.~\eqref{eqn:44} and \eqref{eqn:45} is elaborated in Eq.~\eqref{eqn:hphi}. $\bar{\epsilon}_{kl}$ corresponds to the homogeneous strain in the domain, while the periodic strain is represented by $\epsilon_{kl}$. Moreover, the homogeneous part of the elastic constant $\bar{C}_{ijkl}$ can be written as,
    \begin{equation}
        \label{eqn:47}
        \bar{C}_{ijkl} = \frac{1}{N} \sum_\alpha^N C_{ijkl}^\alpha,
    \end{equation}
    where $C_{ijkl}^\alpha$ are the respective elastic constants of the individual phases, and the heterogeneous part can then be obtained as
    \begin{equation}
        \label{eqn:48}
        \Delta C_{ijkl}^\alpha = C_{ijkl}^\alpha - \bar{C}_{ijkl}.
    \end{equation}
    
    Since the mechanical equilibrium is dictated by
    \[ \frac{\partial \sigma_{ij}}{\partial x_j} = 0, \]
    this implies
    \begin{equation}
        \label{eqn:49}
        \frac{\partial}{\partial x_j} \left[ C_{ijkl}(\vphi)e_{kl}(\vphi) \right] = 0.
    \end{equation}
    \begin{equation}
        \label{eqn:50}
        \Rightarrow \frac{\partial}{\partial x_j} \left[ \left( \bar{C}_{ijkl} + \Delta C_{ijkl}(\vphi) \right) e_{kl}(\vphi) \right] = 0.
    \end{equation}
    \begin{equation}
        \label{eqn:51}
        \Rightarrow \frac{\partial}{\partial x_j} \left[ \bar{C}_{ijkl} e_{kl}(\vphi) \right] = -\frac{\partial}{\partial x_j} \left[ \Delta C_{ijkl}(\vphi) e_{kl}(\vphi) \right].
    \end{equation}
    Writing $\Delta C_{ijkl}(\vphi)e_{kl}(\vphi)$ as $\Delta \sigma_{ij}$ and expanding $e_{kl}(\vphi)$ using equation \eqref{eqn:45} (and, subsequently, substituting $\epsilon_{kl}$ by Eq.\ref{strain}), one obtains,
    \begin{equation}
        \label{eqn:52}
        \bar{C}_{ijkl} \frac{\partial^2 u_k}{\partial x_j \partial x_l} - \bar{C}_{ijkl} \sum_\alpha^N \epsilon_{kl}^{0,\alpha} \frac{\partial h(\phi_\alpha)}{\partial x_j} = - \frac{\partial \Delta \sigma_{ij}}{\partial x_j}.
    \end{equation}
    \begin{equation}
        \label{eqn:53}
        \Rightarrow \bar{C}_{ijkl} \frac{\partial^2 u_k}{\partial x_j \partial x_l} = \bar{C}_{ijkl} \sum_\alpha^N \epsilon_{kl}^{*\alpha} \frac{\partial h_\alpha(\vphi)}{\partial x_j} - \frac{\partial \Delta \sigma_{ij}}{\partial x_j}.
    \end{equation}
    Taking the Fourier transform on both sides of Eq.~\eqref{eqn:53}, one obtains
    \begin{equation}
        \label{eqn:54}
        -\bar{C}_{ijkl}\mathbf{k}_j\mathbf{k}_l \widehat{u}_k = \iota\bar{C}_{ijkl}\sum_\alpha^N \widehat{\epsilon_{kl}^{*\alpha} h_\alpha(\vphi)}\mathbf{k}_j - \iota\widehat{\Delta\sigma}_{ij}\mathbf{k}_j.
    \end{equation}
    \begin{equation}
        \label{eqn:55}
        \Rightarrow G_{ik}^{-1} \widehat{u}_k = -\iota \left[ \bar{C}_{ijkl}\sum_\alpha^N \widehat{\epsilon_{kl}^{*\alpha} h_\alpha(\vphi)}\mathbf{k}_j - \widehat{\Delta\sigma}_{ij}\mathbf{k}_j \right].
    \end{equation}
    \begin{equation}
        \label{eqn:56}
        \Rightarrow \widehat{u}_k = -\iota G_{ik}  \left[ \bar{C}_{ijkl}\sum_\alpha^N \widehat{\epsilon_{kl}^{*\alpha} h_\alpha(\vphi)}\mathbf{k}_j - \widehat{\Delta\sigma}_{ij}\mathbf{k}_j \right].
    \end{equation}
    Here Eq.~\eqref{eqn:56} must be iteratively evolved since $\Delta \sigma_{ij}$ is an unknown which depends on $\widehat{u_k}$. However, for the homogeneous case where $\Delta C_{ijkl}^\alpha=0$, the second term in the brackets on the R.H.S. in the preceding equation is zero. Hence, the value of the periodic displacement can be immediately inverted as, 
    \begin{equation}
     \widehat{u}_k = -\iota G_{ik} \bar{C}_{ijkl}\sum_\alpha^N \widehat{\epsilon_{kl}^{*\alpha} h_\alpha(\vphi)}\mathbf{k}_j
    \end{equation}
    The iterative solution for the inhomogeneous case can also be derived where the homogeneous strain $\bar{\epsilon}_{kl}$ that is contained 
    in the term $\Delta \sigma_{ij}$ becomes relevant and needs to be evaluated. We will reserve that discussion for later once the solvers with CUDA are upgraded with the respective module.

\subsection{Orientation field for multigrain solidification simulations}
\label{orientation}
The OpenFOAM variant of the solver has a provision for the simulation of multiple grains. For this, the GP model \citep{Abhik2012} has been coupled with the orientation field model \citep{henry2012orientation}. Here, the orientation field $\theta$ 
is introduced to represent the variation of the grain orientation in the simulation domain. An additional energetic term $f_{ori}(\theta)$ is added to the integrand at the right-hand side of Eq.~(\ref{eq:grand_eqn}), where, following \citep{henry2012orientation}, $f_{ori}(\theta) = {\epsilon} \nu^2 p(\phi) |\nabla \theta|^2$ has been adopted. Here, $\nu$ is a scaling factor associated with solid-liquid interface and grain boundaries, which is taken as 0.5, and $p(\phi)$ is an interpolation function. Upon taking the variational derivative with the new energetic contribution, the evolution Eq.~(\ref{phi_eqn}) becomes:
\begin{align}
\tau \epsilon \frac{\partial \phi_{\alpha}}{\partial t} &= \epsilon\left(\nabla \cdot \frac{\partial a(\phi, \nabla \phi)}{\partial \nabla \phi_{\alpha}}
-\frac{\partial a(\phi, \nabla \phi)}{\partial \phi_{\alpha}}\right) 
-\frac{1}{\epsilon} \frac{\partial w(\phi)}{\partial \phi_{\alpha}} \nonumber\\ &-\frac{\partial \psi(T, \mu, \phi)}{\partial \phi_{\alpha}}-\lambda
- \epsilon \nu^2 p'(\phi) |\nabla \theta|^2.
\end{align}

\par
Similarly, the evolution equation for the orientation-field parameter can be obtained as,
\begin{equation}
    \dot{\theta} = 2 M_{\theta} \nu^2 \left( \nabla \cdot ( p(\phi) \nabla \theta ) \right),
\end{equation}
where $p(\phi)$ is an interpolation function, and $M_{\theta}$ is a mobility function \citep{henry2012orientation}, given by,
\begin{align}
    p(\phi) &= \frac{7 \phi^3 - 6 \phi^4}{(1-\phi)^2}, \\\        
    M_{\theta} &= M_{scale} \epsilon (1- \phi)^2 ( 1- h(\phi)). 
\end{align}

\par
The parameter $M_{scale}$ is typically $\leq$ 1. $M_{scale}$ influences the time scale of the simulation and can be utilized to tailor the time step. However, reducing $M_{scale}$ below a certain value can deteriorate the accuracy of calculating $\theta$. While the above formulation is appropriate for 2D simulations, for 3D, a separate quaternion-based framework based on \cite{granasy2005el, granasy2005mse, granasy2008} has been implemented. In a later work, we will elaborate on this.

\section{Thermodynamic coupling}
\label{sec:thermodynamic}
Different formulations for the thermodynamics related to the free energy densities of the respective phases exist in the solver modules. These formalisms may be accessed through specific keys in the input file. Here, the module containing the GP formalism using FD with MPI has the most flexibility regarding the choice of thermodynamics. The formalisms may be broadly categorized as approximate and exact. The approximate methods involve the construction of parabolic expressions for the free energy density that are assumed to have the form,
\begin{align}
f^{\alpha}\left(\vc\right) &= \dfrac{1}{V_m}\Big(\sum_{i<j}^{K,K}A_{ij}^{\alpha}c_ic_j + \sum_j^{K} B_j^{\alpha} c_j + C^{\alpha}\Big),
\label{free_energy_density}
\end{align}
where $A^{\alpha}, B^{\alpha}$ and $C^{\alpha}$ are the respective coefficients, and $V_m$ is the molar volume that is assumed equal for all the phases in the present discourse. Here, there are different choices that the key Function\_F can activate with the values (1, 3, 4) in the input file. The coefficients $A_{ij}$ are linked to the curvatures of the free energy curves and hence are related to an effective susceptibility. This matrix can be directly obtained from the databases for real material simulations by using the information of the free energies $G^{\alpha}$ for each of the phases as $\left(\dfrac{\partial^2 G^{\alpha}}{\partial c_i\partial c_j}\right)_{\vc^{\alpha}_{eq}}$ where $\vc^{\alpha}_{eq}$ are the equilibrium compositions of the components in the $\alpha$-phase that is in equilibrium with a given phase, which is presently chosen to be the liquid. Assuming $B_j^{l}$ and $C^{l}$ to be zero for the liquid phase, the conditions of equilibrium (equal diffusion potential and equal grand-potentials) allow us to determine the other coefficients $B_j^{\alpha}$ and $C^{\alpha}$ for any of the solid-phases, such that the equilibrium between the solid phase $\alpha$ and the liquid (just for discussion purposes, in general, any phase can be chosen) phase is reproduced at the respective equilibrium compositions. Thus, one can derive for a given temperature $T_{eq}$:
\begin{align}
 B_{j}^{\alpha} &= 2(A_{jj}^{l}c_{eq,j}^{l} - A_{jj}^{\alpha}c_{eq,j}^{\alpha}) 
 + \sum_{j\neq i}\left(A_{ij}^{\alpha} c_{eq,i}^{\alpha} - A_{ij}^{l} c_{eq,i}^{l}\right)
 \label{Coeff_B}\\
 C^{\alpha}    &= \sum_{i\leq j }\left(A_{ij}^{\alpha} c_{eq,i}^{\alpha}c_{eq,j}^{\alpha} - 
 A_{ij}^{l} c_{eq,i}^{l}c_{eq,j}^{l}\right)
 \label{Coeff_C}
\end{align}

\par
For the value of Function\_F=1 it refers to the case where a parabolic free energy density is chosen of the type described in \citep{CHOUDHURY2015287, PhysRevE.91.022407}. While for Function\_F=1 the inputs are directly $A_{ij}^\alpha$, the equilibrium compositions $c_{eq,i}$ and the slopes of the co-existence lines from which the respective values of $B_i^\alpha$ and $C^\alpha$ are determined as functions of temperature, for Function\_F=3,4 the coupling to thermodynamics is more direct. For Function\_F=3, the unencrypted .tdb file containing the information about the thermodynamic functions is read using the pyCALPHAD \citep{Otis-2017} Python-based tool, and the corresponding functions related to the free energies, diffusion potentials, and the Hessian functions are constructed symbolically using SymPy \citep{10.7717/peerj-cs.103} and converted into header files using codegen module of python that can be directly utilized in the solver modules. Thus, all the information related to evaluating the coefficients in the parabolic formulation happens implicitly, and the only interaction is with the .tdb file.

\par
Similarly, Function\_F=4 creates a direct interface with the thermodynamics, except here, the function executes on information 
extracted from .tdb files using APIs of ThermoCalC and Pandat.
Because the direct information related to the free energy forms is not required in this function, it allows for the treatment of alloys for which only encrypted databases exist. In this case, firstly, the equilibrium compositions between any two phases under consideration are computed from the thermodynamic APIs as a function of temperature by performing a line calculation for the chosen alloy composition and stored in .csv files. Further, the information related to phase-specific Hessian matrix components is computed using the thermodynamic APIs at the equilibrium compositions computed previously for every temperature in the line calculation. For each phase, a separate .csv file is created containing information about the components of the Hessian matrix at different temperatures. These files need to be present at the time of execution of the code. The information read from these files is transformed into continuous temperature functions using spline fits provided by the GSL-GNU scientific libraries. These functions are thereafter utilized for having continuous descriptions of the coefficients of $A_{ij}$, $B_j$, and $C$ as functions of temperature in the solver.

\begin{figure}[!htbp]
    \centering
    \includegraphics[width=\textwidth]{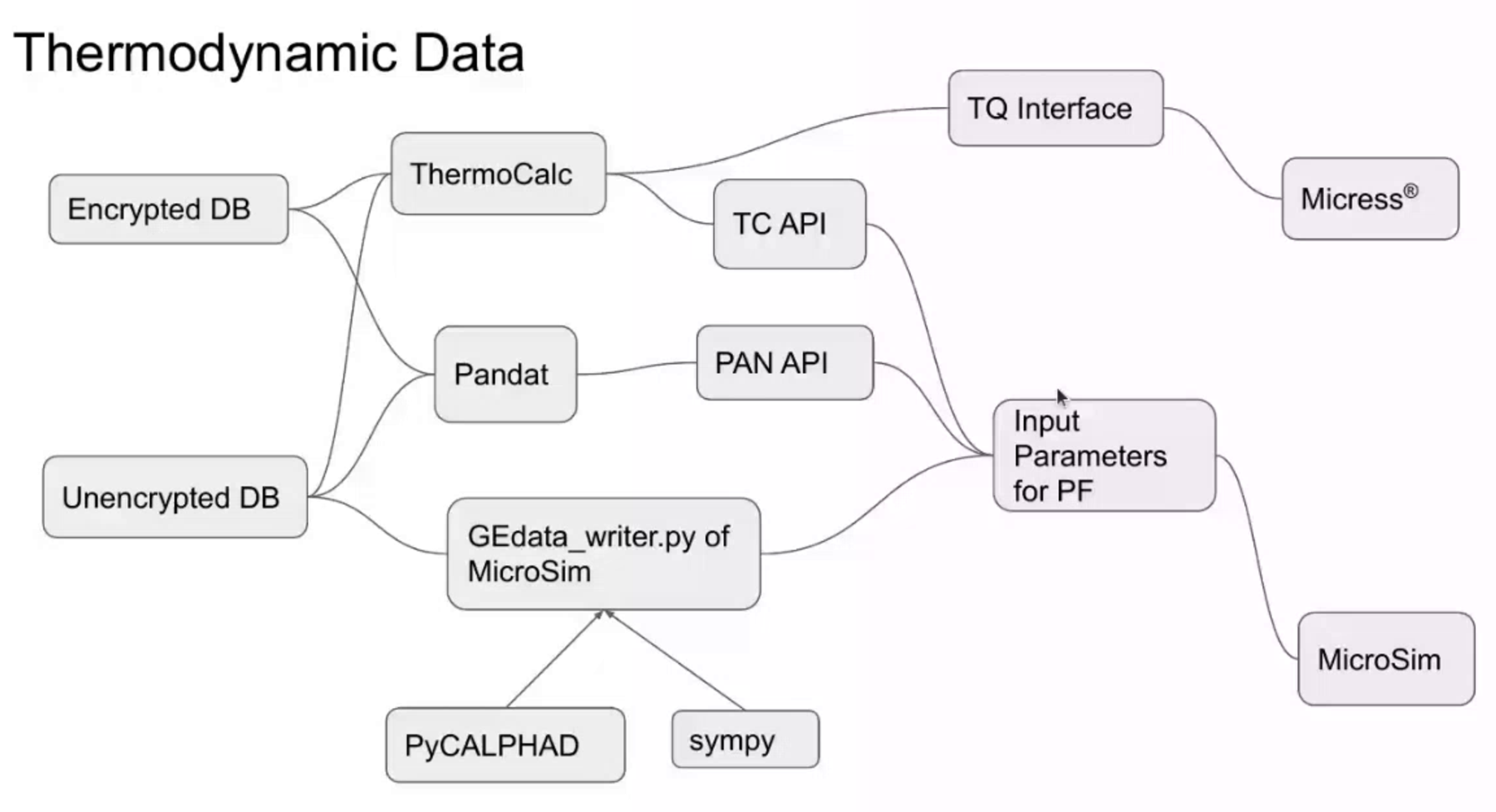}
    \caption{Flowchart of the process of linking thermodynamic database with MICROSIM.}
    \label{thermo}
\end{figure}

\par
Finally, the possibility exists for a direct utilization of thermodynamic information available from unencrypted .tdb files. This is activated using Function\_F=2. The structure of this function is very similar to the Function\_F=3, where the functions related to the free energies, diffusion potentials, and the Hessian are converted symbolically into header files using the pyCALPHAD \citep{Otis-2017} interface. However, here the direct expressions are utilized without any approximations. Since the thermodynamic functions may not be convex through the interval and, in general, are non-linear, it is important to evaluate the function $\vc\left(\vmu\right)$ through a numerical inversion route as an analytical inversion may not be possible. For this, the functions in gsl/gsl\_multiroots.h are utilized as these are heavily optimized for solving these non-linear problems. In addition, processor-specific optimization, such as the intel-oneapi-mkl libraries, may be linked for superior performance. The numerical evaluation of the $\vc\left(\vmu\right)$ allows the function to be locally treated in a convex fashion, and thereby, the driving forces can be calculated. To limit the evaluation of the function $\mathbf{c}\left(\vmu\right)$ to the local equilibrium valley of a given phase, it is necessary to provide a guess for the iteration routines called from the GSL library. These guesses are provided in the tuple "c\_guess" which can be interpreted in the same manner as ``c\_eq" or ``c\_fill" in the input file. Different iterative routines from the gsl\_multiroots library may be chosen based on convergence and the particular problem. The hybridj and hybridsj solvers work the best when it comes to global convergence behavior.

Since these are non-linear functions, in contrast to the parabolic functions, the matrices $\left[\dfrac{\partial c_i}{\partial \mu_j}\right]$ are composition dependent. Therefore, although the $\vmu$ update equation (Eqn. \ref{Mu_explicit_temperature}) is derived from the condition $c_i = \sum_\alpha \vc_i^{\alpha}\left(\mathbf{\vmu}\right)h_\alpha\left(\mathbf{\phi}\right)$ along with the mass conservation equation, the diffusion potential evolution equation linearizes the dependencies on each of the variables. This can introduce errors in the composition constraint for certain cases. To avoid this, an iterative scheme is devised to correct the diffusion potential update if required. The iterative scheme  
 is written as follows and is detailed in FunctionF\_2.h. The mass conservation equation for each of the components follows
 \begin{align}
  \dfrac{\partial c_i}{\partial t} &= \nabla \cdot \left(M_{ij}\nabla\mu_j\right),
  \label{Mass_conservation}
 \end{align}
 
 where $M_{ij}=\sum_{\alpha=1}^{N}\left[D_{ij}^{\alpha}\right]\left[\dfrac{\partial c_i^{\alpha}}{\partial \mu_{ij}}\right]h_\alpha\left(\vphi\right)$. 
 This can be converted into a diffusion potential update equation by
 linearising based on the dependencies on the individual variables as detailed before in Eqn.\ref{Mu_explicit_temperature},
using the composition constraint $c_i = \sum_{\alpha=1}^{N} c_i^\alpha\left(\vmu\right)h_\alpha\left(\vphi\right)$. However, as mentioned when the matrices 
$\left[\dfrac{\partial c_i^{\alpha}}{\partial \mu_{j}}\right]$ become composition dependent, there can be errors in the composition constraint, depending on the changes in the other variables in a given time step. To account for this, a hybrid approach is devised in which the composition is updated using the mass conservation equation in Eqn.\ref{Mass_conservation}, along with the explicit diffusion potential update in Eqn.\ref{Mu_explicit_temperature}
and the following correction scheme is implemented.
 \begin{enumerate}
  \item Step 1: Update composition using Eqn.\ref{Mass_conservation} and diffusion potential using Eqn.\ref{Mu_explicit_temperature}.
  \item Step 2: Calculate the error in each composition of each component as  $\left\lbrace\Delta c_i\right\rbrace = c_i - \sum_{\alpha=1}^{N} c_i^\alpha\left(\vmu\right)h_\alpha\left(\vphi\right)$. For this, the functions that compute the phase compositions using the GSL functions
  are utilized.
  \item Step 3: Compute matrices $\left[\dfrac{\partial c_i^{\alpha}}{\partial \mu_{j}}\right]$ for each phase and calculate the matrix 
  $\left[\dfrac{\partial c_i}{\partial \mu_j}\right] = \sum_{\alpha=1}^{N}\left[\dfrac{\partial c_i^{\alpha}}{\partial \mu_{j}}\right]h_\alpha\left(\vphi\right)$.
  \item Step 3: Compute correction in $\vmu$ by $\left\lbrace\Delta \mu_i\right\rbrace = \left[\dfrac{\partial c_i}{\partial \mu_{j}}\right]^{-1}\left\lbrace\Delta c_i\right\rbrace$.
  \item Update diffusion potential as $\left\lbrace\mu_i\right\rbrace = \left\lbrace\mu_i\right\rbrace + \left\lbrace\Delta \mu_i\right\rbrace$.
  \item Check tolerance and continue from Step 2 if not satisfied.
 \end{enumerate}

One might recognize that the procedure illustrated is a Newton iteration in the variable $\vmu$, starting from the explicit update from Eqn.\ref{Mu_explicit_temperature} as the initial guess. This greatly simplifies the iterations from the classical version in typical Kim-Kim-Suzuki models as the number of iterations is greatly reduced. 
Because of this, it is also prudent for most cases to avoid this iteration altogether, as the magnitude of errors is already very small, and it avoids the computation of 
the matrices $\left[\dfrac{\partial c_i^{\alpha}}{\partial \mu_{j}}\right]$ that involve costly matrix inversion operations.

While this added procedure increases computational costs, the existence of this function (Function\_F=2) provides an avenue for benchmarking the other approximate descriptions and quantifying the magnitude of the errors. A figurative description of the different function formalisms and comparison with other commercial codes is provided in Fig.~\ref{thermo}.

\section{Numerical implementation}
\label{sec:implement}

\subsection{CPU (MPI) implementation}

\begin{figure}[!htbp]
    \centering
    \includegraphics[width=60mm]{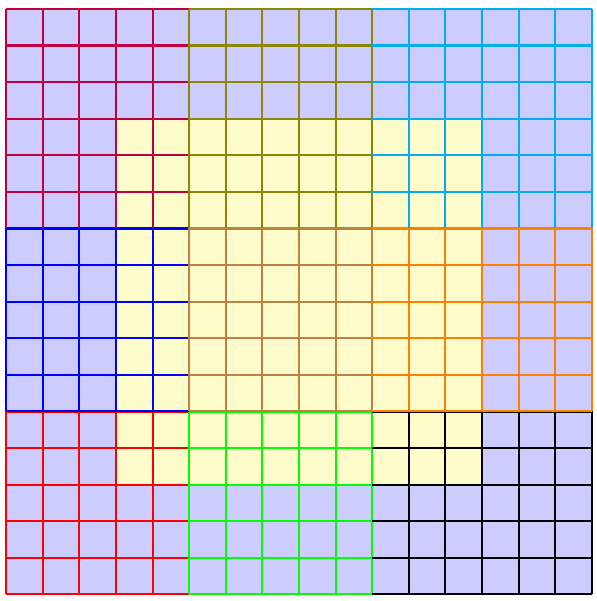}
    \caption{Schematic showing the domain decomposition among the different processors.}
    \label{Domain_decomposition}
\end{figure}

\begin{figure}[!htbp]
    \centering
    \includegraphics[width=100mm]{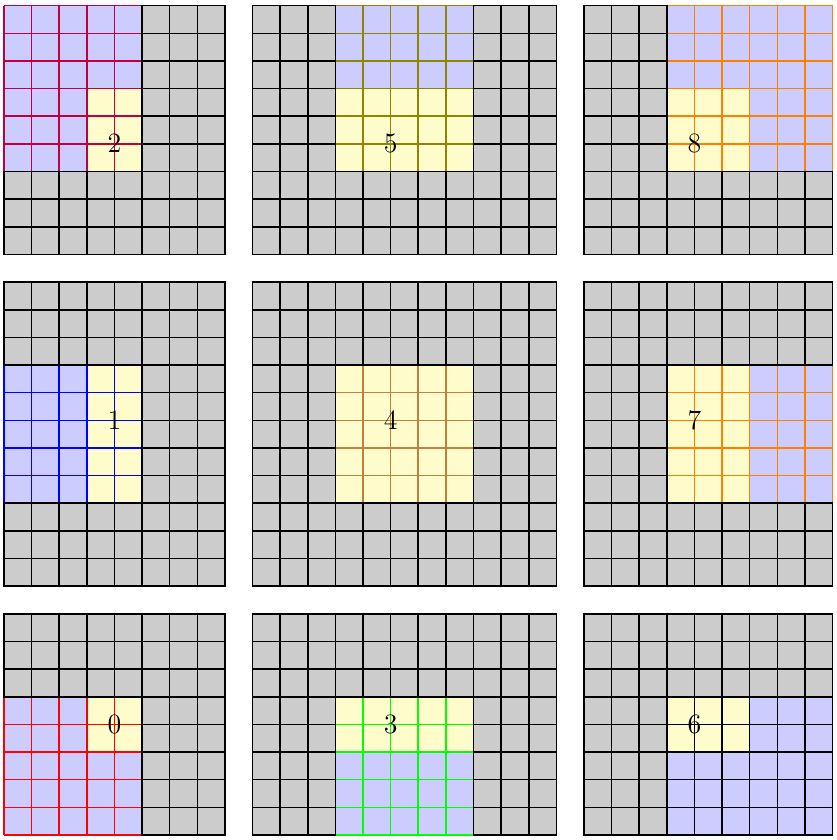}
    \caption{Figure showing the domains with each processor along with the buffer points added for communication between processors.}
    \label{Processor_compute_parts}
\end{figure}

The CPU solver employing the GP formulation is parallelized using a simple domain decomposition using the MPI interface. The domain decomposition scheme is illustrated in Fig.~\ref{Domain_decomposition}. The original domain consists of the real domain points colored in yellow, surrounded by three layers of boundary buffer cells in each direction colored in blue. The decomposition of the domain results in processors that contain boundary cells and processors that consist of only real domain points. Fig.~\ref{Processor_compute_parts} shows the compute domains of each processor along with the buffer points. The colored parts correspond to the grid points in the original domain. The grids colored in black are buffer points added for inter-processor communication. The numbering scheme of the processors is also highlighted where, for a given number of workers, the rank of the processors follows a linear-indexed scheme where the processor count along the y-direction increases the fastest (also known as column-major ordering).

\par
\begin{figure}[!htbp]
 \centering
	\subfloat[\label{Buffer_inter_processor_communication_a}]{\includegraphics[width=40mm]{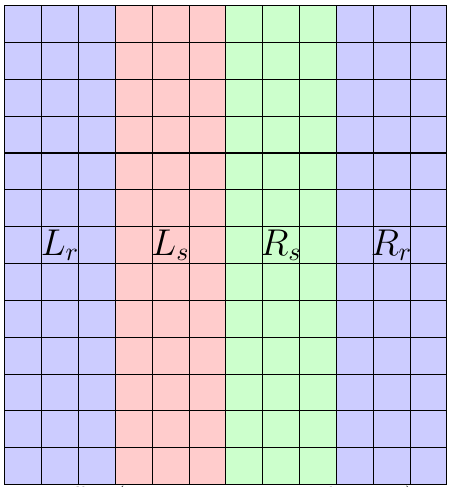}}
	\subfloat[\label{Buffer_inter_processor_communication_b}]{\includegraphics[width=40mm]{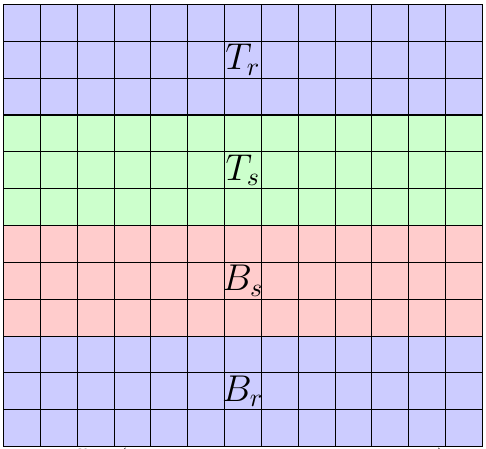}}
	\caption{Temporary buffers for a single pass (a) left-right and (b) top-bottom communication between the processors.}
	\label{Buffer_inter_processor_communication}    
\end{figure}

Fig.~\ref{Buffer_inter_processor_communication} shows the structure of the buffer created for the transfer of information between processors as a single chunk of memory. Since the data structure is in the column-major format, the communication between processors in the x-direction (Left-Right communication) is performed by creating contiguous MPI\_derived\_type containing all the elements of the structure (struct fields) and the amount of memory corresponding to the grids colored in red and green in  Fig.~\ref{Buffer_inter_processor_communication}(a) are sent to the left and right processors respectively. For the communication between the top and bottom processors, similarly, a buffer is created as shown in Fig.~\ref{Buffer_inter_processor_communication}(b). Since the data to be transferred is now in terms of rows instead of columns, a separate MPI\_derived\_type using the MPI\_Vector type is created with the appropriate data stride, such that the entire chunks of memory as shown in colors red and green, can be transferred to the bottom and top processors respectively in a single communication. The colors depicted in blue in Fig.~\ref{Buffer_inter_processor_communication}(a) and (b) are utilized for receiving information from the left/right and top/bottom processors, respectively. The letters $L_r$ $R_r$ stand for Left-receive and Right-receive respectively. Similarly $B_r$ and $T_r$ are labels for grids that will receive data from the bottom and top processors, respectively. A similar terminology is used for the letters $L_s$,$R_s$,$B_s$, and $T_s$ that stand for the sending operations corresponding to the subscript ``s". A similar strategy is adapted for domain decomposition in 3D, where the domain is decomposed into cuboids 
of different sizes depending on the number of allocated processors.

\subsection{GPU implementation}
MICROSIM offers multiple solvers that can leverage GPUs as hardware accelerators for offloading computations from the CPU. GPUs are extensive collections of low-powered cores that can communicate with one another at a faster rate and speed than a group of CPU cores. This feature lends itself well to calculations on stencils and ``vectorizable" operations such as for-loop iterations. To use this acceleration, three modes of implementation have been used - CUDA, OpenCL, and AMReX. 

\subsubsection{CUDA implementation}
CUDA is a programming framework maintained by NVIDIA and is an efficient platform for developing GPU-enabled scientific code. Using CUDA, the movement of data from the CPU (termed the host-side) to the GPU (termed the device-side) and back, the parallelization of algorithms to run on device-side data, and optimization of the code to run on different generations of GPUs are all streamlined. 

\par
A direct, programmer-friendly method to offload calculations to the GPU is to convert domain-wide iterators to `kernels,' which are C functions that are executed in parallel by numerous threads. Each thread operates only on a single element of the domain, after which it is killed. CUDA calls large blocks of such threads in succession to iterate through the entire data. Most modern GPUs are capable of launching blocks of up to 2048 such threads. 

\begin{figure}[!htbp]
    \centering
    \includegraphics[width=100mm]{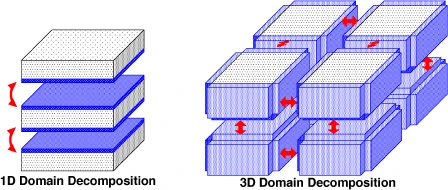}
	\caption{Schematic of domain decomposition for the CUDA solver. The white blocks are computational cells. The blue layers serve as buffers for communication between GPUs, and also double up as boundary layers at the physical boundary of the simulation. \citep{gpudecompfigure}}
	\label{gpudecompFigure}    
\end{figure}

\par
The amount of memory available per GPU is limited, necessitating either pooling multiple GPUs together or distributing work efficiently between them. MICROSIM uses the latter approach, using a distributed approach as shown in Fig.~\ref{gpudecompFigure}. To enable communication between GPUs on separate nodes, CUDA-aware MPI is necessary. In MICROSIM, the CUDA solver has been built and tested using OpenMPI-4.0.5. In addition to OpenMPI, UCX is required for direct GPU-to-GPU data transfer. It is best to compile using the OpenMPI libraries packaged in NVIDIA's HPC-SDK for ease of usage.

\subsubsection{OpenCL implementation}
OpenCL is a cross-platform language compatible across CPUs and GPUs. It can be executed on various devices, providing a level of abstraction from the underlying hardware architecture. This makes it suitable for heterogeneous systems. OpenCL implementation has host code and device code. Host code consists of the API calls for device selection, buffer allocation and kernel calls. The device code consists of kernels, which are functions that execute in parallel.

\par
\begin{figure}[!htbp]
    \centering
    \includegraphics[width=70mm]{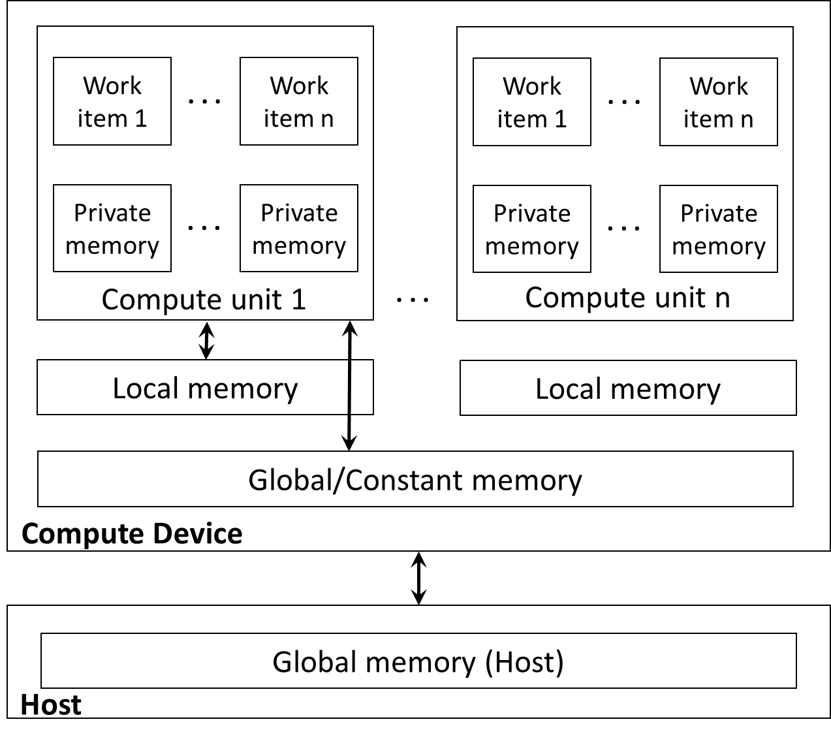}
    \caption{OpenCL memory hierarchy model \citep{OpenCL}.}
    \label{openclmemory}
\end{figure}

The memory model of the OpenCL is represented in Fig.~\ref{openclmemory}. The memory model supports different types of memory spaces, such as global, local, and private. The computational domain is loaded onto the GPU, and memory management is left to the GPU. The work items (equivalent to threads in Fig.~\ref{openclmemory}) execute in work groups, and these can execute in parallel on the available processing units. The same operation is performed on different elements of data concurrently. For file writing, the data is written to disk after reading it from the GPU, which involves a GPU to CPU communication.

\begin{figure}[!htbp]
    \centering
    \includegraphics[width=80mm]{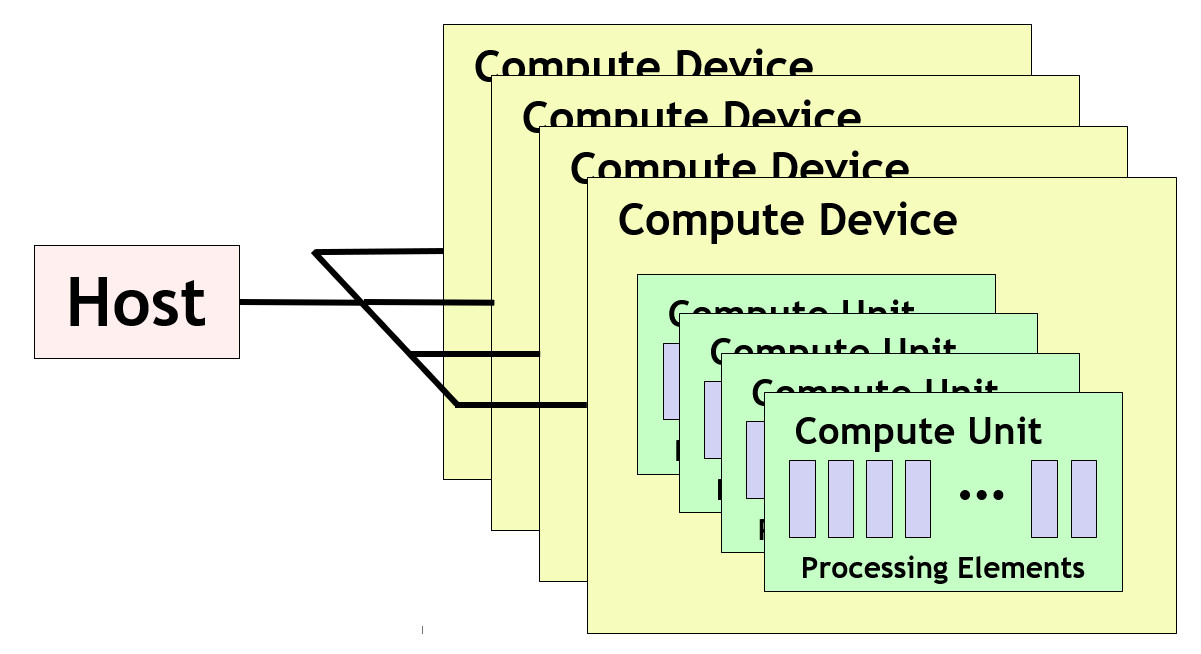}
    \caption{OpenCL platform Model \citep{OpenCL}.}
    \label{openclplatform}
\end{figure}

\begin{figure}[!htbp]
    \centering
    \includegraphics[width=\textwidth]{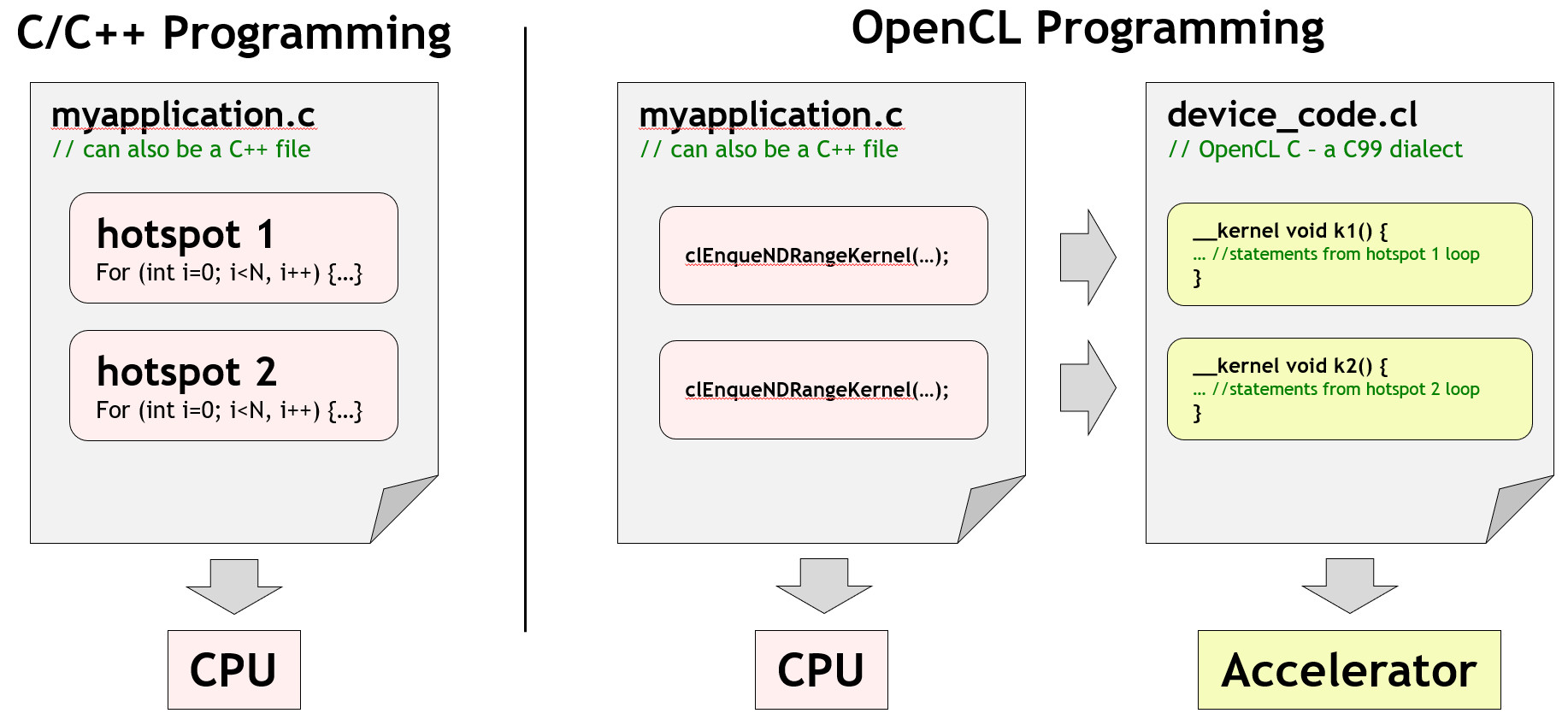}
    \caption{Traditional programming vs OpenCL programming model \citep{OpenCL}.}
    \label{openclkernels}
\end{figure}

\begin{figure}[!htbp]
    \centering
    \includegraphics[width=\textwidth]{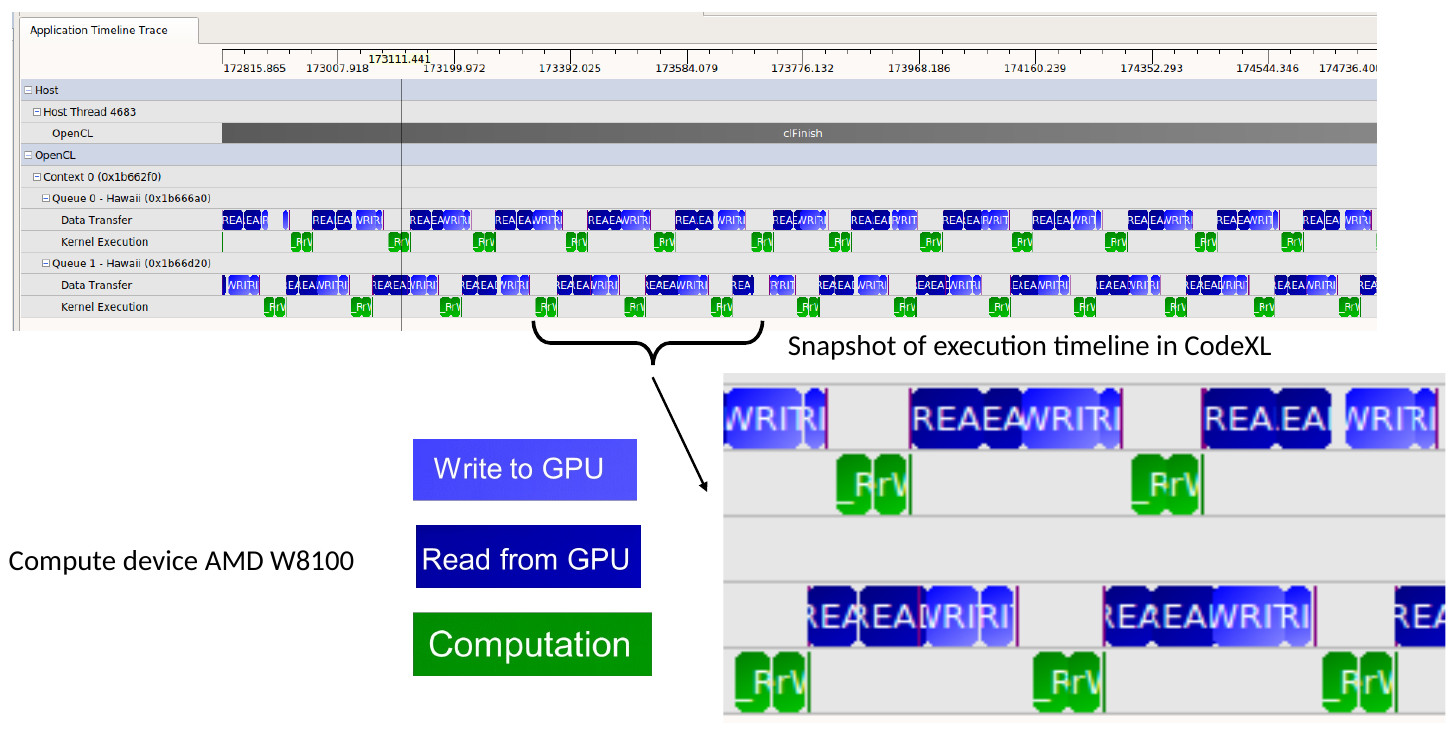}
    \caption{Optimization of OpenCL tasks by interspersing offloading of information with execution operations.}
    \label{openclwrite}
\end{figure}

\par
Further, we elaborate on the subroutines utilized for the input/output and execution operations. Firstly, the list of platforms (CPUs and GPUs) has been obtained using \texttt{clGetPlatformIDs}. The list of GPUs within each platform has been acquired using \texttt{clGetDeviceIDs}. This allows the allocation of specific tasks to the appropriate GPUs. The computational domain has been decomposed into one-dimensional subdomains (using MPI) using the CPUs available. Those subdomains are offloaded onto multiple GPUs from each CPU assigned to them. A schematic showing the connection between the CPU (host) and the GPUs (compute devices) has been displayed in Fig.~\ref{openclplatform}. \texttt{clCreateBuffer} allocates the buffer required for arrays. To write to a buffer object from host memory \texttt{clEnqueueWriteBuffer} has been used. The kernels can be created using \texttt{clCreateKernel} from existing functions, and arguments are set using \texttt{clSetKernelArg}. To execute a kernel on a device, \texttt{clEnqueueNDRangeKernel} command is utilized. A comparison between traditional and OpenCL programming is shown in Fig.~\ref{openclkernels}. The computation over each subdomain is performed as a work group on the assigned GPU. The command \texttt{clEnqueueReadBuffer} is used for reading from a buffer object onto the host memory. The computed data is sent from the GPU to the assigned CPU. The CPUs exchange information between them using MPI, and the data is written on the disk. Since the communication between the CPU and GPU is time-consuming, an optimization scheme for offloading of information interspersed with execution operations (see Fig.~\ref{openclwrite}), leading to maximum utilization of the available hardware is under development and will soon be part of MICROSIM.

\subsection{AMReX implementation}
This section covers the solver developed using the AMReX software framework \cite{amrex_link}. The available abstraction layer in AMReX (through handles) effectively hides the details of the architecture (hence the programming model) from the application, which enables the development of solvers targeting heterogeneous architectures on next-generation computing systems.
\par
\begin{figure}[!htbp]
    \centering
    \includegraphics[width=10cm]{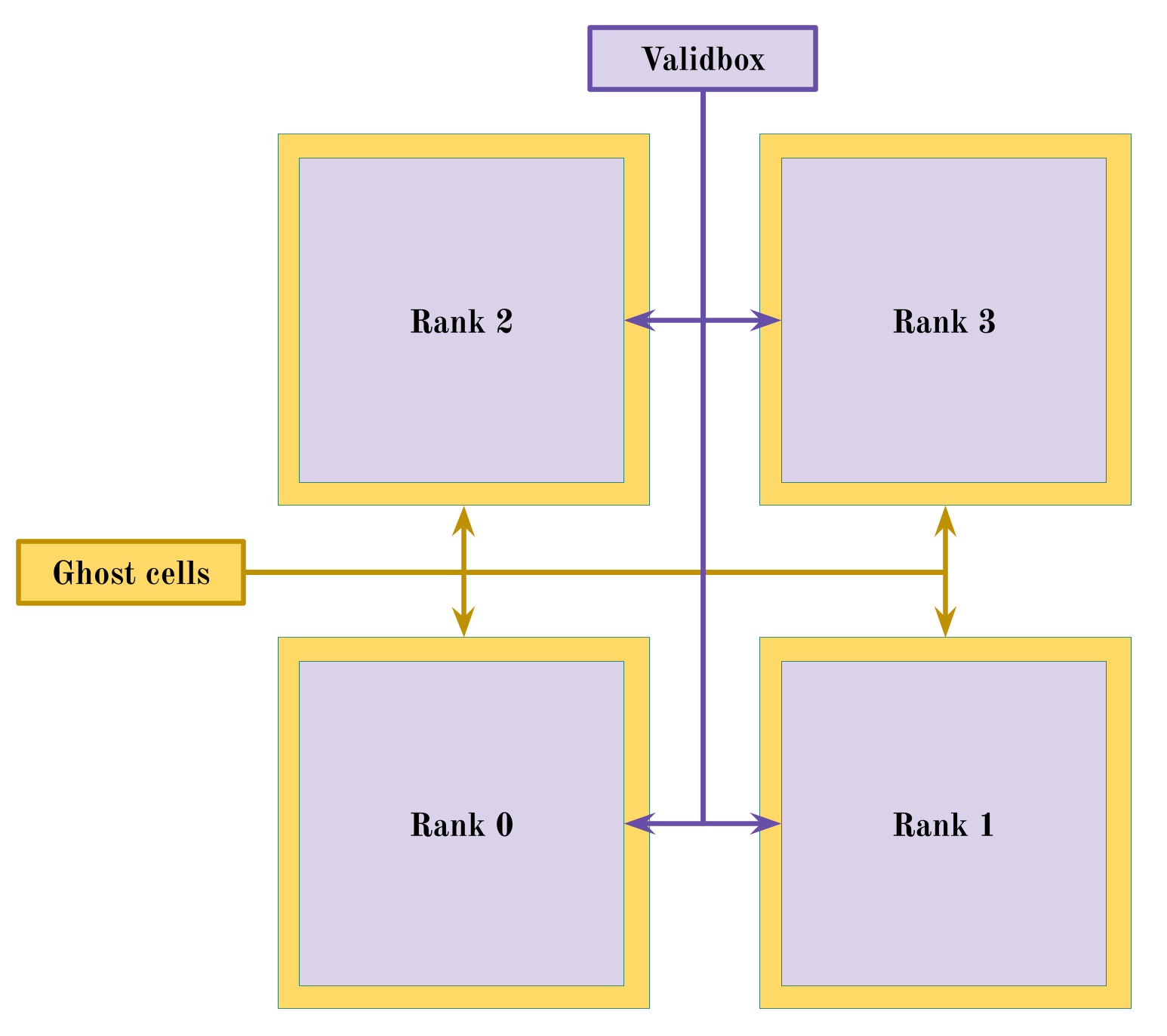}
    \caption{Distribution of a MultiFab across different ranks for 2D domain}
    \label{AMReX_MultiFab}
\end{figure}

AMReX provides handles for defining the details of the computational domain, such as index type, discretization, and coordinate system, using the Box and Geometry,  which results in the index space mapped to the problem domain. Additionally, AMReX implements the domain boundary condition through BCRec. The BoxArray is the outcome of splitting the original index space (box) into multiple boxes and holding the collection of boxes on a single level. DistributionMapping manages the requirements of the parallel solver by handling domain decomposition (through BoxArray) alongside the mapping of sub-domains to the process topology. A \texttt{MultiFab} stores the distributed data along with the ghost regions. The data access, processing, and update in the MultiFab is through the associated iterator \texttt{MFIter} by the respective processes that own the sub-domain data (its Validbox). The computation involves the ValidBox, and the ghost regions are only to access the data from the neighboring sub-domain. Fig.~\ref{AMReX_MultiFab} highlights the data distribution to respective processes, the ValidBox, and the ghost regions for a 2D domain. The \texttt{FillBoundary()} function updates the interior, periodic, and physical boundary ghost cells between successive iterations. In the MPI implementation, the ghost cells update requires the mechanism for data exchange across the neighbors in process topology through non-blocking calls executed by all the processes involved in the computation. AMReX handles this by a simple call to the \texttt{FillBoundary()} function, thus relieving the developer’s effort. In the data parallel region, AMReX provides a function template, \texttt{ParallelFor}, which handles the iteration space corresponding to the valid box as an argument, along with \texttt{AMREX\_GPU\_DEVICE}. When built with GPU support, \texttt{AMREX\_GPU\_DEVICE} acts as a macro and launches GPU kernels for the computations. For codes running only on CPUs, ParallelFor acts as a standard nested loop, thus making the code portable across heterogeneous architectures. 
 
\begin{figure}[!htbp]
    \centering
    \includegraphics[width=\textwidth]{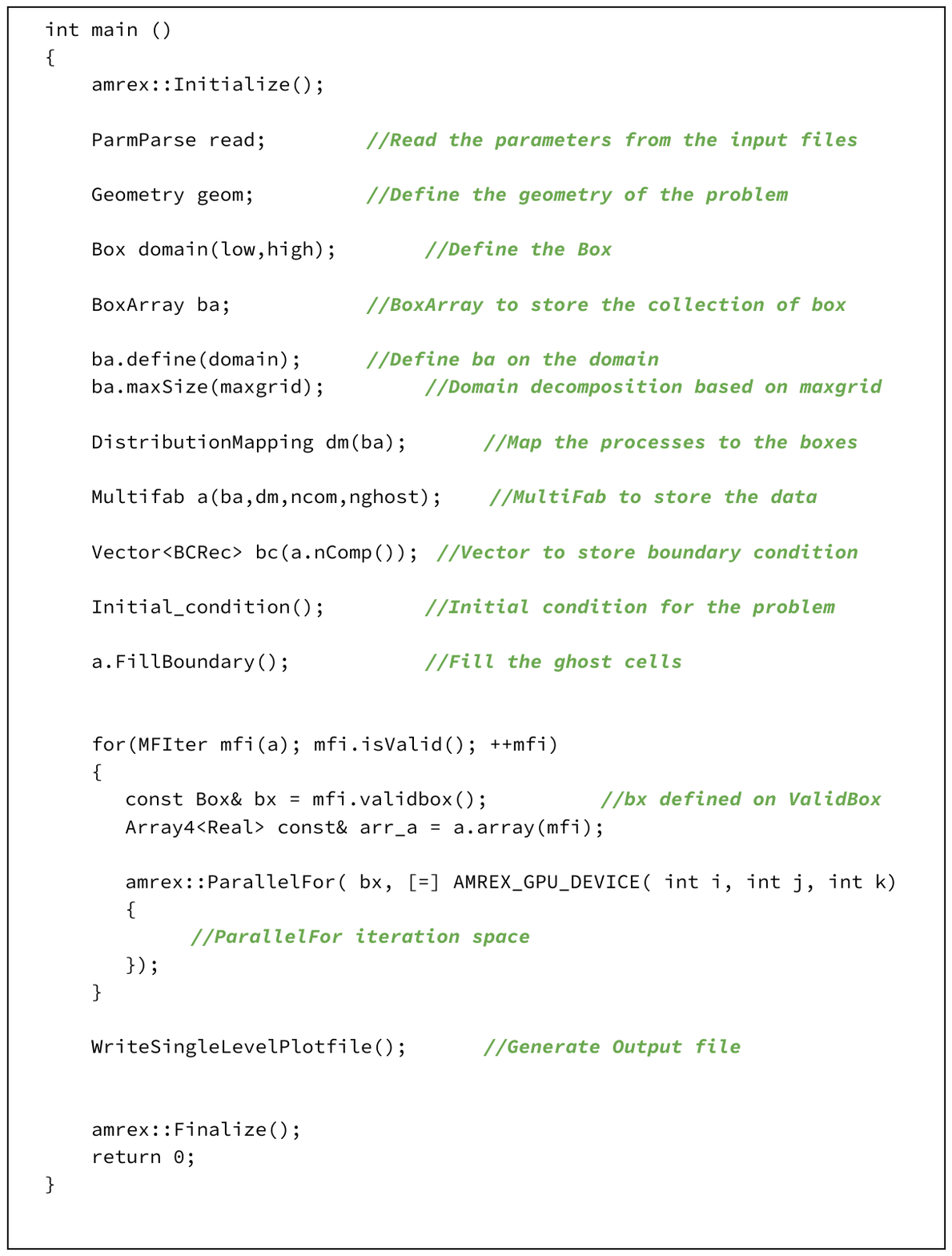}
    \caption{Pseudo-code for Level-0 AMReX Solver}
    \label{AMReX_pseudo}
\end{figure}
The AMReX phase-field solver in the MICROSIM repository does not involve the AMR algorithm. This computation is level `0’ as the keyword level is associated with the hierarchy of levels of refinement of the portions of the computational grid for enabling the adaptive mesh computation. Fig.~\ref{AMReX_pseudo} highlights the pseudo-code for the Level-0 AMReX solver. 
  
\subsection{OpenFOAM implementation}
\label{OF_impl}
OpenFOAM \citep{openfoam} is an open-source library for solving multi-physics problems using the finite volume method. It is written in C++, and the versions used in this work are compiled using either GCC (GNU Compiler Collection) or Intel OneAPI. The finite volume-based solvers have been constructed within the framework of OpenFOAM \citep{openfoam}. The governing equations have been solved using an implicit scheme.
\par
OpenFOAM comes with built-in parallelization strategies using MPI, with different possibilities for domain decomposition in 3D that can be activated using keywords `simple' and `scotch' in the decomposeParDict file present in the system folder of the relevant case file. This feature provides immense flexibility as the user is just concerned with implementing the respective physics-based model formulations rather than parallelization. 

\section{Input files}
\label{sec:infile}
The models discussed in Section~\ref{sec:formulation} have various parameters that need to be specified for running the solvers. The parameters are assigned their values within an input file, such as shown below:  

\begin{lstlisting}
##Geometrical dimensions of the simulation domain
DIMENSION = 2;
MESH_X = 200;
MESH_Y = 200;
MESH_Z = 1;
##Discretization, space and time
DELTA_X = 1e-8;
DELTA_Y = 1e-8;
DELTA_Z = 1e-8;
DELTA_t = 18e-9;
##Number of phases and composition
NUMPHASES = 2;
NUMCOMPONENTS = 2;
#Running and saving information
NTIMESTEPS = 2000000;
NSMOOTH = 10;
SAVET = 200;
STARTTIME = 0;
RESTART = 0;
numworkers = 4;
## Component and Phase names
# COMPONENTS = {Al,Cu,B};
COMPONENTS = {Al, Zn};
PHASES = {alpha, liquid};
##Material properties
##GAMMA={12, 13, 14, 23, 24...}
GAMMA = {0.1};
# Diffusivity = {Diagonal:0/1, phase, 11,22,33, 12, 13, 23...};
DIFFUSIVITY = {1, 0, 0};
DIFFUSIVITY = {1, 1, 1e-9};
##Gas constant and molar volume
R = 8.314;
V = 10e-6;
##Elasticity related parameters
EIGEN_STRAIN = {0, 0.01, 0.01, 0.0, 0.0, 0.0, 0.0};
EIGEN_STRAIN = {1, 0.01, 0.01, 0.0, 0.0, 0.0, 0.0};
VOIGT_ISOTROPIC = {0, 29.24e9, 20.19e9, 13.5e9};
VOIGT_ISOTROPIC = {1, 58.48e9, 40.38e9, 27e9};
#VOIGT_CUBIC = {phase, c11, c12, c44};
#VOIGT_TETRAGONAL = {phase, c11, c12, c13, c33, c44, c66};
##Boundary conditions
#0: Free, 1: Neumann, 2: Dirichlet, 3: Periodic, 4: Complex
#Boundary = {phase, X+, X-, Y+, Y-, Z+, Z-}
BOUNDARY = {phi, 1, 1, 1, 1, 0, 0};
BOUNDARY = {mu, 1, 1, 1, 1, 0, 0};
BOUNDARY = {c, 1, 1, 1, 1, 0, 0};
BOUNDARY = {T, 1, 1, 1, 1, 0, 0};
# Boundary = {phi, 1, 1, 0};
# Boundary = {"u", 3, 3, 2, 2};
#Boundary_value = {Value X+, Value X-, Value Y+, Value Y-, Value Z+, Value Z-}
BOUNDARY_VALUE = {phi, 0, 0, 0, 0, 0, 0};
BOUNDARY_VALUE = {mu, 0, 0, 0, 0, 0, 0};
BOUNDARY_VALUE = {c, 0, 0, 0, 0, 0, 0};
BOUNDARY_VALUE = {T, 0, 0, 0, 0, 0, 0};
##Type of simulation
ISOTHERMAL = 0;
BINARY = 1;
#TERNARY
DILUTE = 0;
T = 857;
##FILEWRITING and OUTPUTTING TO SCREEN
## WRITEFORMAT ASCII/BINARY/HDF5(Only for MPI)
##TRACK_PROGRESS: interval of writing out the progress of the simulation to stdout. 
WRITEFORMAT = ASCII;
WRITEHDF5 = 1;
TRACK_PROGRESS = 10;
##Model-specific parameters: Grand-potential model
##Phase-field parameters; epsilon:interface width; it is not the gradient energy coefficient
epsilon = 4e-8;
tau = 1.31;
Tau = {0.28};
##Anisotropy functions
##Anisotropy mode, FUNCTION_ANISOTROPY=0 is isotropic
Function_anisotropy = 1;
Anisotropy_type = 4; 
dab = {0.01};
#Rotation_matrix = {0, 1, Euler_x(ang), Euler_y(ang), Euler_z(ang)};
Rotation_matrix = {0, 1, 0, 0, 45};
##Potential function
Function_W = 1;
Gamma_abc = {};
#Shifting of domain for infinite domain simulations
Shift = 0;
Shiftj = 30;
#Writing of composition fields along with the chemical potential fields
Writecomposition = 1;
#Noise
Noise_phasefield = 1;
Amp_Noise_Phase = 0.0;
##Temperature
Equilibrium_temperature = 870; 
Filling_temperature = 857;
#TEMPGRADY={BASETEMP, DELTAT, DISTANCE, OFFSET, VELOCITY}
Tempgrady = {856.95, 0.1, 10e-6, 0, 320e-6};
##Function_F
Function_F = 3;
A = {0, 1};
A = {1, 1};
#ceq = {0, 0, 0.83};
#ceq = {0, 1, 0.55};
#ceq = {1, 1, 0.55};
#ceq = {1, 0, 0.55};
#cfill = {0, 0, 0.83};
#cfill = {0, 1, 0.55};
#cfill = {1, 1, 0.55};
#cfill = {1, 0, 0.55};
#slopes = {0, 0, 645};
#slopes = {0, 1, 238};
#slopes = {1, 0, 238};
#slopes = {1, 1, 238};
ceq = {0, 0, 0.926};
ceq = {0, 1, 0.817};
ceq = {1, 1, 0.817};
ceq = {1, 0, 0.817};
cfill = {0, 0, 0.926};
cfill = {0, 1, 0.817};
cfill = {1, 1, 0.817};
cfill = {1, 0, 0.817};
c_guess = {0, 0, 0.909653};
c_guess = {0, 1, 0.7727};
c_guess = {1, 1, 0.7727};
c_guess = {1, 0, 0.7727};
slopes = {0, 0, 666};
slopes = {0, 1, 333};
slopes = {1, 0, 333};
slopes = {1, 1, 333};
num_thermo_phases = 2;
tdbfname = alzn_mey.tdb;
tdb_phases = {FCC_A1, LIQUID};
phase_map = {FCC_A1, LIQUID};
\end{lstlisting}

\par
The filling file defines the initial condition required for solving the governing equations discussed in Section~\ref{sec:formulation}. An example of a filling file is given below:
\begin{lstlisting}
#FILLCUBE = {0, 10, 10, 0,  20, 20, 0};
#FILLCUBE = {0, 10, 10, 0,  20, 20, 0};
##FILLCYLINDER = {phase, x_centre, y_centre, z_start, z_end, radius}
FILLCYLINDER = {0, 0, 0, 0, 0, 5};
###FILLELLIPSE = {phase, x_center, y_center, major_axis, eccentricity, rotation_angle_deg}
#FILLELLIPSE = {0, 50, 50, 0, 10, 0.1, 10};
#FILLSPHERE = {0, 50, 50, 0, 10};
\end{lstlisting}

\par
The details about the keys present in the input file and filling file are described in the repository of \href{https://github.com/ICME-India/MicroSim/blob/main/Grand_potential_Finite_difference_2D_MPI/Manual.pdf}{MICROSIM}.

\section{Validation and benchmarking}
\label{single}
In this subsection, we present a test case where the same real alloy is simulated using the different solver modules, and the results are compared. A single dendrite is simulated toward the Al-rich side of the phase diagram with an alloy composition whose melting point is 870K. For the thermodynamics, we utilize the open-source ``alzn\_mey.tdb" file that is attached in the supplementary and also available elsewhere \citep{Otis-2017}. The model-specific parameters discussed in Section~\ref{sec:formulation} have been kept uniform throughout all the simulations for different solvers. The mean solid-liquid interfacial energy is assumed to be $\gamma_{\alpha\beta} =$ 0.1 J-m$^{-2}$ and a four-fold cubic anisotropy of the form described in Eq.~(\ref{eq:aniso_eqn}) is assumed with a strength $\delta_{\alpha\beta}$=0.01. The interface width corresponding to the simulations at the different undercooling is assumed such that the interface consists of at least 10 points. The interface width is also chosen to be large enough such that the dendrite tip is resolved efficiently. The diffusivity of solid, $D^{\alpha}$ has been considered to be negligible while that of liquid, $D^{\beta} =$ 10$^{-9}$ m$^2$-s$^{-1}$. The simulations are started at the equilibrium temperature $T_{eq}$ of 870K, the solid phase composition, $c^{\alpha}_{eq}$ is initialized with 0.927, and the liquid phase with $c^{\beta}_{eq}$=0.817 which correspond to the equilibrium values of Aluminium at 870K. Thereafter, the simulations are performed at 4 K and 13 K undercoolings.

\begin{figure}[!htbp]
	\subfloat[\label{fig:interfacea}]{\includegraphics[width=87.5mm]{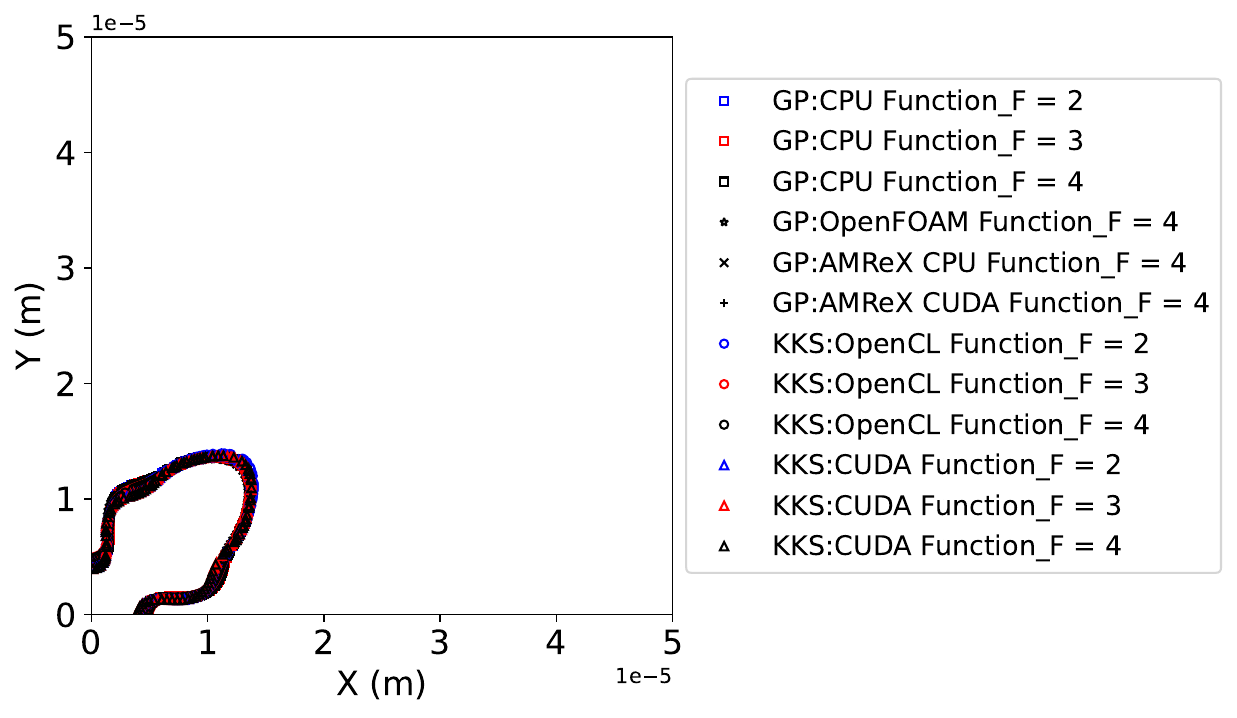}}
	\subfloat[\label{fig:interfaceb}]{\includegraphics[width=52.5mm]{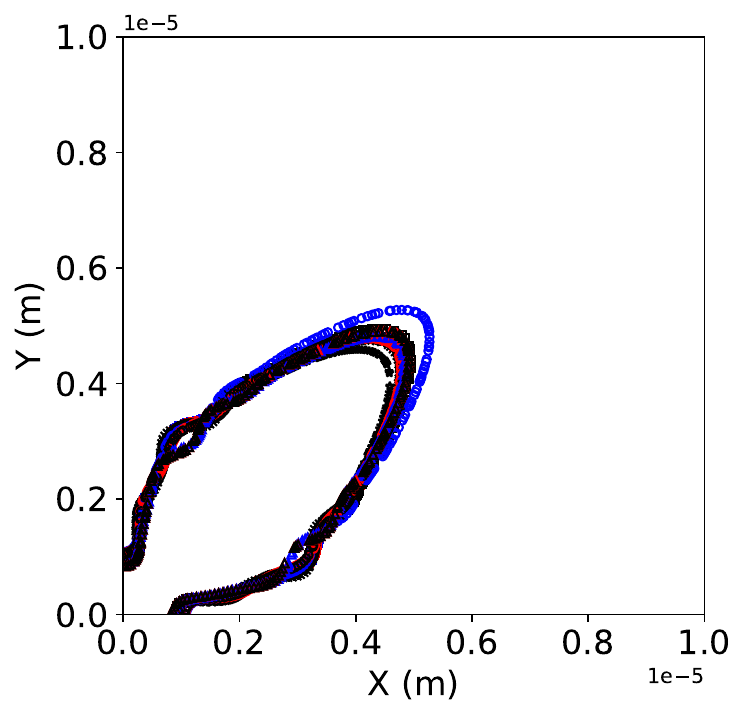}}
	\caption{Contour plots of the solid-liquid interface (at $\phi =$ 0.5) of AlZn alloy for different free energy functions and solvers at undercooling temperature, $\Delta T$ of (a) 4K (at simulation time of 0.9 s) and (b) 13K (at simulation time of 0.0126 s).}
	\label{fig:interface}       
\end{figure}

\par
The cell dimension, $\Delta x$, for the two cases, is assumed to be the largest value that admits a numerically stable and physically correct dendritic tip. For $\Delta T =$ 4K and 13K, $\Delta x$ has been evaluated to be 5$\times$10$^{-8}$ m and 10$^{-8}$ m, respectively. The time step for the simulations pertaining to different solvers has been chosen to obtain a stable and accurate solution. To simulate the growth of a single dendrite, a 1000$\times$1000 grid size has been found to be sufficient for both $\Delta T$ cases. The initial solid seed radius has been taken as five times the cell dimension, which is the minimum required value to avoid melt-down of the seed.

\par
For comparing the results between the solvers, we first plot the dendritic shapes at the same instance of time. For this, the solid-liquid interface at $\phi =$ 0.5 has been compared among the solvers for $\Delta T$ of 4K at a simulation time of 0.9 s in Fig.~\ref{fig:interface}(a), and $\Delta T$ of 13K at a simulation time of 0.0126 s in Fig.~\ref{fig:interface}(b). For the GP solvers using FD with MPI, the plots contain a comparison between the different possibilities for coupling with the thermodynamic databases. We observe that at both undercoolings, the differences between shapes are small, revealing that the parabolic approximations for the thermodynamics would be an efficient strategy. This same conclusion is reached upon comparing the corresponding results with the KKS solver using CUDA. Further, the AMReX and OpenFOAM solvers results with Function\_F=4 reveal a good match with the other CPU and GPU-based solvers. Regarding the KKS solver using the OpenCL formulation, the results from simulations using Function\_F=3,4 depict a reasonable match with the other solvers. However, one of the results for the case of Function\_F=2 stands out. This is due to the difference in the Newton iterator that was adopted for the solution to the equations of equality in the diffusion potentials (Eq.~(\ref{eqmu})) and the imposition of the composition constraint (Eq.~(\ref{cint})). While the other solvers could utilize the optimized multi-root routines available from the GSL libraries, it was not possible to do so in the OpenCL framework. We believe that a more optimized routine will reduce the deviation between the OpenCL and the other implementations.
\par
\begin{figure}[!htbp]
	\subfloat[\label{fig:velocitya}]{\includegraphics[width=84mm]{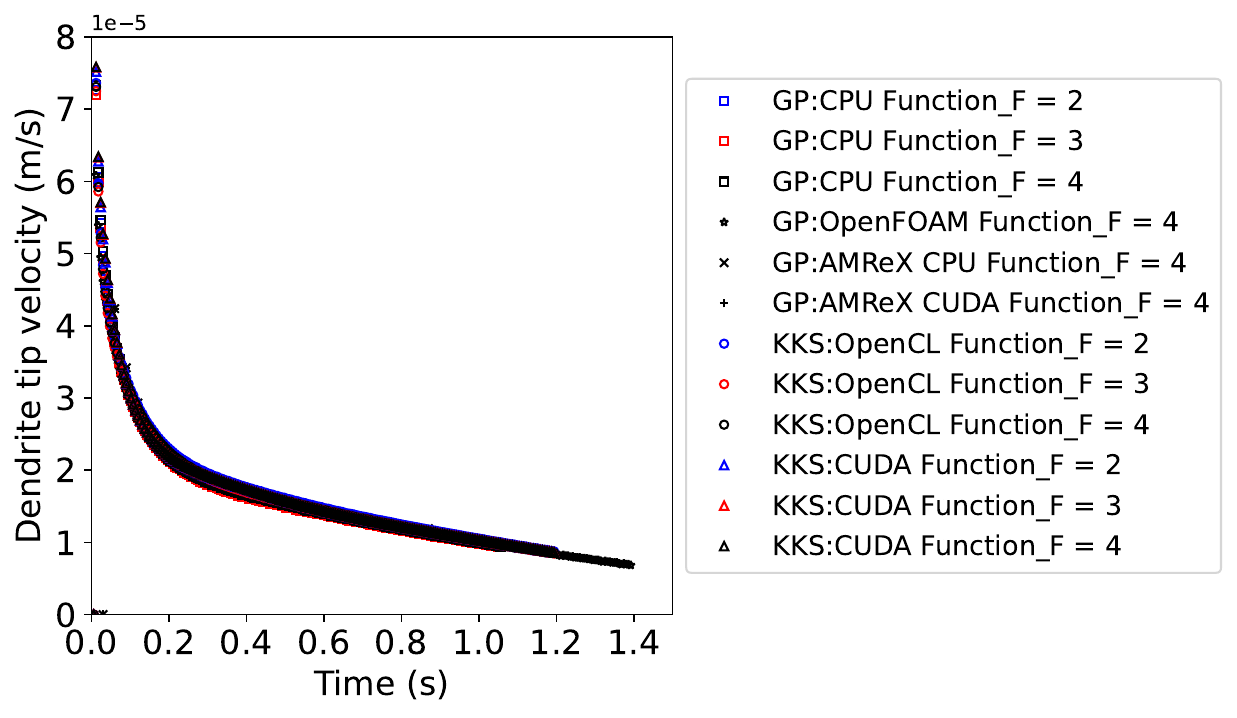}}
	\subfloat[\label{fig:velocityb}]{\includegraphics[width=56mm]{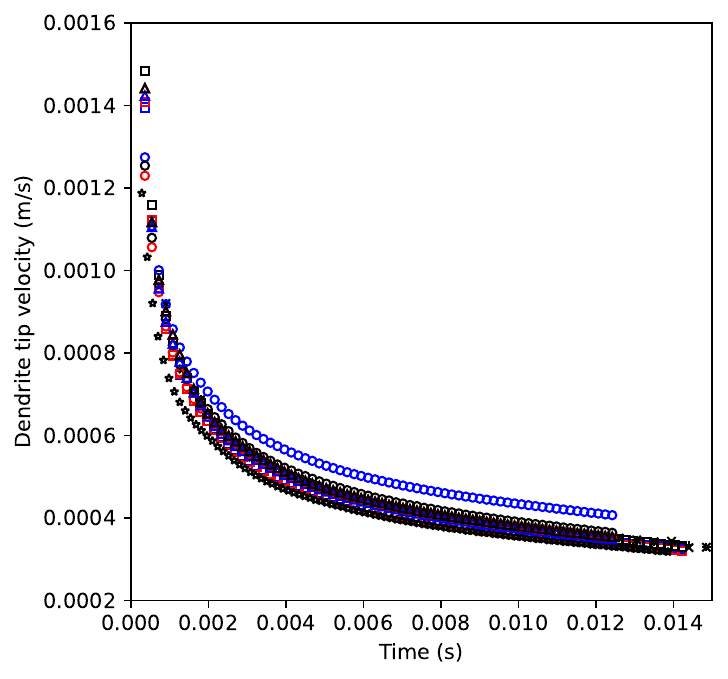}}
	\caption{Evolution of velocity of AlZn dendrite tip during solidification obtained for different free energy functions and solvers at undercooling temperature, $\Delta T$ of (a) 4K and (b) 13K.}
	\label{fig:velocity}    
\end{figure}

Similarly, the evolution of velocity of dendrite tips has been compared among the solvers for $\Delta T$ of 4K in Fig.~\ref{fig:velocity}(a), and $\Delta T$ of 13K in Fig.~\ref{fig:velocity}(b). The dendrite tip decelerates rapidly till about 0.2 s for 4K, while it is about 0.002 s for 13K. Afterward, the velocity tends to saturate for both cases, thus beginning the steady-state growth. The same conclusions with regard to the comparison between the solvers and the thermodynamic formulations can also be reached through a visualization of the plots.
\par
In summary, the results reveal that notwithstanding the differences in the model formulations, discretizations, solution methodologies, parallelization strategies, thermodynamic formulations, and implementation, the solver modules in MICROSIM provide similar results for the same materials problem. This validates and benchmarks the solver for the simulation of real alloys. For the reader's ease, the input file utilized for the comparison is shared in the supplementary material along with the .tdb file and .csv files for the Function\_F=4.

\begin{figure}[!htbp]
    \centering
	\subfloat[\label{fig:OF_3D}]{\includegraphics[width=84mm]{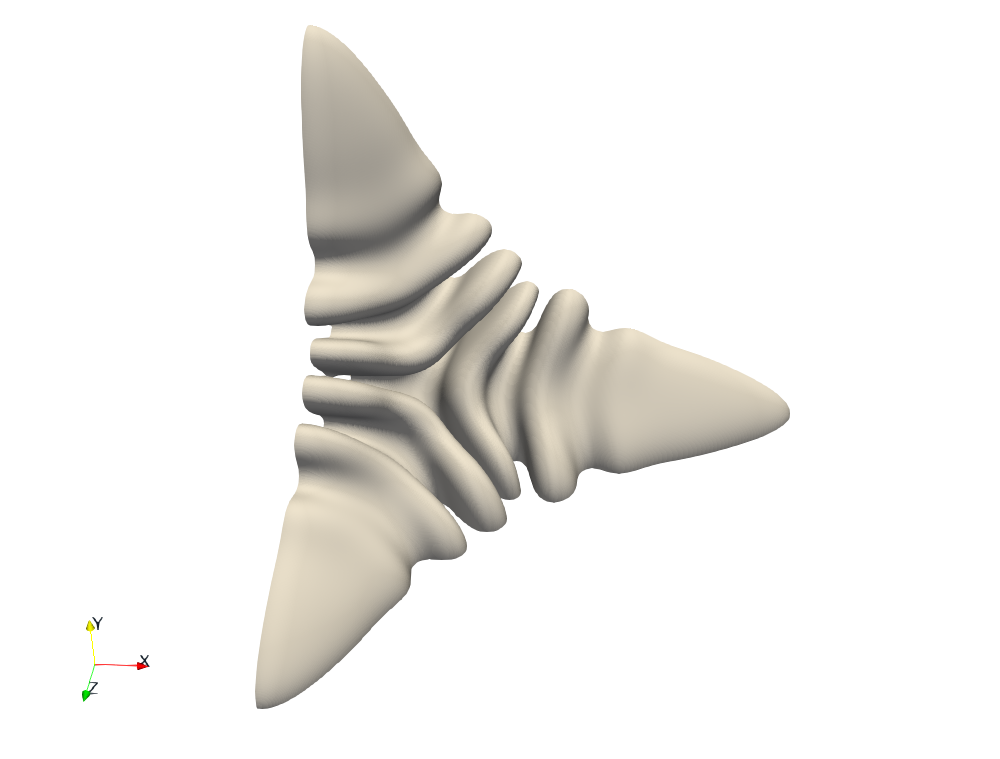}}
	\subfloat[\label{fig:AMReX_3D}]{\includegraphics[width=70mm]{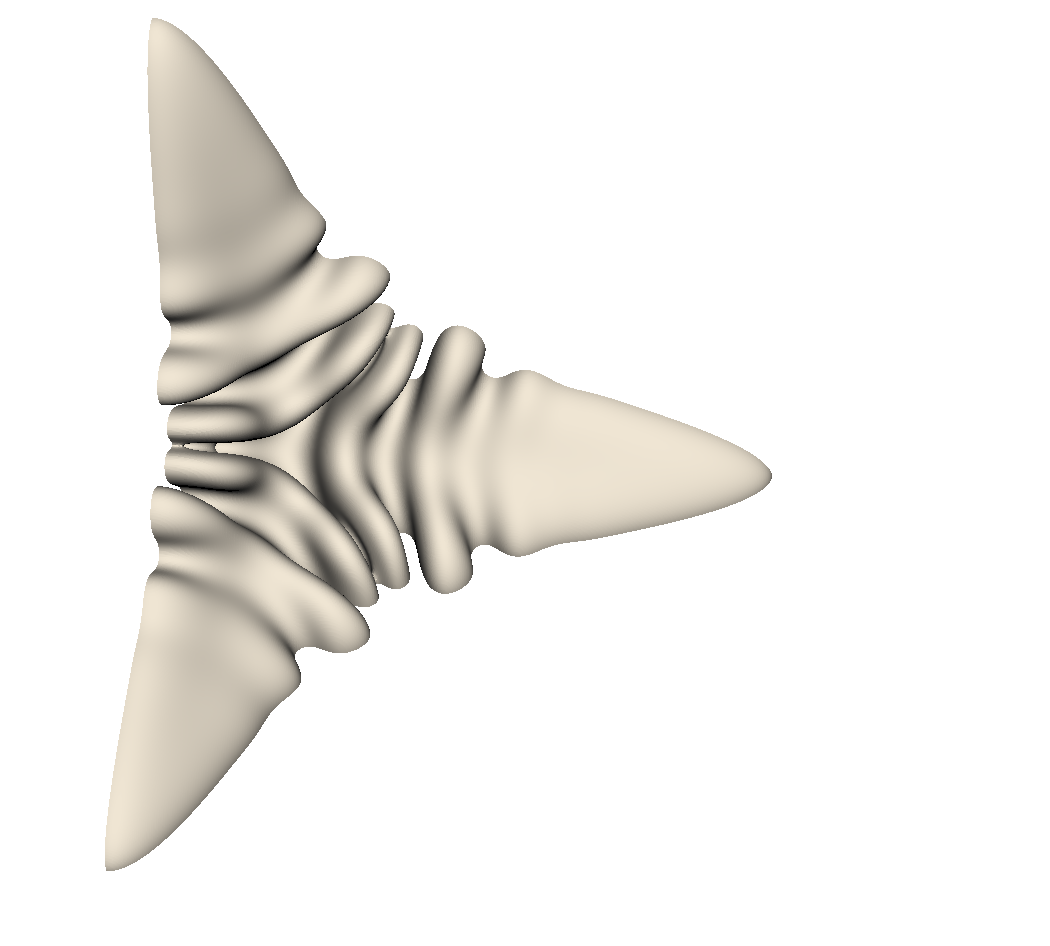}}
    \caption{Solid-liquid interface at one of the octants of an AlZn dendrite obtained from three-dimensional solidification simulation at (a) $1.6\times10^{-5}$ s using the OpenFOAM solver, and (b) $1.75\times10^{-5}$ s using the AMReX solver at undercooling temperature, $\Delta T$ of 30K.}
    \label{OF_AMReX_3D}
\end{figure}
\par
All the solvers are capable of simulating in three dimensions. Such a result has been obtained from a simulation with a grid size of 500$\times$500$\times$500 and an undercooling of 30K using the OpenFOAM and AMReX solvers. In this simulation, the length scale of solidification has been reduced due to a higher undercooling ($\Delta x$ =2$\times$10$^{-9}$ m). Due to the symmetry of the problem and to minimize the use of computational resources, only an octant of a dendrite has been simulated. The solid-liquid interface around the dendrite is displayed at $1.6\times10^{-5}$ s in Fig.~\ref{OF_AMReX_3D}(a), and at $1.75\times10^{-5}$ s in~\ref{OF_AMReX_3D}(b) for OpenFOAM and AMReX solvers, respectively. The results from both solvers appear identical. The time step, $\Delta t$ for OpenFOAM solver is 2$\times$10$^{-9}$ s, while for AMReX solver is 3.5$\times$10$^{-10}$ s. It must be noted that OpenFOAM employs an implicit finite volume scheme (Subsection~\ref{OF_impl}), allowing for a larger time step.
\par
Finally, the execution times of all the solvers for $\Delta T =$ 13K are represented in Table~\ref{tab:exectime}, which compares the performance. While the OpenFOAM solvers might seem slower than the rest, it is the only module at present that permits the utilization of parallelized implicit solution routines in MICROSIM (through built-in OpenFOAM modules). Because of this feature, the time steps can be substantially upscaled compared to the explicit routines.  
\begin{table}[!htbp]
\caption{Comparison of execution times for Function\_F $=$ 4 and $\Delta T =$ 13K. The CPU core comparison is for simulations performed on the PARAM-PRAVEGA supercomputer at IISc, India.}
\label{tab:exectime}
\centering
\begin{tabular}{lll}
\hline
Model         & Domain size        & Execution time \\ \hline
GP-MPI        & $1000 \times 1000$ & 1860 seconds using 96 CPU cores          \\
GP-AMReX (CPU)      & $1000 \times 1000$ & 3991 seconds using 96 CPU cores          \\
GP-AMReX (GPU) & $1000 \times 1000$ &  1633 seconds using 2 NVIDIA V100 GPUs   \\
GP-OpenFOAM   & $1000 \times 1000$ & 3444 seconds using 480 CPU cores        \\
KKS-CUDA      & $1000 \times 1000$ & 1577 seconds using 2 NVIDIA V100 GPUs  
\\\hline
\end{tabular}

\end{table}

In the following section, we explicitly report the performance and scaling behavior of the solvers.

\section{Performance and scaling}
The solvers have been used for running some simulations over different sets of computational resources. The number of CPUs for MPI, AMReX, and OpenFOAM-based solvers, or GPUs for CUDA, OpenCL, and AMReX-based solvers, has been varied. This provides an estimation of the modules' strong and weak scaling behavior.
\subsection{MPI}
\label{MPI}
\begin{figure}[!htbp]
    \centering
    \includegraphics[width=0.8\textwidth]{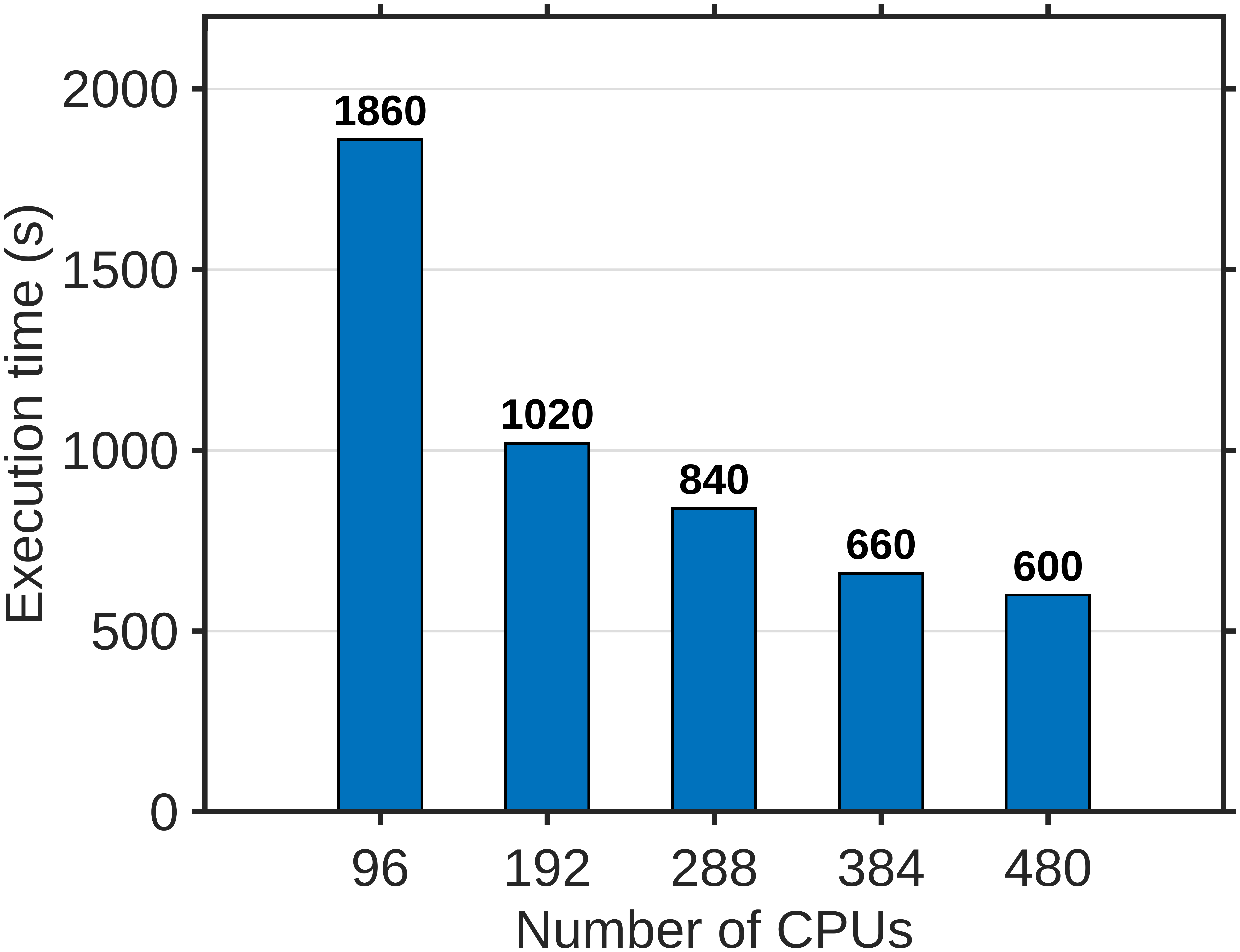}
    \caption{Results from strong scaling tests of the MPI solver.}
    \label{MPI_scale}
\end{figure}

Solidification simulations of a two-phase AlZn alloy have been conducted using the grand-potential MPI solver in two dimensions on a 1000$\times$1000 grid size. The same simulation has been conducted using different CPU (strong scaling test) numbers, such as 96, 192, 288, 384, and 480 with each CPU assigned to a subdomain. The strong scaling results are presented in Fig.~\ref{MPI_scale} till 400000 iterations. It can be observed from Fig.~\ref{MPI_scale} that a two times increase in the number of CPUs leads to $\sim$ 1.82 times speed-up, a three times increase in the number of CPUs leads to $\sim$2 times speedup, while five times increase in the number of CPUs led to a $\sim$3 times speedup.

\par
\begin{figure}[!htbp]
    \centering
    \includegraphics[width=0.8\textwidth]{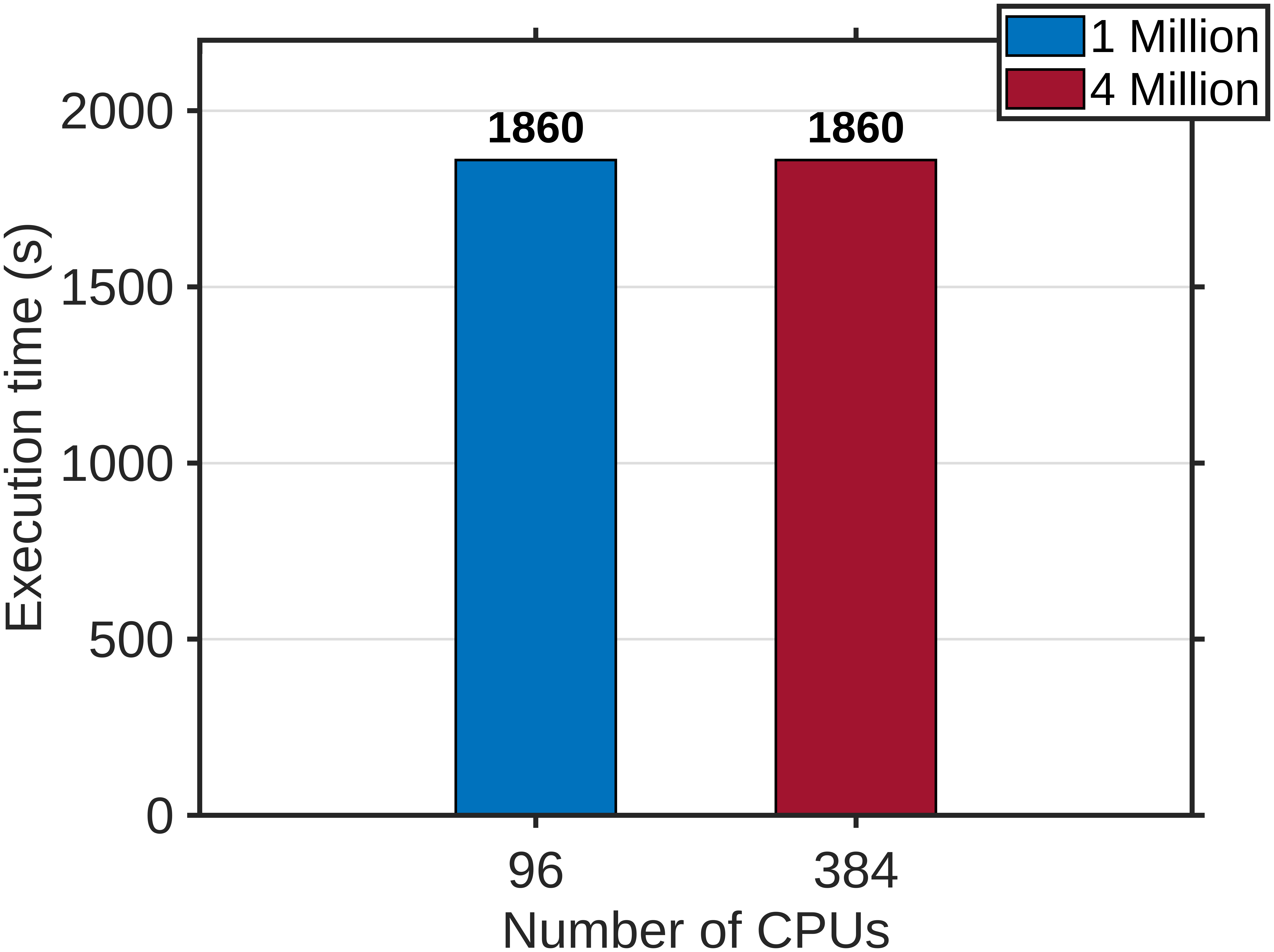}
    \caption{Results from weak scaling tests of the MPI solver.}
    \label{MPI_weak}
\end{figure}

The weak scaling tests for the grand-potential-based MPI solver on 1000 $\times$ 1000 and 2000 $\times$ 2000 grid size are conducted, where the number of CPUs has been increased by four times (96 to 384) when the domain is enhanced proportionately. The duration of simulation for both cases till 400000 iterations have been compared in Fig.~\ref{MPI_weak}, which shows identical execution times for both simulations, highlighting the excellent scalability of the grand-potential solver using FD parallelized using MPI.

\subsection{CUDA}

\begin{figure}[!htbp]
	\subfloat[\label{fig:CUDA_scaling_2D}]{\includegraphics[width=68mm]{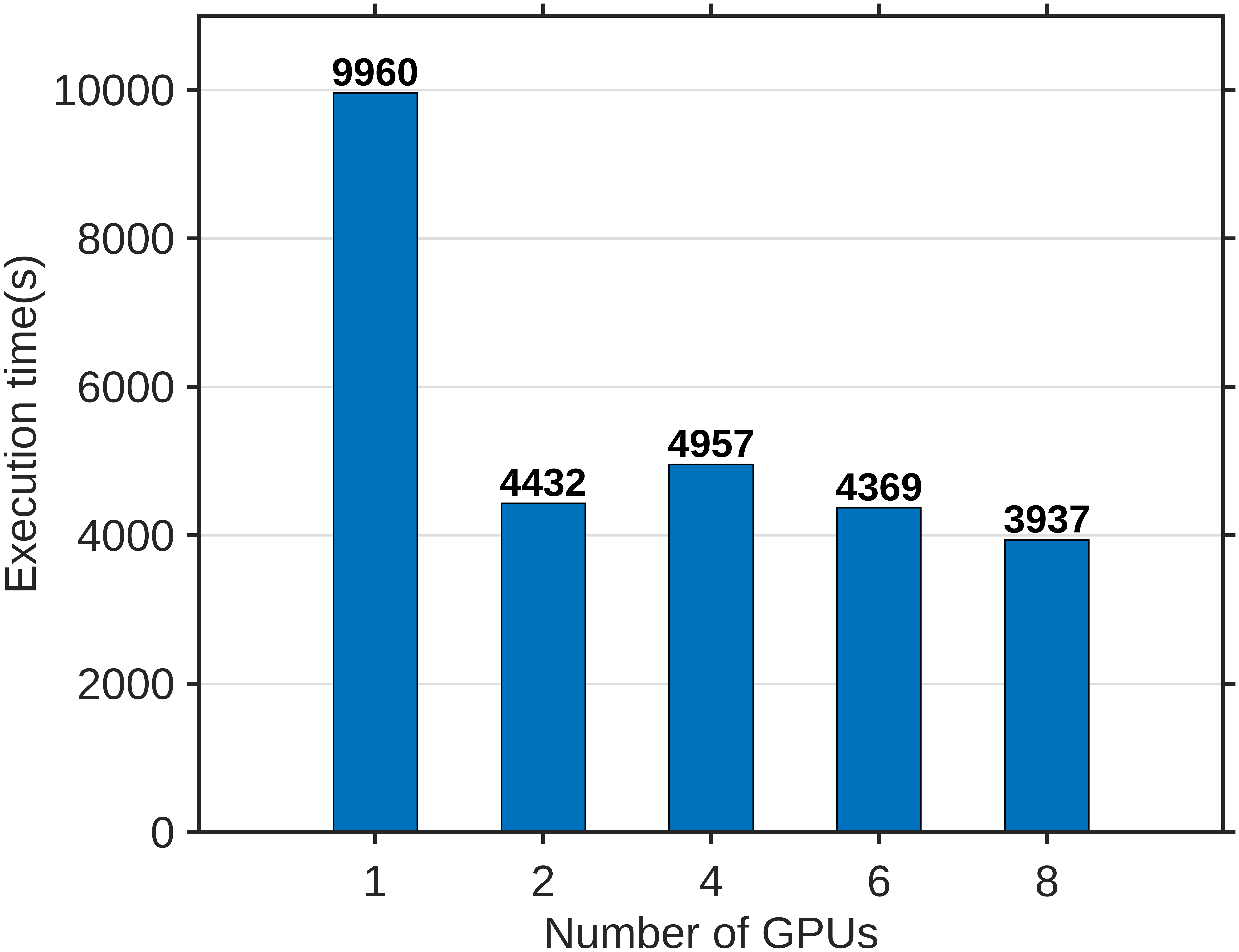}}
	\subfloat[\label{fig:CUDA_scaling_3D}]{\includegraphics[width=68mm]{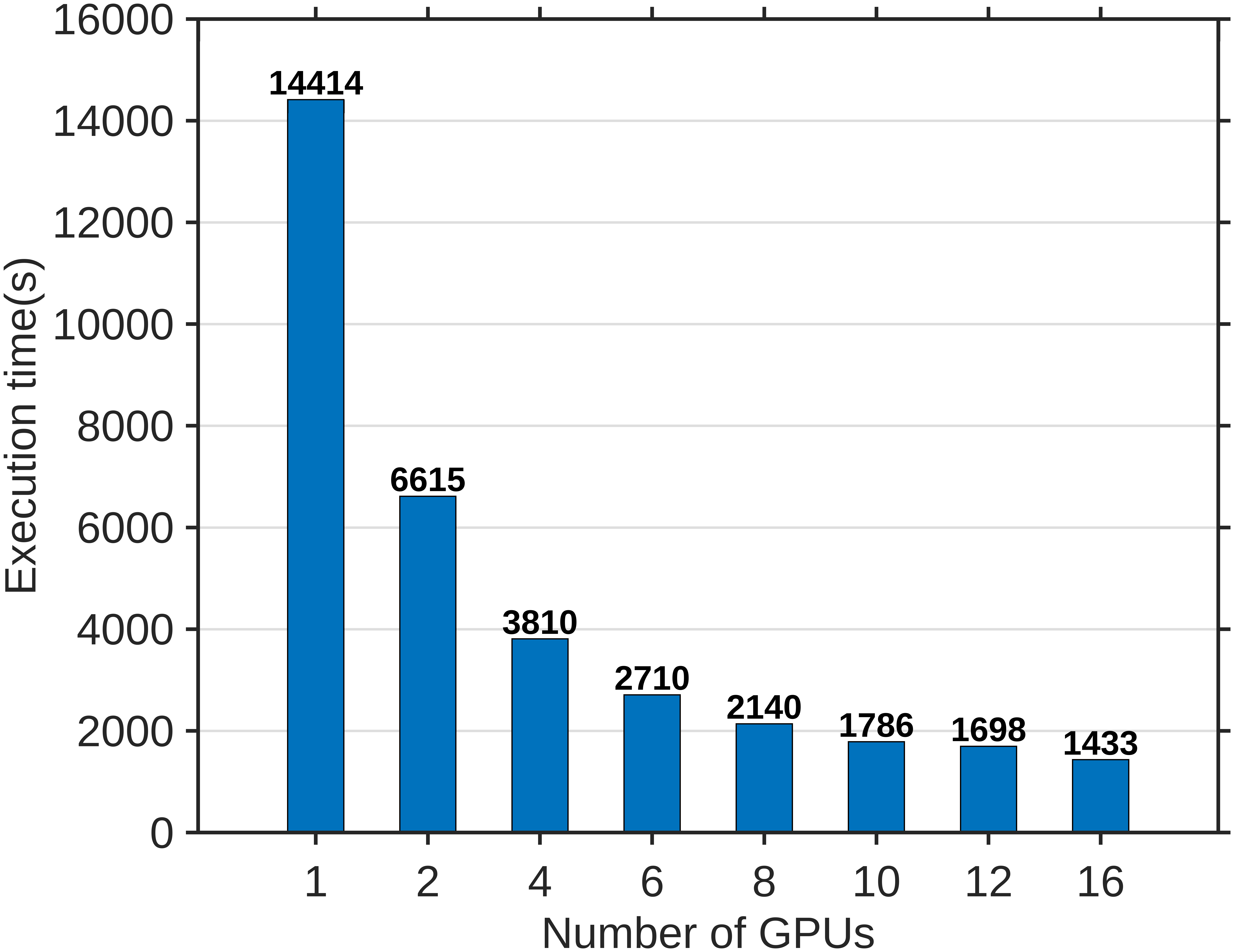}}
	\caption{Results from strong scaling tests using CUDA MPI for (a) 2D simulations and (b) 3D simulations.}
	\label{fig:CUDA_scale}    
\end{figure}

The KKS CUDA solver has been used for carrying out solidification simulations of a two-phase AlZn alloy in two dimensions on a 1000$\times$1000 grid and in three-dimensions on a 500$\times$500$\times$500 grid. Both two-dimensional and three-dimensional simulations have been conducted using different numbers of GPUs (strong scaling test), such as 1, 2, 4, 6, and so on, with each GPU assigned to a subdomain. The strong scaling results from two-dimensional simulations are presented in Fig.~\ref{fig:CUDA_scale}(a) till 2000000 iterations, while Fig.~\ref{fig:CUDA_scale}(b) is for three-dimensional simulations continued till 30000 iterations. It can be observed from Fig.~\ref{fig:CUDA_scale}(a), a 2 times increase in the number of GPUs led to $\sim$2.25 times speedup, while increasing the number of GPUs further does not speed up the simulation. However, Fig.~\ref{fig:CUDA_scale}(b) shows a monotonous trend in terms of the speedup of the simulation.

\subsection{AMReX}
\begin{figure}[!htbp]
	\subfloat[\label{fig:amrex_1000}]{\includegraphics[width=66mm]{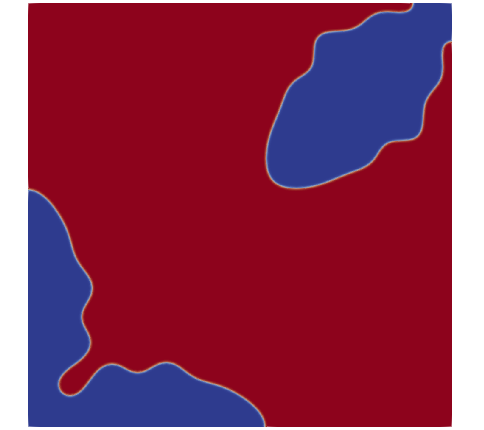}}
	\subfloat[\label{fig:amrex_2000}]{\includegraphics[width=84mm]{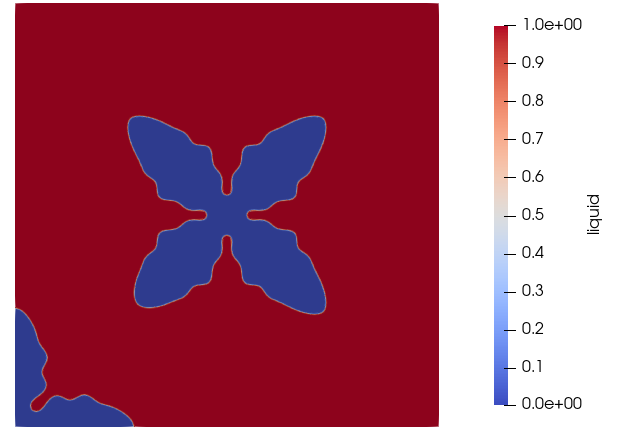}}
	\caption{Simulation of a three-phase AlZn alloy at $\Delta T$ = 13K with two different grid sizes, (a) 1000$\times$1000 (1 Million cells), and (b) 2000$\times$2000 (4 Million cells).}
	\label{fig:AMReX_Dendrite}    
\end{figure}

\begin{figure}[!htbp]
	\subfloat[\label{fig:Cpu_scal}]{\includegraphics[width=68mm]{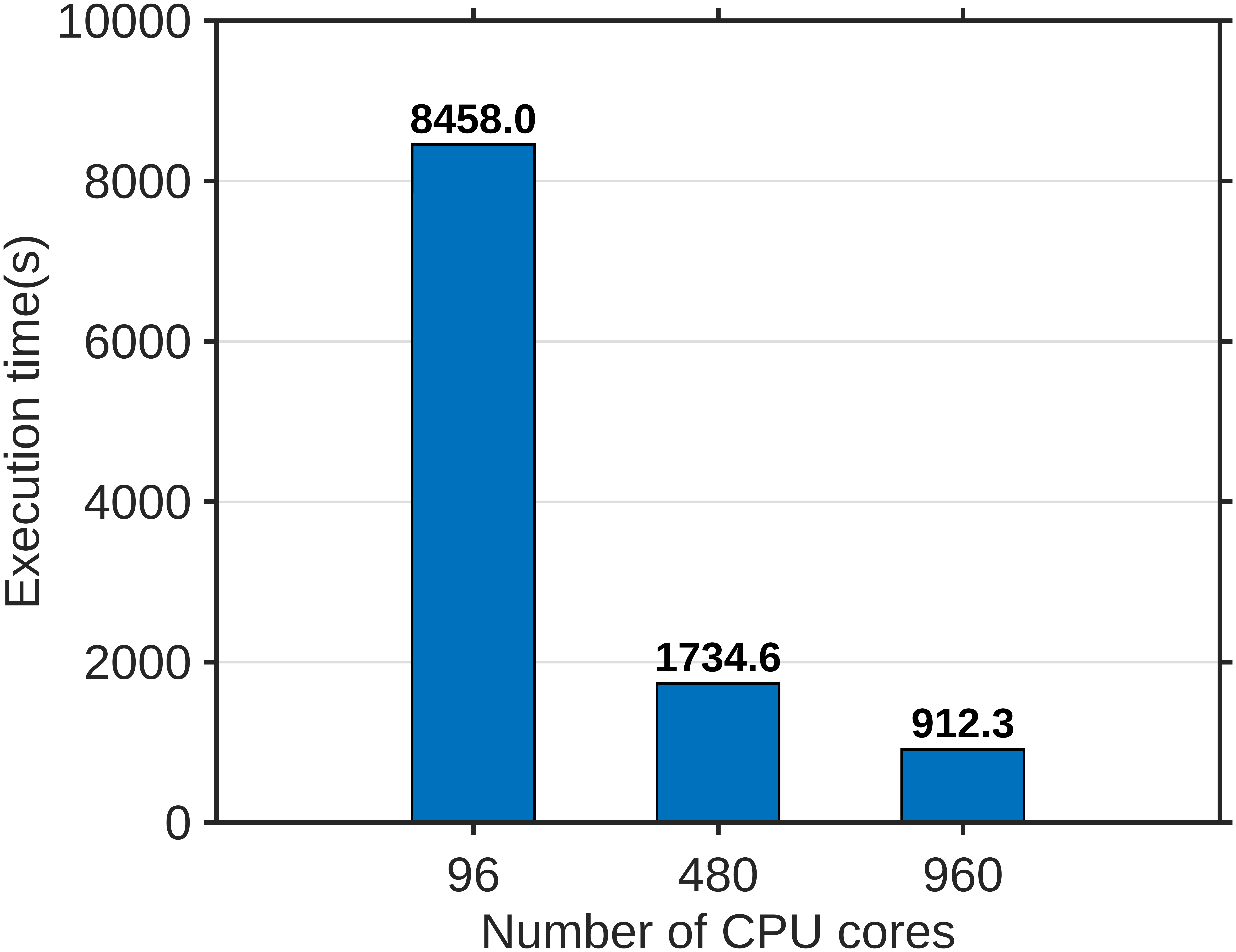}}
 \hspace{2mm}
	\subfloat[\label{fig:Gpu_scal}]{\includegraphics[width=68mm]{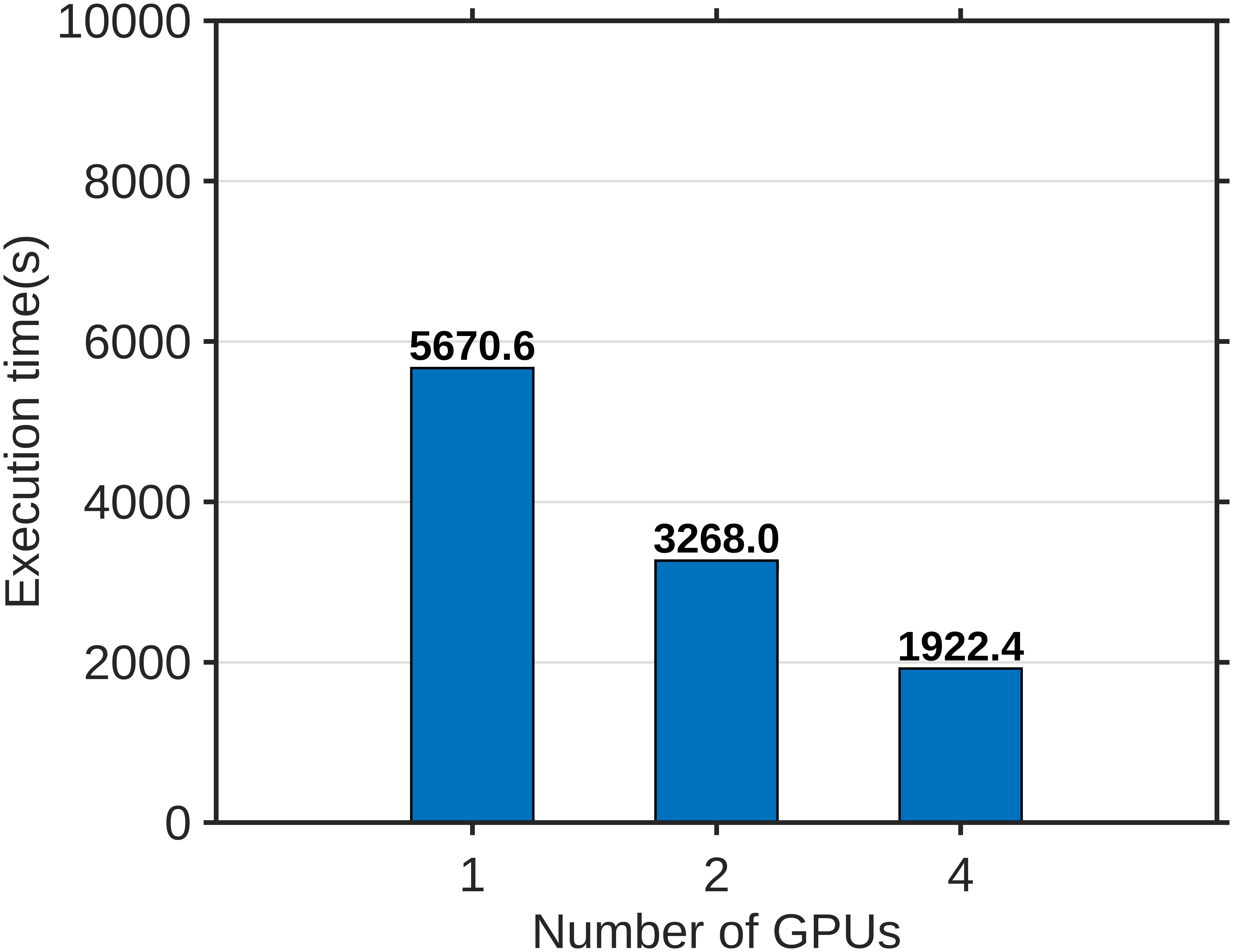}}
	\caption{Results from strong scaling tests on (a) CPU, and (b) GPU.}
	\label{fig:Strong_scal}    
\end{figure}

Fig.~\ref{fig:AMReX_Dendrite} is a snapshot of the dendritic growth for AMReX-based two-dimensional simulation for two different grid sizes for a three-phase AlZn system for the grand potential model. The scaling exercise is for the computation on a 1000×1000 grid for 700000 iterations. The strong scaling results for CPU and GPU are presented in Fig.~\ref{fig:Strong_scal}(a) and ~\ref{fig:Strong_scal}(b), respectively. In Fig.~\ref{fig:Strong_scal}(a), a five times increase in computation power led to a 4.87 times speedup, while a further two times increase in computation power led to an additional 1.9 times speedup. For computations on the GPU, each MPI rank offloaded its work to a single GPU. In Fig.~\ref{fig:Strong_scal}(b), using two GPUs, the execution time is reduced by a factor of 1.74. Further increasing the usage to four GPUs reduced the execution time by another factor of 1.69.

\begin{figure}[!htbp]
	\subfloat[\label{fig:Cpu_scal_1&4}]{\includegraphics[width=69mm]{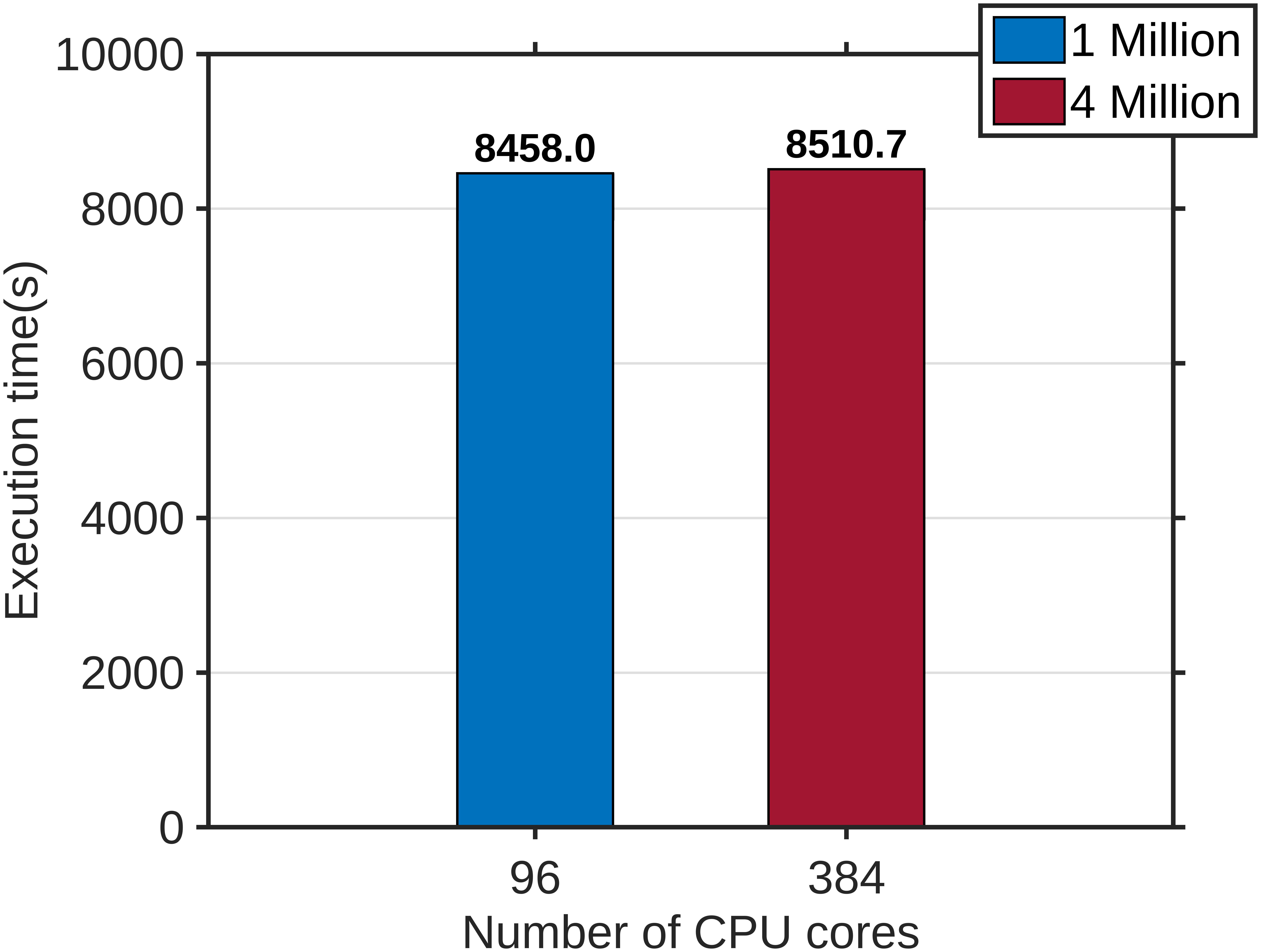}}
 \hspace{2mm}
	\subfloat[\label{fig:Gpu_scal_1&4}]{\includegraphics[width=69mm]{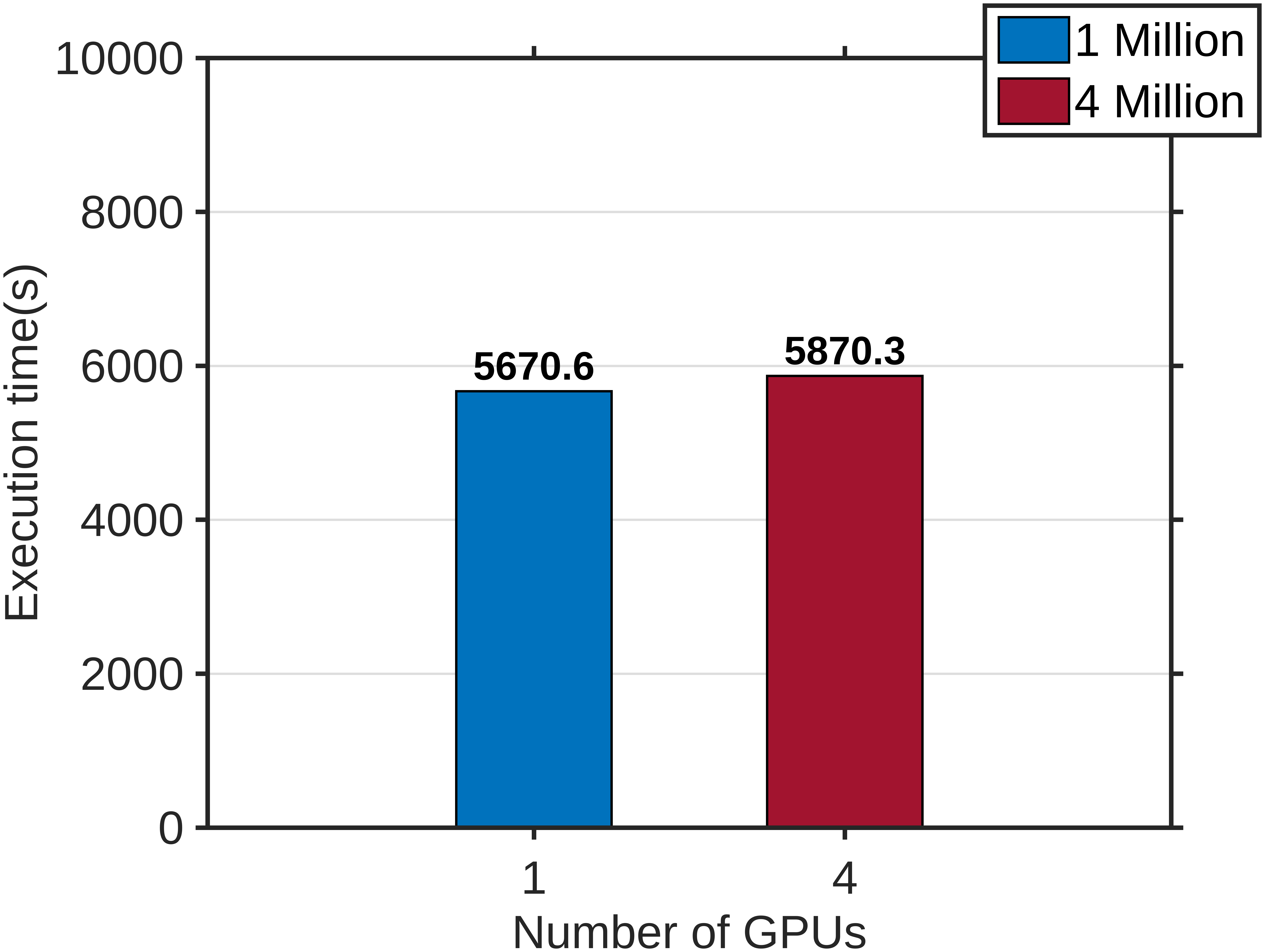}}
	\caption{Results from weak scaling test on (a) CPU, and (b) GPU.}
	\label{fig:Weak_scal}    
\end{figure}

The weak scaling results for CPU and GPU on 1000$\times$1000 and 2000$\times$2000 grid sizes are presented in Fig.~\ref{fig:Weak_scal}(a) and ~\ref{fig:Weak_scal}(b), respectively. In Fig.~\ref{fig:Weak_scal}(a), keeping the computational load constant on each core by increasing both the domain size and the number of cores four times, the execution time observed in this case is more or less the same. The exercise on the GPU reflects a similar trend as highlighted in Fig.~\ref{fig:Weak_scal}(b).

\subsection{OpenFOAM}
\begin{figure}[!htbp]
    \centering
    \includegraphics[width=0.8\textwidth]{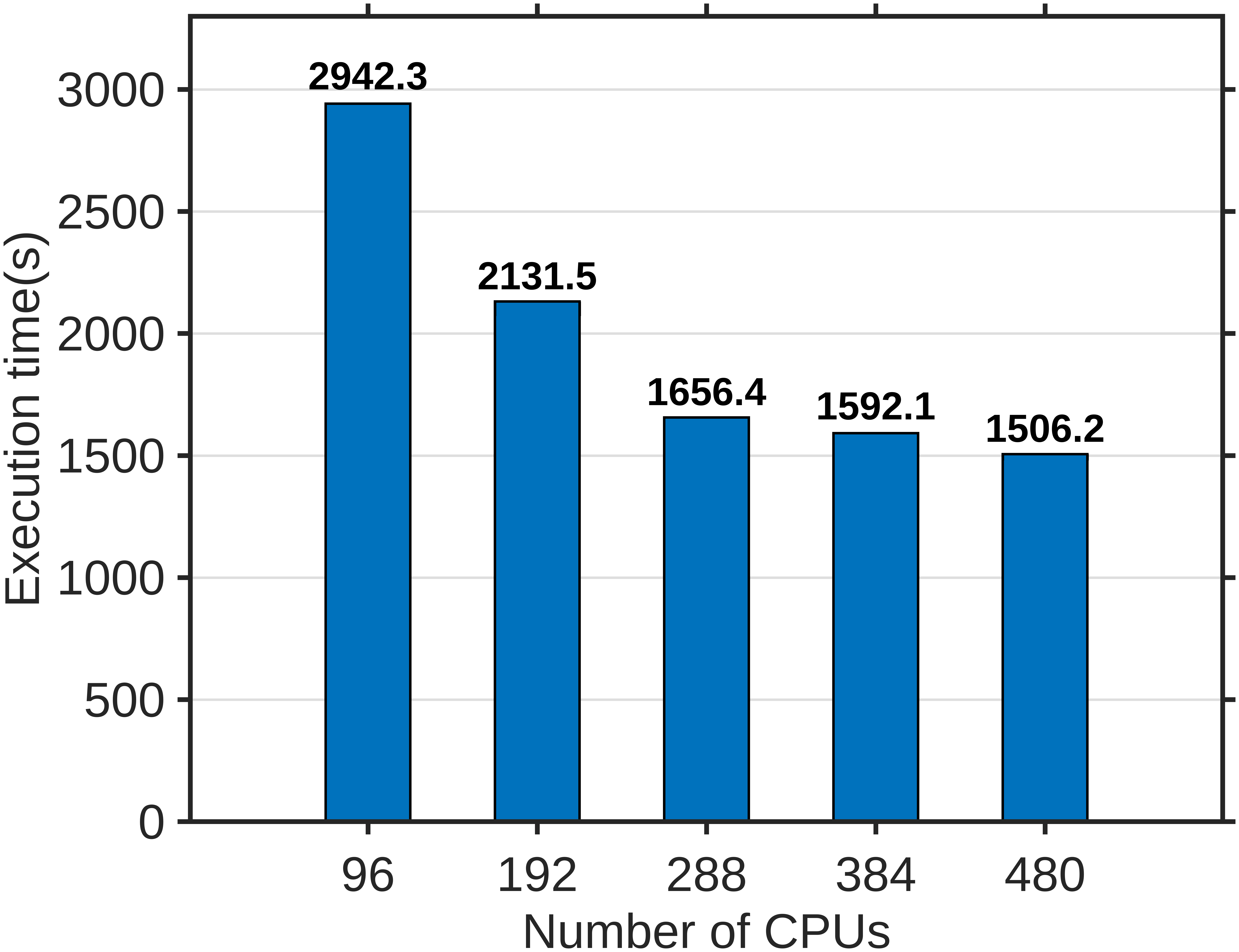}
    \caption{Results from strong scaling tests of the OpenFOAM solver.}
    \label{OF_scale}
\end{figure}

OpenFOAM simulations of two-dimensional solidification of a two-phase AlZn alloy have been conducted using the grand-potential model on a 1000 $\times$ 1000 grid size.  Similar to the scaling test of the MPI solver (see Subsection~\ref{MPI}), the strong scaling results are presented in Fig.~\ref{OF_scale}. The simulations have been continued till 400000 iterations, which is the same as the MPI solver. In Fig.~\ref{OF_scale}, a 3 times increase in the number of CPUs led to a 1.77 times speedup, while a 5 times increase in the number of CPUs led to a 1.95 times speedup.

\begin{figure}[!htbp]
    \centering
    \includegraphics[width=0.8\textwidth]{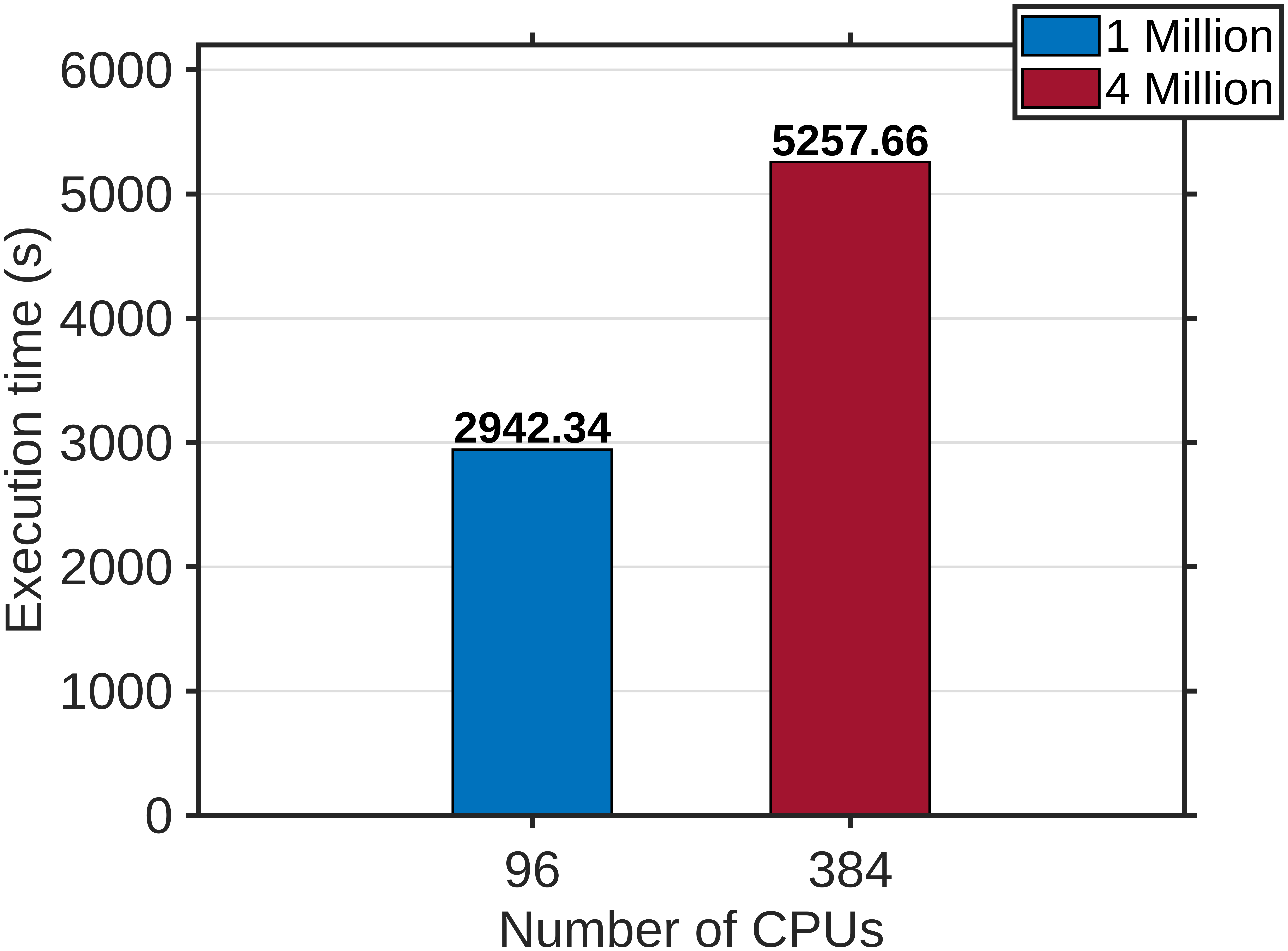}
    \caption{Results from weak scaling tests of the OpenFOAM solver.}
    \label{OF_weak}
\end{figure}

The weak scaling results for OpenFOAM on 1000$\times$1000 and 2000$\times$2000 grid size are demonstrated in Fig.~\ref{OF_weak}. The number of CPUs has been kept proportional to the domain size as in Subsection~\ref{MPI}. In Fig.~\ref{OF_weak}, the execution time observed for the 2000$\times$2000 case is much higher than the 1000$\times$1000 case. The results from Fig.~\ref{OF_scale} and~\ref{OF_weak} can be compared with the MPI solver in Fig.~\ref{MPI_scale} and~\ref{MPI_weak}, respectively. It can be seen that the MPI solver scales much better for the same amount of iterations compared to the OpenFOAM solver. However, as mentioned in Subsection~\ref{OF_impl}, OpenFOAM employs an implicit finite volume scheme, allowing for a larger time step than the explicit scheme.

\section{Illustration of the capabilities of the solver modules}
In this section, we show snippets of the capabilities of the solver modules, highlighting the ease with which simulations of complicated microstructural problems are possible.

\subsection{Multi-phase solidification}
\label{multiphase}

\subsubsection{Multi-grain solidification}
\begin{figure}[!htbp]
    \centering
    \includegraphics[width=12cm]{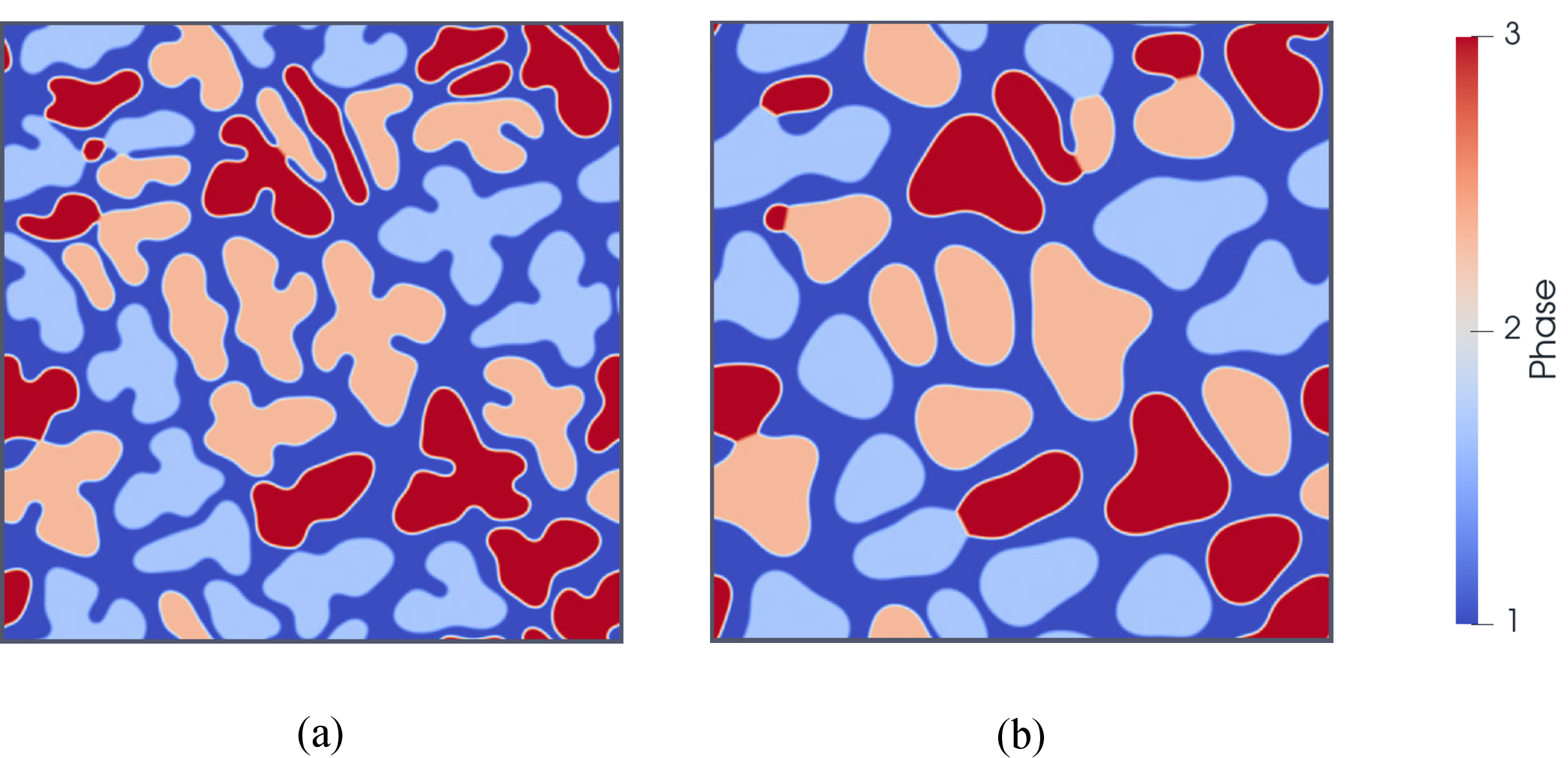}
    \caption{Evolution of three grains in an AlZn alloy at two different instances during (a) beginning of dendrite growth, and (b) coarsening of dendrites.}
    \label{fig:multiphase_AlZn_MPI}
\end{figure}

While the simulation results in Section~\ref{single} depict capabilities for the simulation of single-phase solidification of real alloys, the same solver modules can be utilized for the simulation of multiple grains through just a modification of the input file. The supplementary file contains the input file (Infile\_multigrain\_AlZn.in), which can be utilized for the generation of the following results with three grains of the FCC\_A1 phase in the AlZn system growing at an undercooling of 13K as depicted in Fig. \ref{fig:multiphase_AlZn_MPI}. 

\begin{figure}[!htbp]
    \centering
    \includegraphics[width=0.8\textwidth]{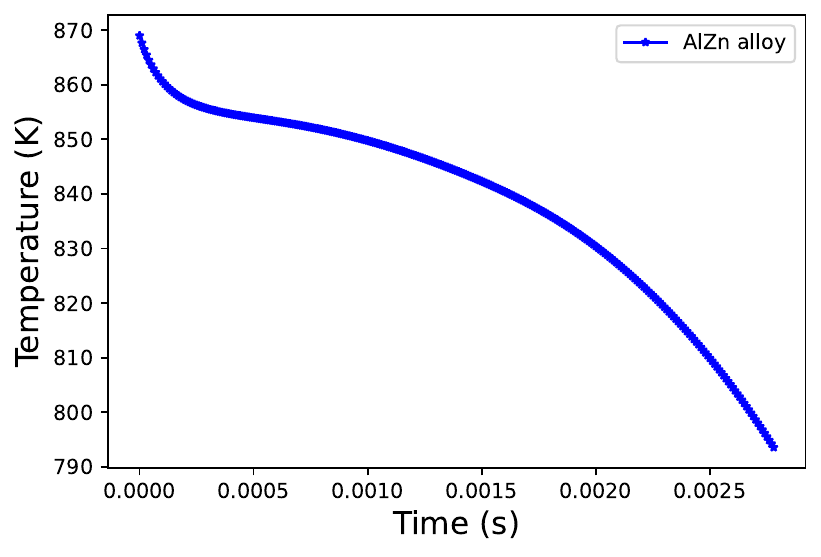}
    \caption{Temperature (K) evolution between liquidus and solidus at a cooling rate of 4$\times$10$^6$ J-mol$^{-1}$-s$^{-1}$ for an AlZn alloy.}
    \label{Temperature_AlZn}
\end{figure}

In the preceding simulations, the system's temperature is constant, which is representative of a simulation at constant undercooling. For performing simulations with a cooling rate imposed on the domain that is representative of real-world casting situations, we have OpenFOAM modules for performing multigrain solidification. Here, the orientation-field module is utilized, as described before in the formulation (Subsection~\ref{orientation}). A multigrain solidification simulation has been carried out for the AlZn alloy from Section~\ref{single} at a constant cooling rate of 4$\times$10$^6$ J-mol$^{-1}$-s$^{-1}$. The solid (FCC) phase seeds have been nucleated at random positions with random $\theta$ at the beginning of this simulation. The temperature evolution between liquidus and solidus over the total simulation time is represented in Fig.~\ref{Temperature_AlZn}. The flattening of the temperature curve occurs due to the balance between the heat-extraction rate and the heat generated as a result of growth. This heating of the domain due to growth is referred to as recalescence.

\begin{figure}[!htbp]
	\subfloat[\label{fig:phiAlZna}]{\includegraphics[width=45mm]{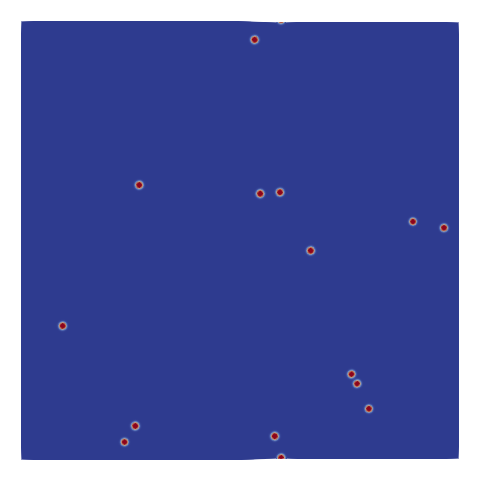}}
	\subfloat[\label{fig:phiAlZnb}]{\includegraphics[width=45mm]{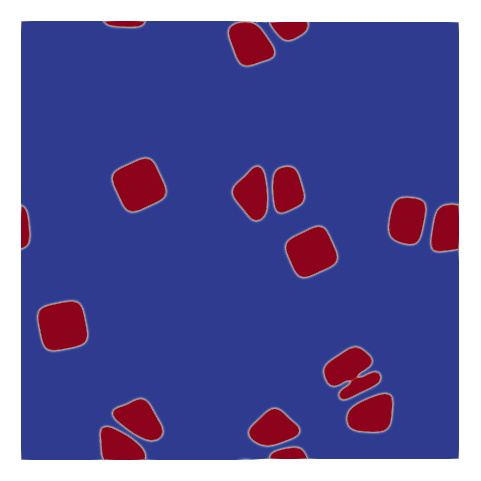}}
	\subfloat[\label{fig:phiAlZnc}]{\includegraphics[width=60mm]{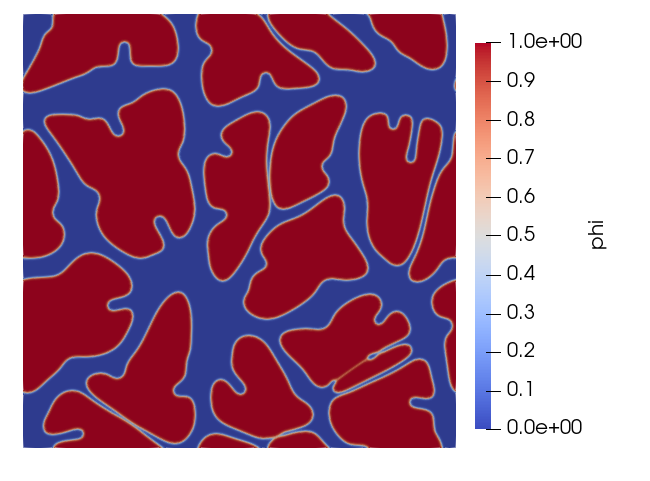}}
	\caption{Evolution of phase fraction, $\phi$ in an AlZn alloy at a cooling rate of 4$\times$10$^6$ J-mol$^{-1}$-s$^{-1}$ during (a) nucleation, (b) beginning of dendrite growth, and (c) grain formation.}
	\label{fig:phiAlZn}
\end{figure}

\begin{figure}[!htbp]
	\subfloat[\label{fig:thetaAlZna}]{\includegraphics[width=45mm]{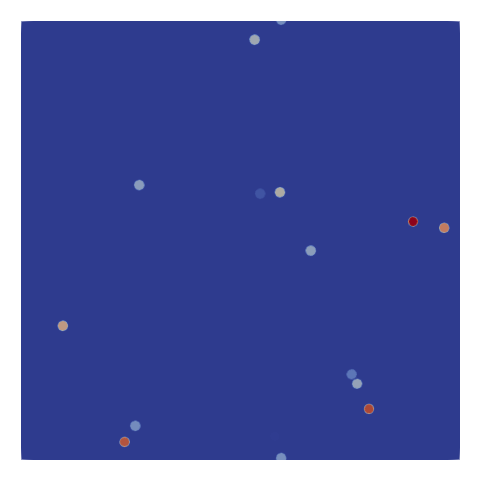}}
	\subfloat[\label{fig:thetaAlZnb}]{\includegraphics[width=45mm]{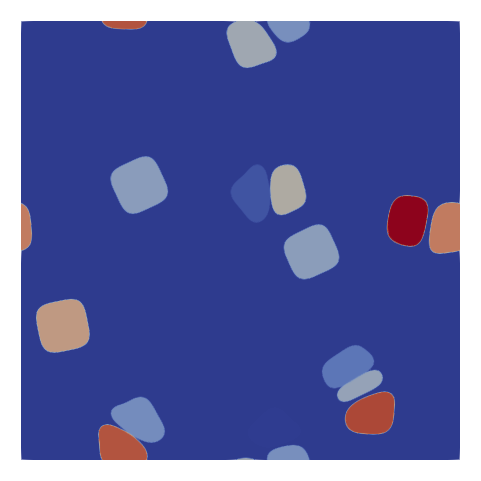}}
	\subfloat[\label{fig:thetaAlZnc}]{\includegraphics[width=65mm]{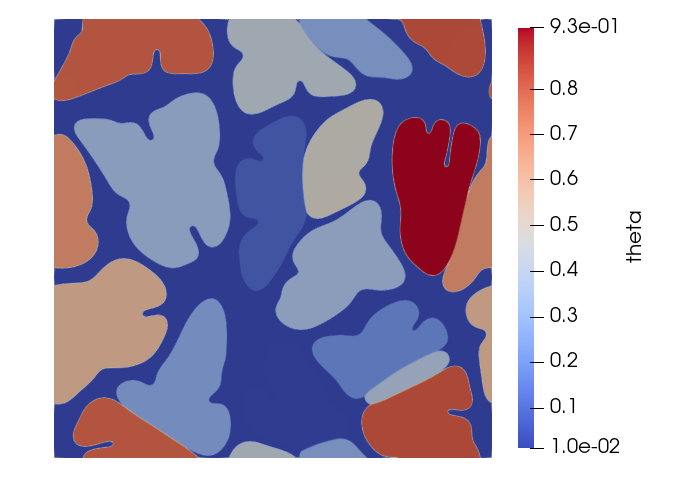}}
	\caption{Evolution of grain orientation, $\theta$ in an AlZn alloy at cooling rate of 4$\times$10$^6$ J-mol$^{-1}$-s$^{-1}$ during (a) nucleation, (b) beginning of dendrite growth, and (c) grain formation.}
	\label{fig:thetaAlZn}
\end{figure}

\begin{figure}[!htbp]
    \centering
    \includegraphics[width=0.8\textwidth]{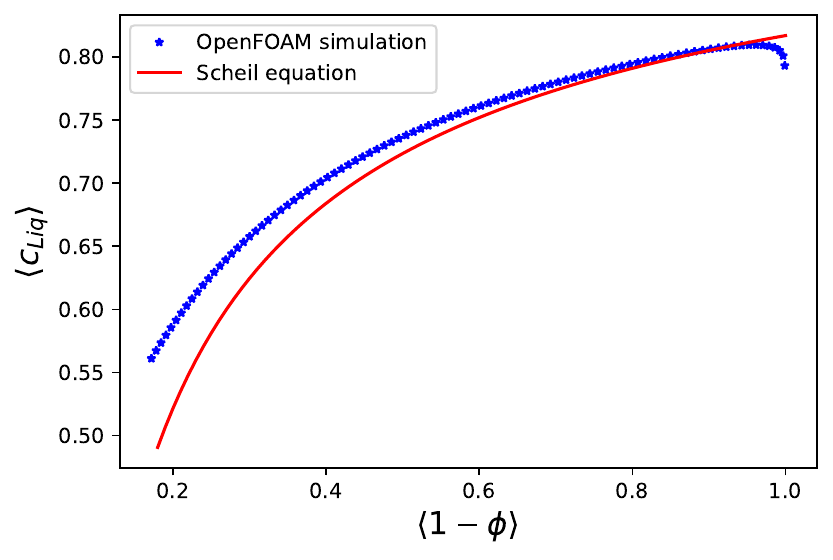}
    \caption{Comparison of average liquidus composition, $\langle c_{Liq} \rangle$ vs average liquidus volume fraction, $\langle 1 - \phi \rangle$ obtained from OpenFOAM simulation and Scheil equation.}
    \label{Scheil_AlZn}
\end{figure}

The contours of $\phi$ and $\theta$ are shown in Fig.~\ref{fig:phiAlZn} and~\ref{fig:thetaAlZn}, respectively, at three different times.  Figure~\ref{fig:phiAlZn}(a) represents the nucleation of the FCC phase within the liquid phase. Figure~\ref{fig:phiAlZn}(b) captures the FCC phase beginning to form a dendrite shape. Figure~\ref{fig:phiAlZn}(c) shows the FCC phase engulfing the domain while forming grains. Such results were also reported in \citep{granasy2005el}. It can be expected that secondary phases may begin forming beyond a certain volume fraction of the FCC phase. The evolution of the average liquidus composition, $\langle c_{Liq} \rangle$ with the average liquidus volume fraction, $\langle 1 - \phi \rangle$ is displayed in Fig.~\ref{Scheil_AlZn} from the simulation as well as the Scheil equation. It can be seen from Fig.~\ref{Scheil_AlZn} that $\langle c_{Liq} \rangle$ obtained from the simulation is larger than the Scheil equation. It is possible that the Scheil equation underestimates $\langle c_{Liq} \rangle$, since it is calculated over a domain with the assumption of instantaneous diffusion in the liquid.

This demonstrates the potential capability of the solver stack for simulating multigrain solidification. Further discussions about the multigrain solidification module will be reported in future work.

\subsubsection{Solidification involving physically distinct phases}
\begin{figure}[!htbp]
    \centering
    \includegraphics[width=\textwidth]{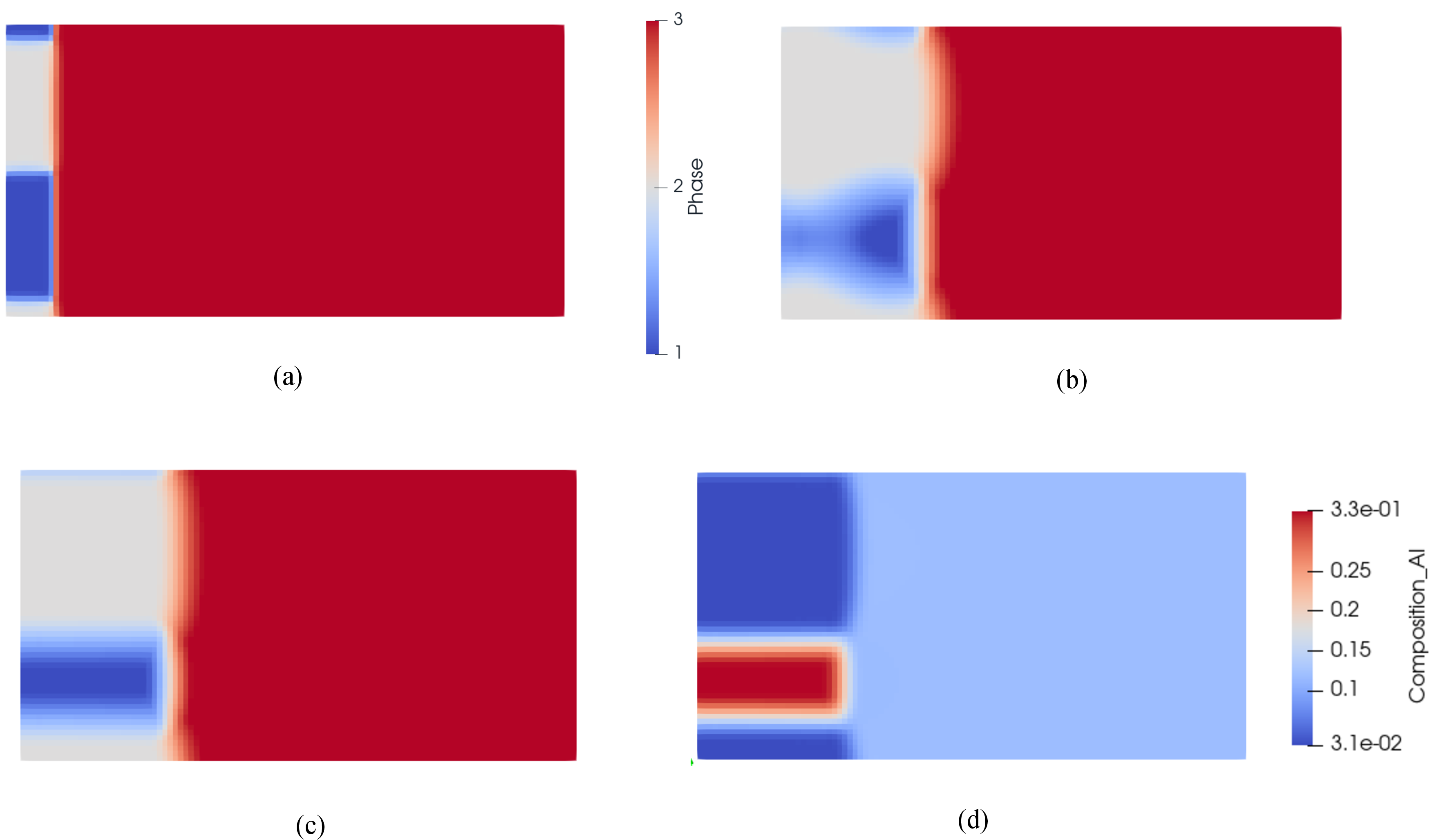}
    \caption{Evolution of FCC\_A1 and HCP\_A3 phases in binary AlZn alloy during eutectic solidification at (a) initial stage, (b) intermediate stage, and (c) final stage, along with (d) the Al compositions of the phases.}
    \label{eutectic}
\end{figure}

Next, we explore situations where physically different phases are involved in a phase transformation, for example, in a eutectic or a peritectic reaction. These simulations can also be performed using the solver modules in MICROSIM. In the following, we highlight the simulation performed for the eutectic reaction in the AlZn alloy, which is performed in a directional solidification setting. Here, the solvers based on MPI, OpenCL, CUDA, and AMReX can simulate directional solidification in the presence of a temperature gradient. Fig.~\ref{eutectic} represents the evolution from the beginning, where the phase fractions of the FCC\_A1 and the HCP\_A3 phases are very different from the equilibrium volume fractions given by the phase diagram. However, as depicted in the simulation, after the initial transient, the volume fractions close to the one predicted by the phase diagram are selected at steady-state. Along with this, the compositions (see Fig.~\ref{eutectic}(d)) of the phases are also comparable to the values at the eutectic temperature. 

The input file (Input\_AlZn\_eutectic.in) used for this simulation is attached for reference in the supplementary material of this paper. This particular simulation also highlights the solver's capability for executing long runs where the growth of phases is primarily in a given direction. For phase transformations such as solidification, the kinetics in the solid are rather low. Therefore, part of the already solidified domain may be removed, and a corresponding size of the new liquid of the original alloy composition can be appended at the far end. This way, while the domain's size remains the same, the advance of the solidification front is captured akin to a situation where a moving camera translates along with the solidification front. This feature in the solver modules may be activated using the ``SHIFT" key in the input file. 

\subsubsection{Solidification in a temperature gradient}
\begin{figure}[!htbp]
    \centering
    \includegraphics[width=\textwidth]{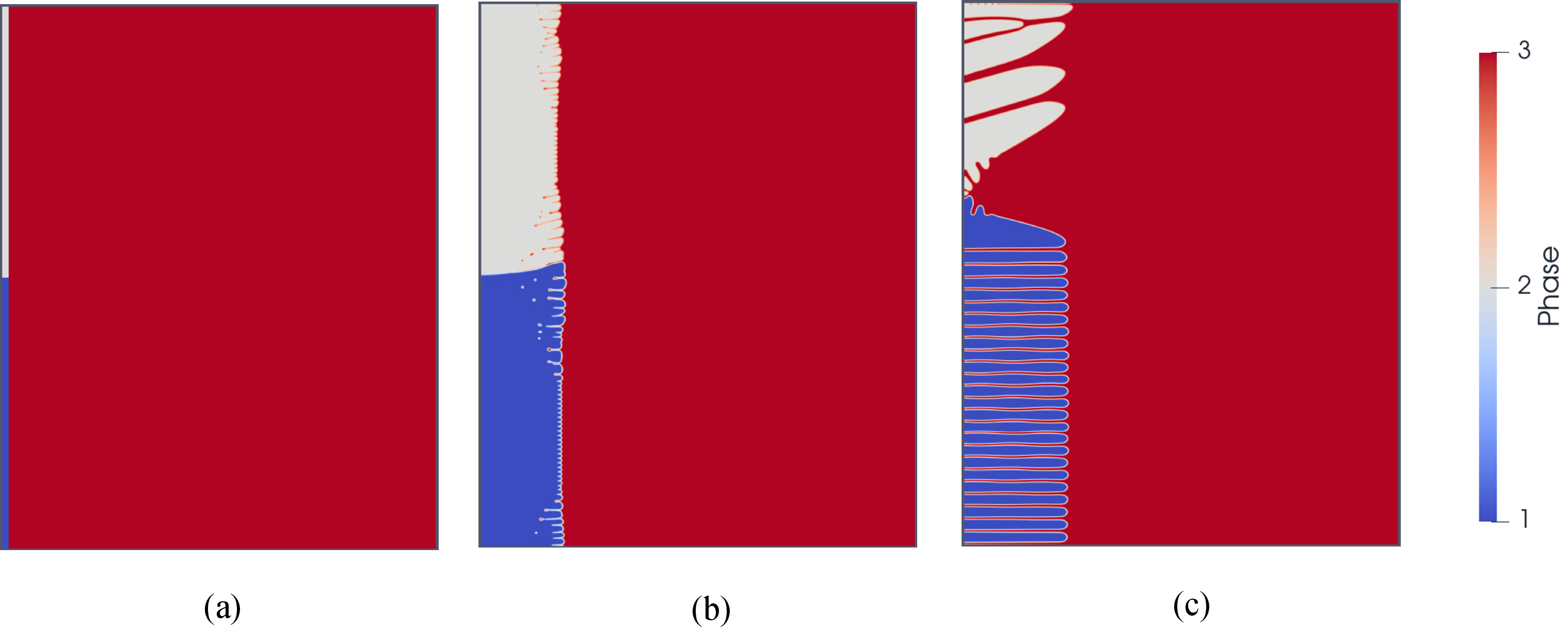}
    \caption{Evolution of phases in three-phase binary AlZn alloy during directional solidification at three different instances.}
    \label{fig:MPI_directional_2}
\end{figure}

One of the common situations for solidification simulations is where a frozen linear temperature gradient translating with a uniform velocity is imposed for replicating directional solidification conditions. This can be activated in the solver modules of MICROSIM through the ``Tempgrady" key in the input file. Once activated, a temperature gradient of the desired value is initialized in the y-direction of the domain, which can be translated with a given velocity to affect the desired cooling rate. Simulations have been performed for a model alloy using the thermodynamic Function\_F=1 for the solidification of bi-crystal depicted in Fig.~\ref{fig:MPI_directional_2}. In addition, the ``SHIFT" key in the input file is switched on.

\subsubsection{Multicomponent solidification}
\begin{figure}[!htbp]
	\subfloat[\label{fig:phiNiAlMoa}]{\includegraphics[width=48mm]{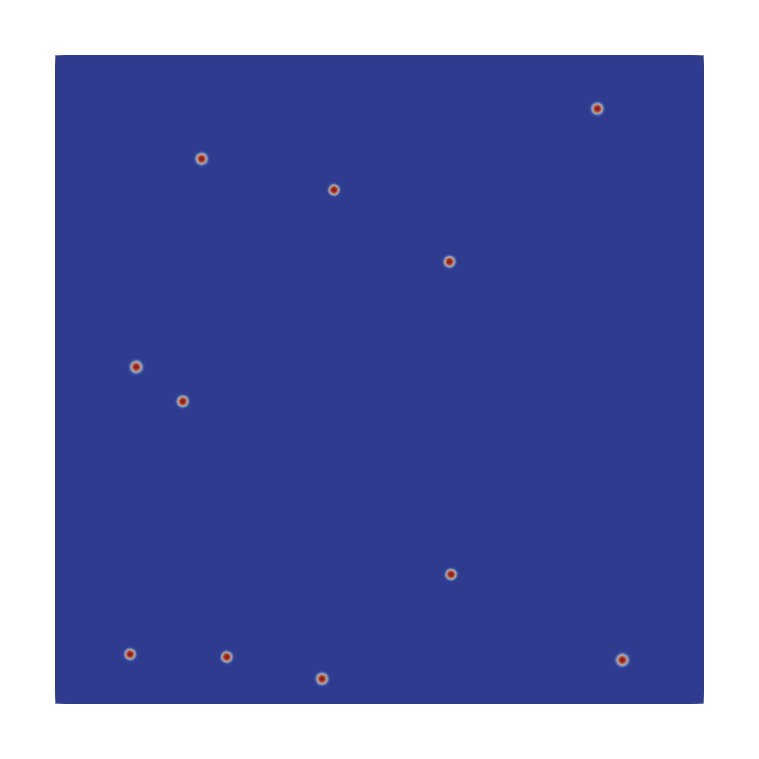}}
	\subfloat[\label{fig:phiNiAlMob}]{\includegraphics[width=48mm]{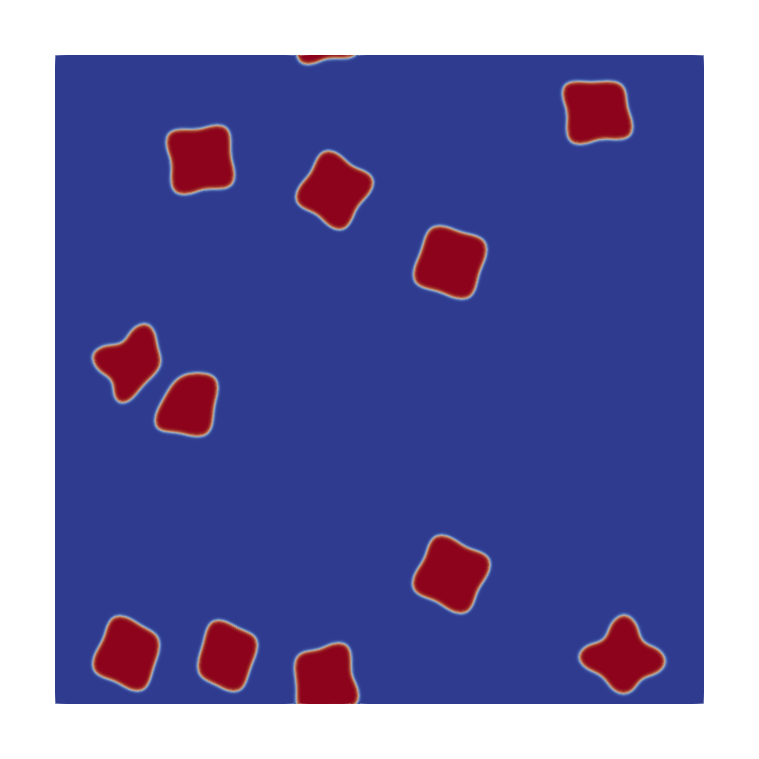}}
	\subfloat[\label{fig:phiNiAlMoc}]{\includegraphics[width=63mm]{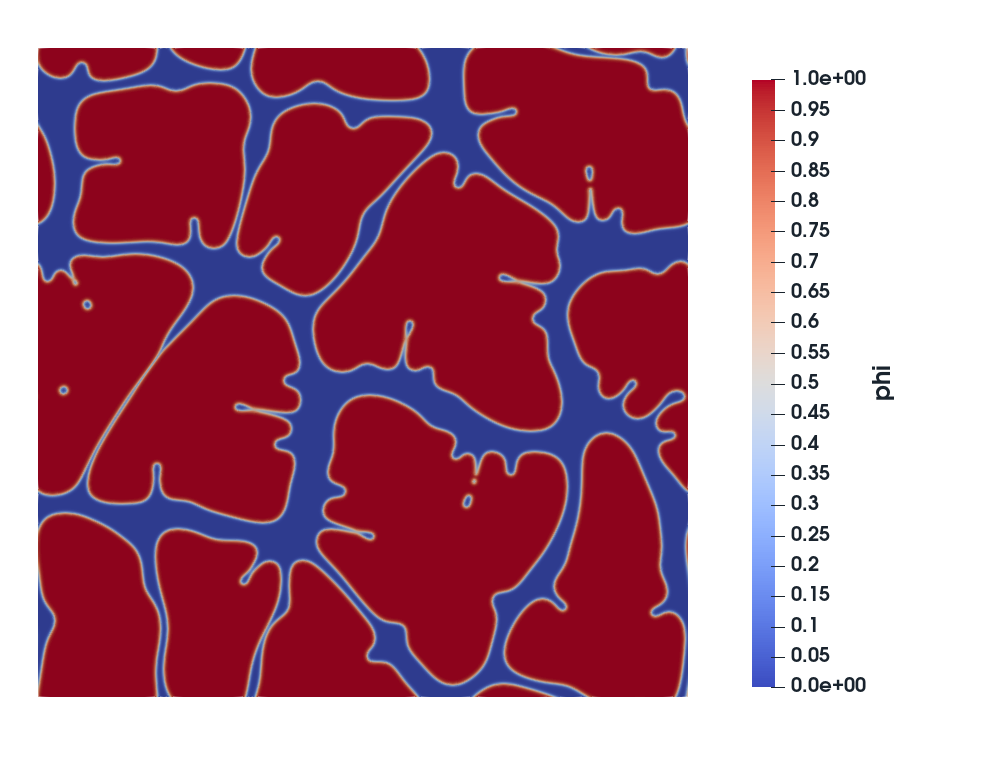}}
	\caption{Evolution of phase fraction, $\phi$ in NiAlMo alloy at cooling rate of 4$\times$10$^5$ J-mol$^{-1}$-s$^{-1}$ during (a) nucleation, (b) beginning of dendrite growth, and (c) grain formation.}
	\label{fig:phiNiAlMo}       
\end{figure}

\begin{figure}[!htbp]
	\subfloat[\label{fig:thetaNiAlMoa}]{\includegraphics[width=48mm]{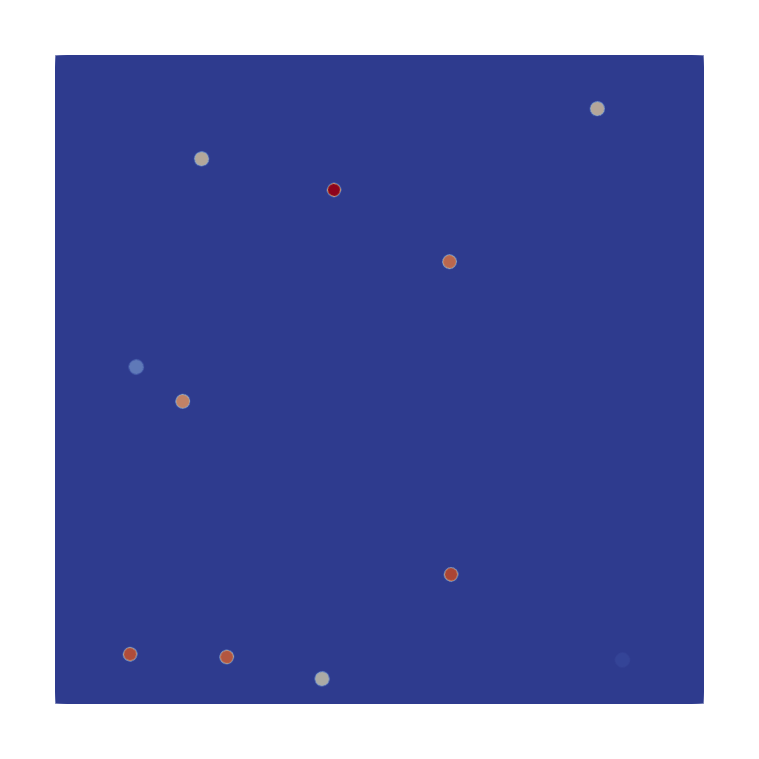}}
	\subfloat[\label{fig:thetaNiAlMob}]{\includegraphics[width=48mm]{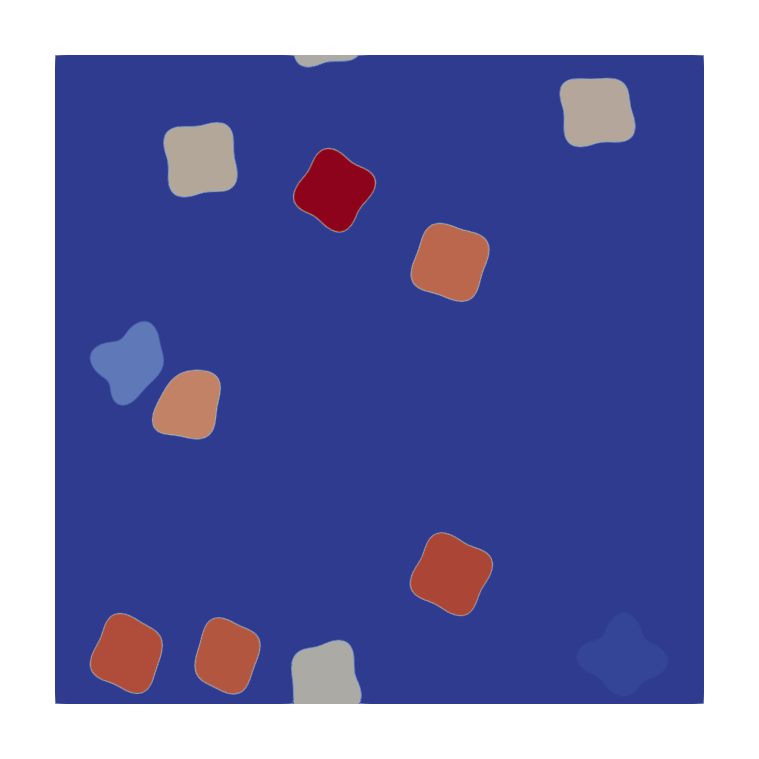}}
	\subfloat[\label{fig:thetaNiAlMoc}]{\includegraphics[width=63mm]{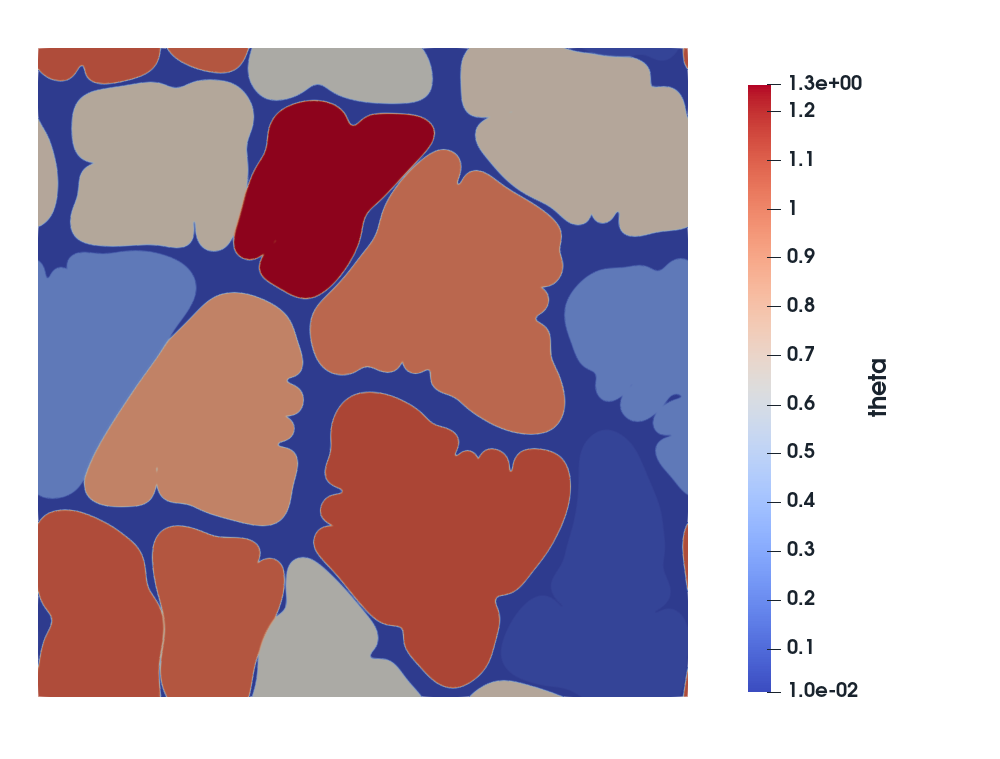}}
	\caption{Evolution of orientation field, $\theta$ in NiAlMo alloy at a cooling rate of 4$\times$10$^5$ J-mol$^{-1}$-s$^{-1}$ during (a) nucleation, (b) beginning of dendrite growth, and (c) grain formation.}
	\label{fig:thetaNiAlMo}       
\end{figure}

\begin{figure}[!htbp]
    \centering
    \includegraphics[width=0.8\textwidth]{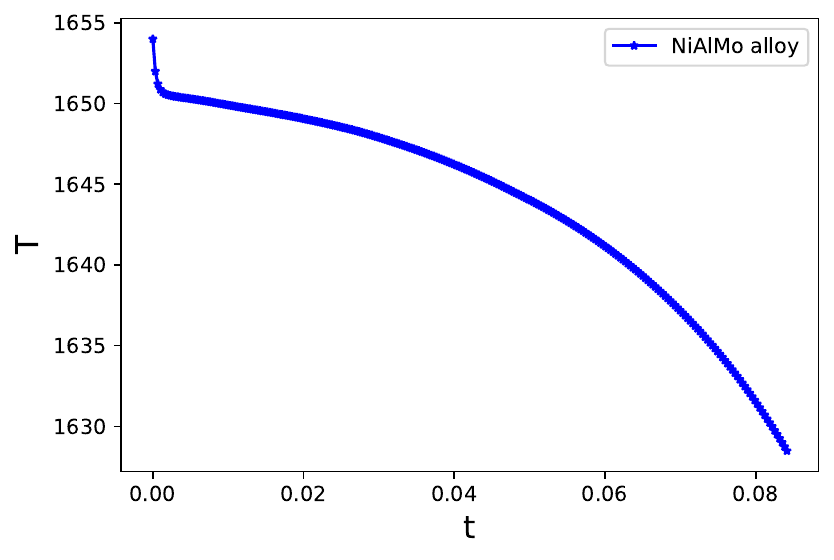}
    \caption{Temperature (K) evolution between liquidus and solidus at a cooling rate of 4$\times$10$^5$ J-mol$^{-1}$-s$^{-1}$ for a NiAlMo alloy.}
    \label{Temperature_NiAlMo}
\end{figure}

The solver modules are also capable of the simulation of multi-grain and multi-component alloys. Fig.~\ref{fig:phiNiAlMo} depicts a simulation performed with the OpenFOAM modules of the solver for a cooling rate of 4$\times$10$^5$ J-mol$^{-1}$-s$^{-1}$. The corresponding grain orientations are represented in Fig.~\ref{fig:thetaNiAlMo}. Similar to the binary simulations, the temperature history is recorded in Fig.~\ref{Temperature_NiAlMo}. 

\subsection{Precipitation under the influence of elastic stresses}
The MPI, OpenFOAM, and CUDA solvers contain modules for the simulation of precipitation under the influence of coherency stresses. The formulations relating to the different implementations are detailed in Subsection~\ref{subsec:elst}.

\subsubsection{Multicomponent precipitation}
\begin{figure}[!htbp]
    \centering
    \includegraphics[width=\textwidth]{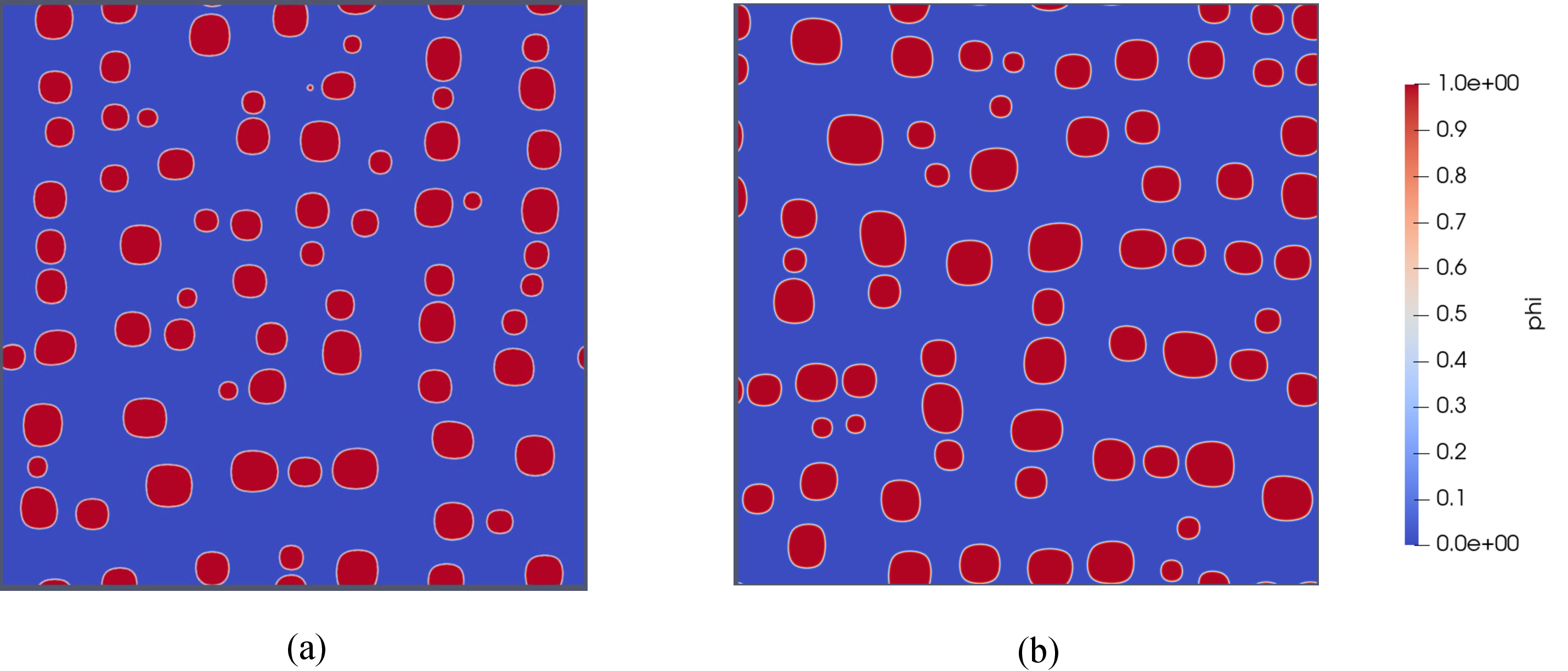}
    \caption{Phase fraction, $\phi$ in two-phase NiAlMo alloy with Zener anisotropy, $A_z =$ 3, and a dilatational misfit at undercooling temperature, $\Delta T$ of 100K, obtained from (a) the MPI solver, and (b) the CUDA solver. The simulation results are compared after 6 million iterations.}
    \label{fig:ppt_MPI_CUDA}
\end{figure}

\begin{figure}[!htbp]
  \subfloat[\label{MPI_CUDA_radius}]{\includegraphics[width=0.45\textwidth]{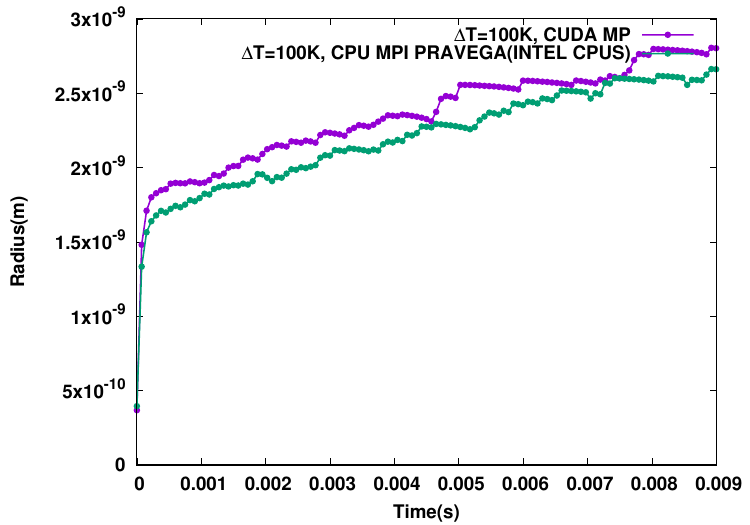}}
	\subfloat[\label{MPI_CUDA_number}]{\includegraphics[width=0.45\textwidth]{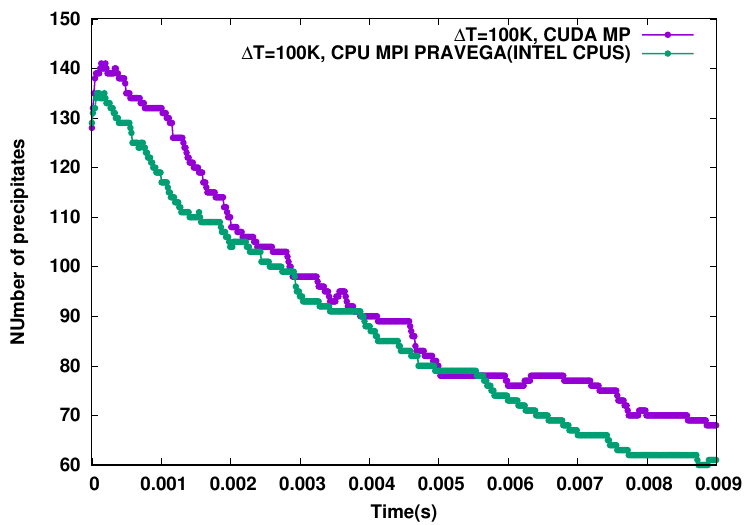}}
  \caption{Plots showing a comparison of the average radius in (a) and the number of precipitates in (b) for the simulations performed with the MPI and CUDA solvers.}
  \label{MPI_CUDA_comparison}
\end{figure}

Figs.~\ref{fig:ppt_MPI_CUDA} depict precipitation simulations in the NiAlMo system with the MPI and CUDA-based solvers. The temperature and composition result in a volume fraction of approximately 20\%. The coupling to the thermodynamic database is performed by using Function\_F=4. The material is chosen to be homogeneous and anisotropic with Zener anisotropy, $A_z =$ 3. The misfit strain at the interface between precipitate and matrix is dilatational, i.e., $E_{ij}^{*1} = \begin{bmatrix} 0.01 & 0 \\ 0 & 0.01 \end{bmatrix}$. As a comparison between the solvers, we have measured the average radius of the precipitates as a function of time and the number of precipitates in Fig.~\ref{MPI_CUDA_comparison}.

We find that although the models and implementations are different, the predictions regarding the size and distribution of the precipitates are similar between the MPI and CUDA-based solvers. We remark here that the key differentiation between the solvers here is the way mechanical equilibrium is calculated, wherein in the MPI-based solvers, the equilibrium equation is solved in real space while in the CUDA solvers, the solution of the displacement field is derived in Fourier-space. In spite of these differences in the formulation, models, and discretization, the simulation results are similar. While the MPI and the CUDA-based solvers have modules for precipitation at a given temperature, the precipitation module in OpenFOAM provides an additional feature. The solver can also be used to simulate precipitation growth at a constant cooling rate over larger temperature intervals. This is quite useful for simulations involving multi-stage heat treatments or heat treatments with a prescribed cooling rate. 

\subsubsection{Multi-variant precipitation}
\begin{figure}[!htbp]
    \centering
    \includegraphics[width=\textwidth]{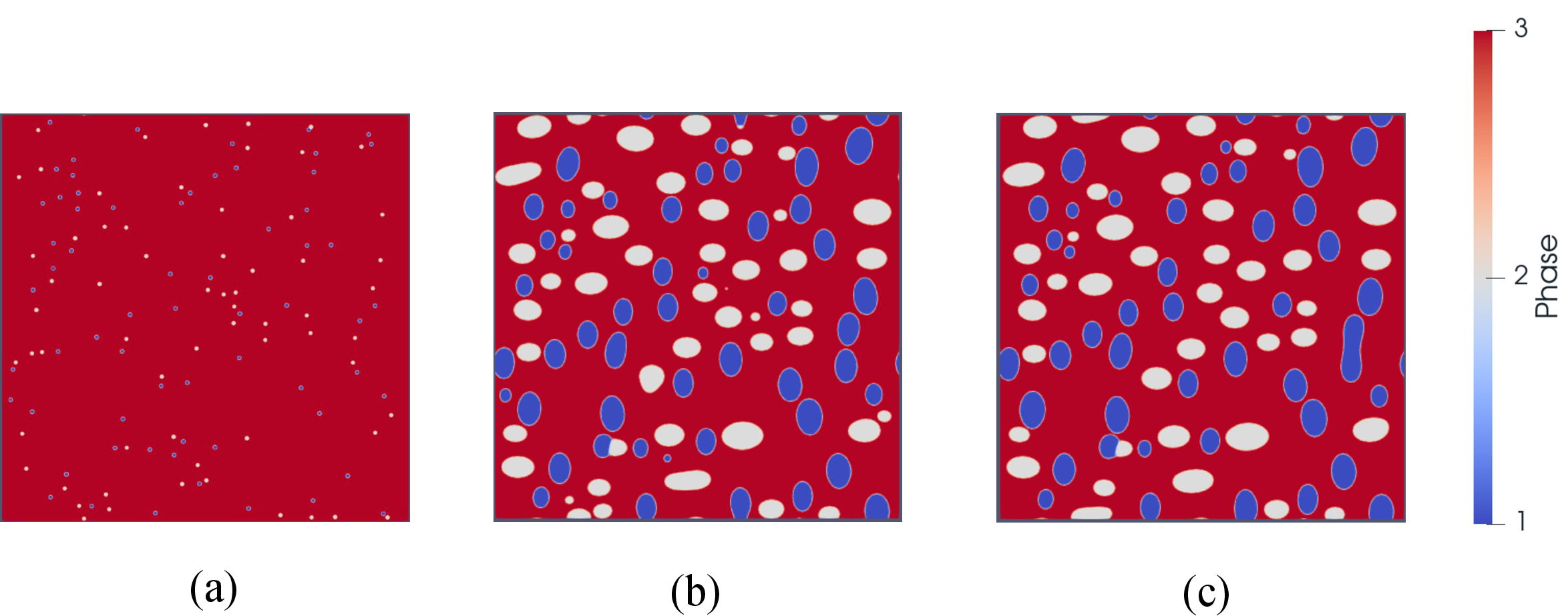}
    \caption{Evolution of phases in three-phase NiAlMo alloy with Zener anisotropy, $A_z =$ 3 at undercooling temperature, $\Delta T$ of 100K at three different instances during cubic–tetragonal transformation.}
    \label{fig:MPI_NiAlMo_2}
\end{figure}

Leveraging the multi-phase and multi-component structure of the solvers, we present some interesting test cases in this section. The first is a case of a cubic to tetragonal transition that gives rise to two variants in 2D. The two variants are simulated by respectively attributing the following eigenstrain matrices $\left( E_{ij}^{*1,2} = \begin{bmatrix} 0.002 & 0 \\ 0 & -0.01 \end{bmatrix},~ \begin{bmatrix} -0.01 & 0 \\ 0 & 0.002 \end{bmatrix} \right)$, that leads to the elongation either in the x- or the y- directions during evolution. Simulations are performed with the GP solvers using FD with MPI for the case of Zener anisotropy $A_z$=3 with a homogeneous stiffness matrix for the matrix and the precipitate variants, and the results are displayed in Fig. \ref{fig:MPI_NiAlMo_2}. The same thermodynamic description as the NiAlMo alloy in the preceding section is utilized for these simulations. 

\begin{figure}[!htbp]
	\subfloat[\label{fig:multi_NiAlMoa}]{\includegraphics[width=45mm]{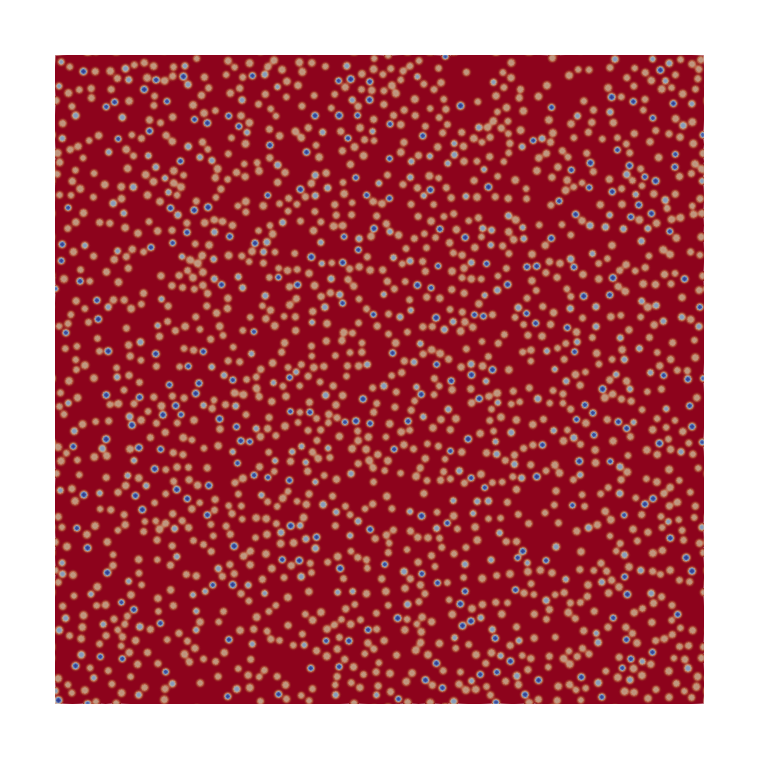}}
	\subfloat[\label{fig:multi_NiAlMob}]{\includegraphics[width=45mm]{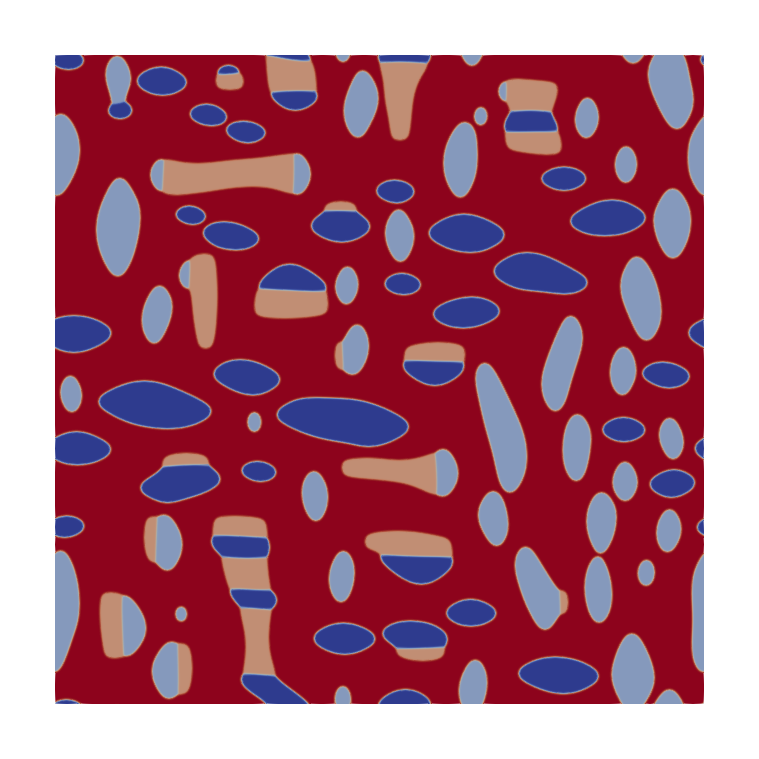}}
    \subfloat[\label{fig:multi_NiAlMoc}]{\includegraphics[width=60mm]{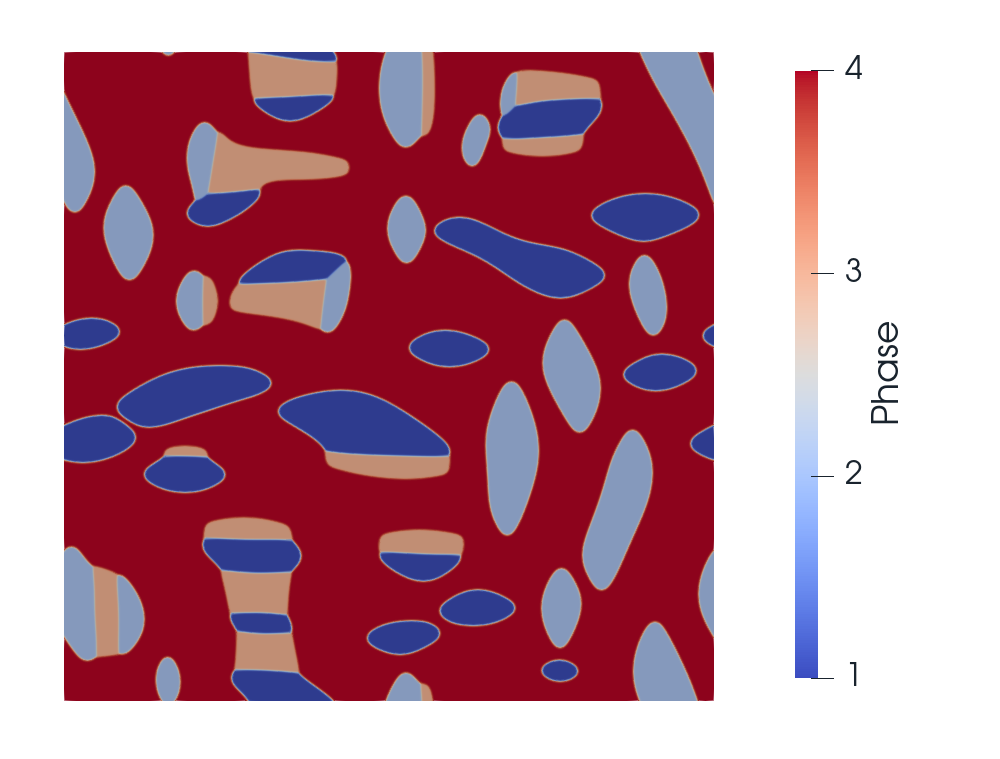}}
	\caption{Evolution of phases in four-phase NiAlMo alloy with Zener anisotropy, $A_z =$ 3 at undercooling temperature, $\Delta T$ of 100K at three different instances during cubic–tetragonal transformation.}
	\label{fig:multi_NiAlMo}    
\end{figure}

For the next case, we present simulations where three variants nucleate from the matrix, which include two tetragonal variants as in the preceding case and an additional cubic dilatational variant. The material is again chosen to be homogeneous and anisotropic with Zener ratio, $A_z =$ 3. The simulation has been carried out with an undercooling of 100K using the OpenFOAM solver. The misfit strain at the interface between precipitate phases and the matrix is either tetragonal $\left( E_{ij}^{*1,2} = \begin{bmatrix} 0.002 & 0 \\ 0 & -0.01 \end{bmatrix},~ \begin{bmatrix} -0.01 & 0 \\ 0 & 0.002 \end{bmatrix} \right)$ or dilatational $\left( E_{ij}^{*3} = \begin{bmatrix} 0.01 & 0 \\ 0 & 0.01 \end{bmatrix} \right)$ during the cubic–tetragonal transformation. The evolution of different phases is displayed in Fig.~\ref{fig:multi_NiAlMo} for the four-phase NiAlMo alloy at three different times, where we notice the formation of precipitate aggregates that involve the dilatational and the tetragonal precipitates.

\begin{figure}[!htbp]
    \centering
    \includegraphics[width=\textwidth]{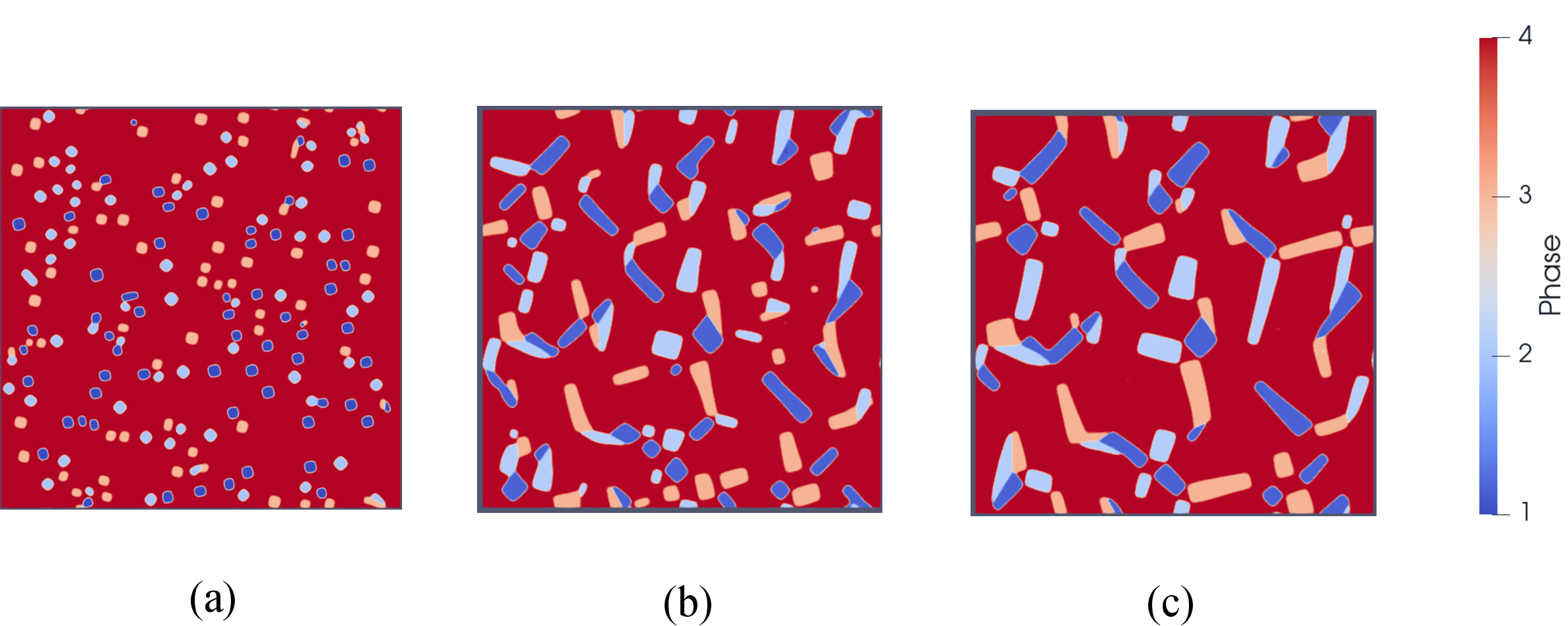}
    \caption{Evolution of phases in four-phase NiAlMo alloy with Zener anisotropy, $A_z =$ 3 at undercooling temperature, $\Delta T$ of 100K at three different instances during hexagonal to orthorhombic transformation transformation.}
    \label{fig:MPI_hexortho_NiAlMo}
\end{figure}

Further, we simulate a hexagonal to orthorhombic transition that is relevant for certain precipitation reactions in materials. During hexagonal to orthorhombic transformation, the misfit strain at the interface between precipitate phases and matrix are assumed to be $E_{ij}^{*1,2} = \begin{bmatrix} 0.01 & 0 \\ 0 & -0.01 \end{bmatrix}$, $\begin{bmatrix} -0.005 & 0.00866 \\ 0.00866 & 0.005 \end{bmatrix}$ and $E_{ij}^{*3} = \begin{bmatrix} -0.005 & -0.00866 \\ -0.00866 & 0.005 \end{bmatrix}$ for the three orthorhombic variants precipitating from the hexagonal matrix. The stiffness matrix is assumed to be isotropic with a Zener anisotropy of $A_z$=1 and homogeneous between the precipitation and the matrix. The evolution of different phases is displayed in Fig.~\ref{fig:MPI_hexortho_NiAlMo}, and the simulations are performed using the GP-solvers using FD with MPI, with the same thermodynamic description for the precipitate and matrix phases as before. We notice self-organization of the precipitates arising out of the matrix that inherits the symmetry of the eigenstrain matrices of the respective variants.

\subsection{Grain-growth}
\begin{figure}[!htbp]
    \centering
    \includegraphics[width=0.8\textwidth]{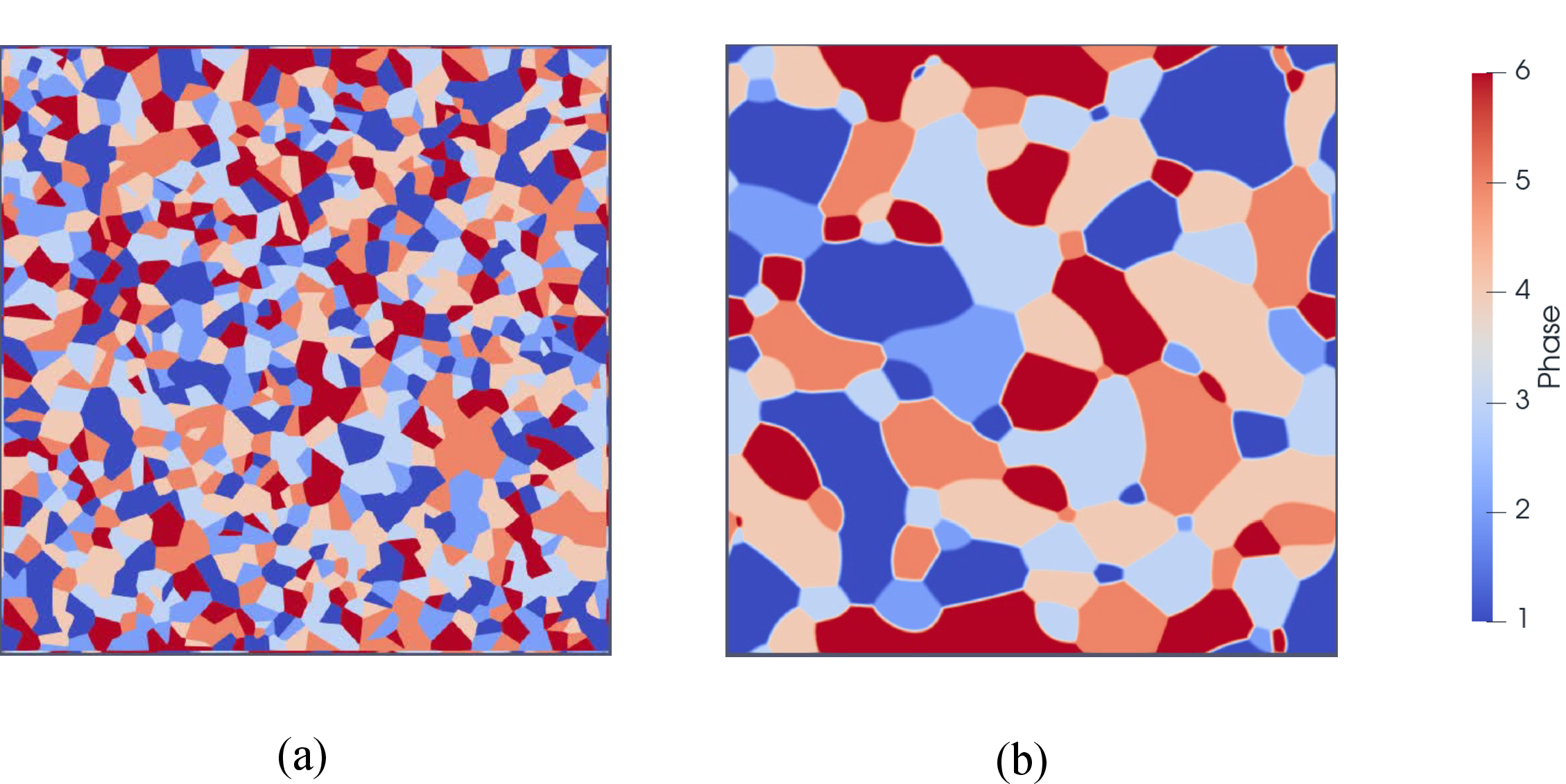}
    \caption{Evolution of a multi-grain microstructure at two different instances during a 2D isotropic grain-growth simulation.}
    \label{fig:grain_growth_1000_MPI}
\end{figure}

\begin{figure}[!htbp]
    \centering
    \includegraphics[width=0.8\textwidth]{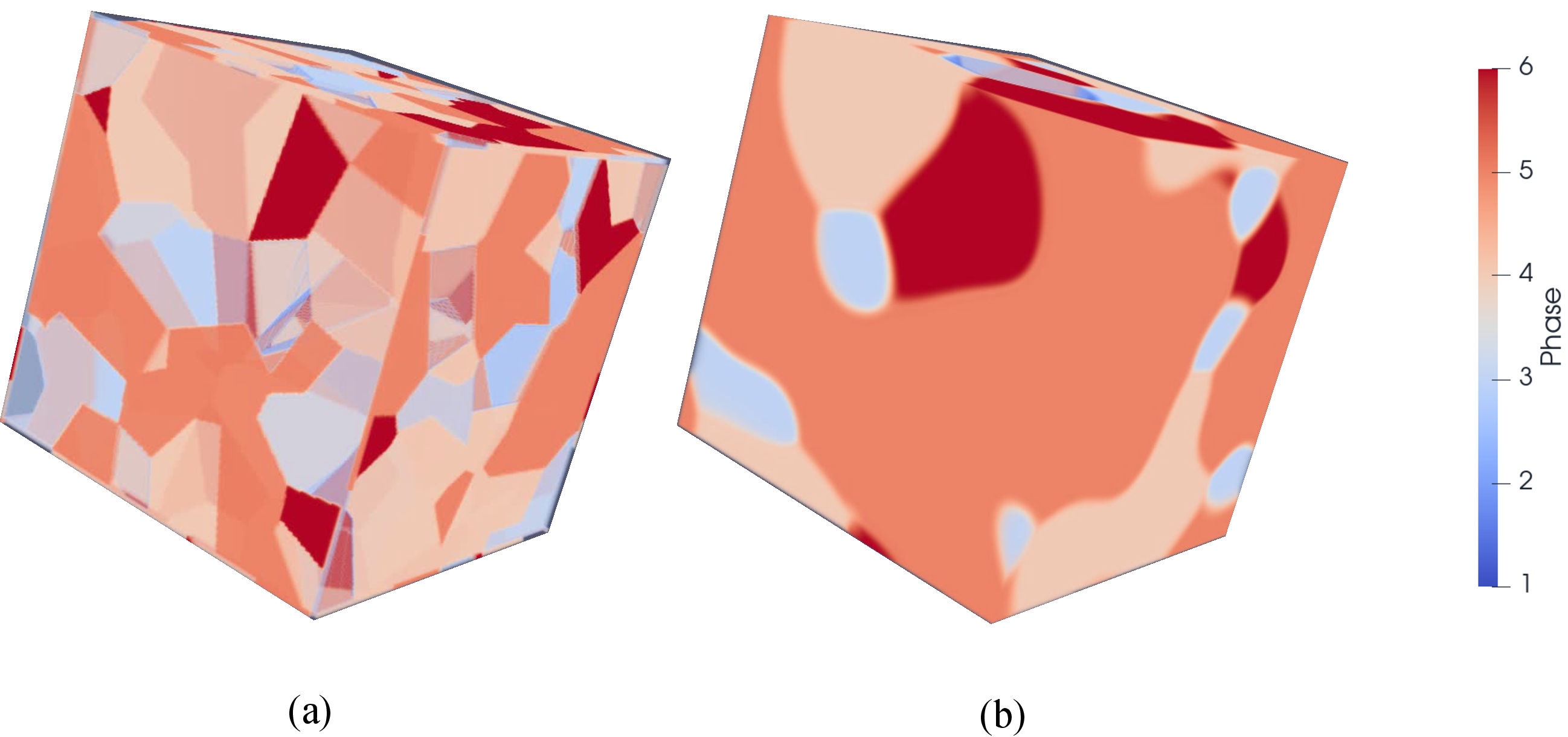}
    \caption{Evolution of grains in two different instances during a 3D grain-growth simulation.}
    \label{fig:grain_growth3D_MPI}
\end{figure}

Finally, with regards to the capabilities of the solver, we present cases of isotropic grain-growth in two and three dimensions depicted in Fig.~\ref{fig:grain_growth_1000_MPI} and Fig.~\ref{fig:grain_growth3D_MPI}, respectively. This module can be initiated with the input file key GRAINGROWTH=1. Presently, only the GP-solver with FD using MPI has this feature, and it will be enabled in the other solver modules in the future.

\section{Graphical user interface}
The graphical user interface (GUI) of MICROSIM has been implemented in Python 3 (v3.9) and uses the modules such as pyqt5, scikit-image, vtk, tinydb, SymPy v1.8, pyCALPHAD v0.9.2, PyMKS \cite{pymks} and yt. The two functionalities of the GUI are generation of input files and post-processing of results, which are discussed below.

\subsection{Input file generator}
\begin{figure}[!htbp]
    \centering
    \includegraphics[width=0.8\textwidth]{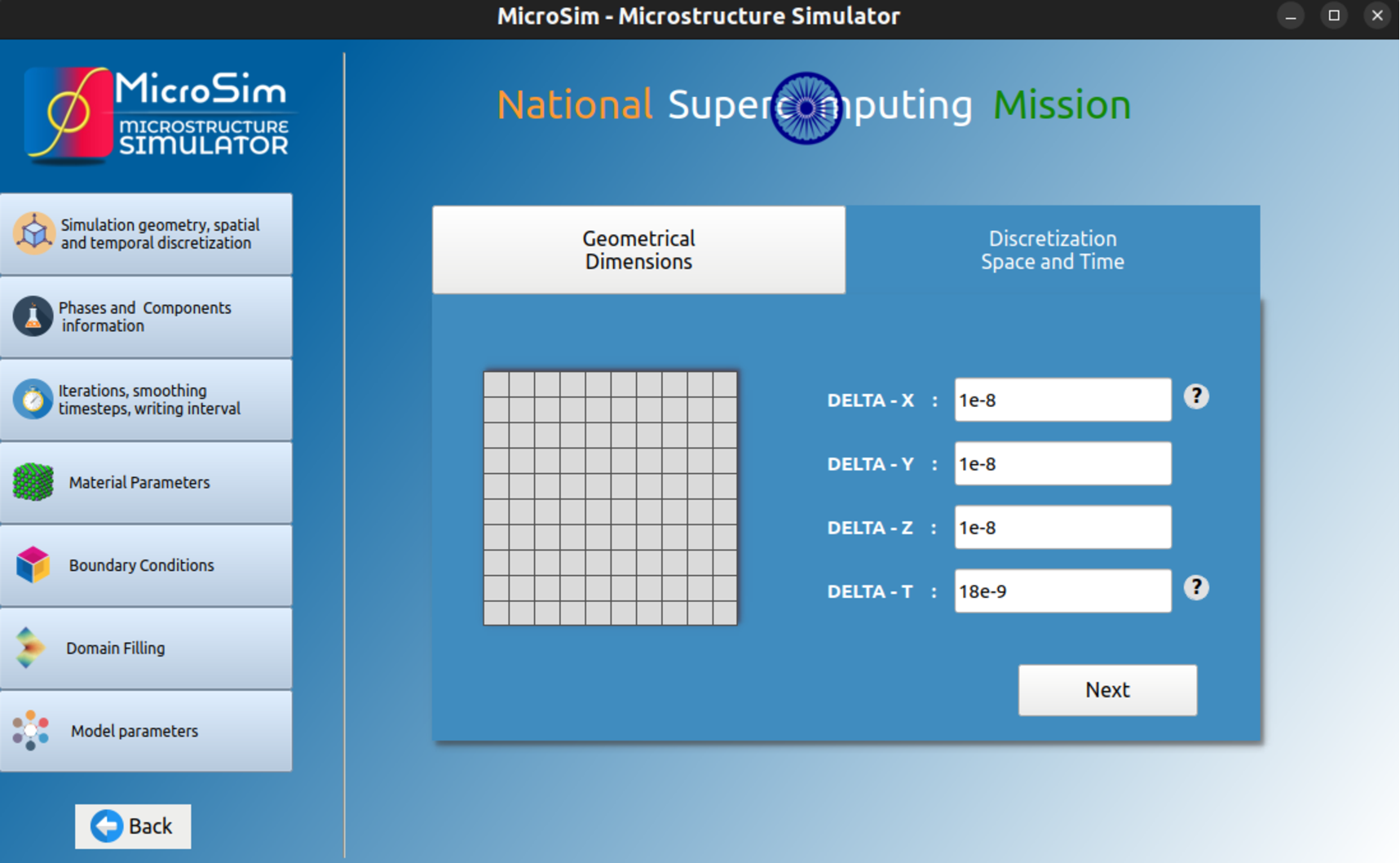}
    \caption{The graphical user interface of MICROSIM for generating the input file and accessing the post-processing tools.}
    \label{infile_gui}
\end{figure}

The input file displayed in Section~\ref{sec:infile} can be generated after launching the GUI and choosing a solver. A screenshot of the input file generator is shown in Fig.~\ref{infile_gui}. The panel on the left side of Fig.~\ref{infile_gui} contains buttons for different sections of the input file for adding values to respective keys. The button `Simulation geometry, spatial and temporal discretization' lets the user specify grid size and discretization values as shown on the right side of Fig.~\ref{infile_gui}. The `Phases and Components information' button sets the number of phases and components present in the alloy. The `Iterations, smoothing timesteps, writing interval' button specifies the number of time steps for running the simulations, smoothing the fields before solving the equations, and printing the output data. The `Material Parameters' button includes diffusivity, surface energy, thermodynamic database details, and for solving displacement fields, eigenstrain, and elastic moduli. The `Boundary Conditions' button specifies the type of boundary condition assigned to each variable and their values at the boundary. The `Domain Filling' button sets the initial condition by defining different phases at different regions. The `Model parameters' button specifies the value of parameters present in a model.

\subsection{Post-processing tools}
\begin{figure}[!htbp]
    \centering
    \includegraphics[width=0.8\textwidth]{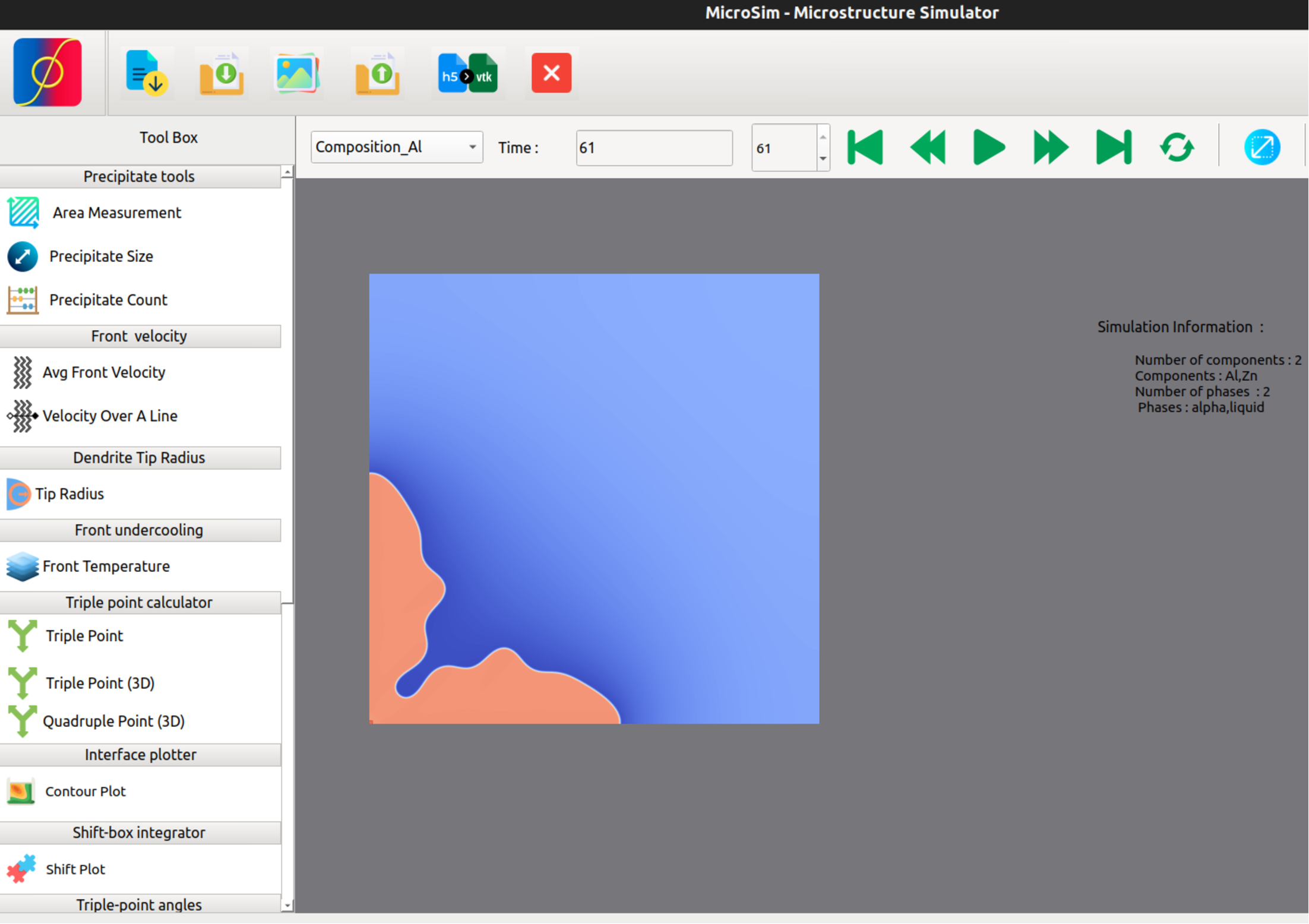}
    \caption{Post-processing tools available at the left panel in the graphical user interface of MICROSIM. The plot of the composition of Al for a single dendrite solidification case is also shown.}
    \label{pptools-dendrite}
\end{figure}

\begin{figure}[!htbp]
    \centering
    \includegraphics[width=0.8\textwidth]{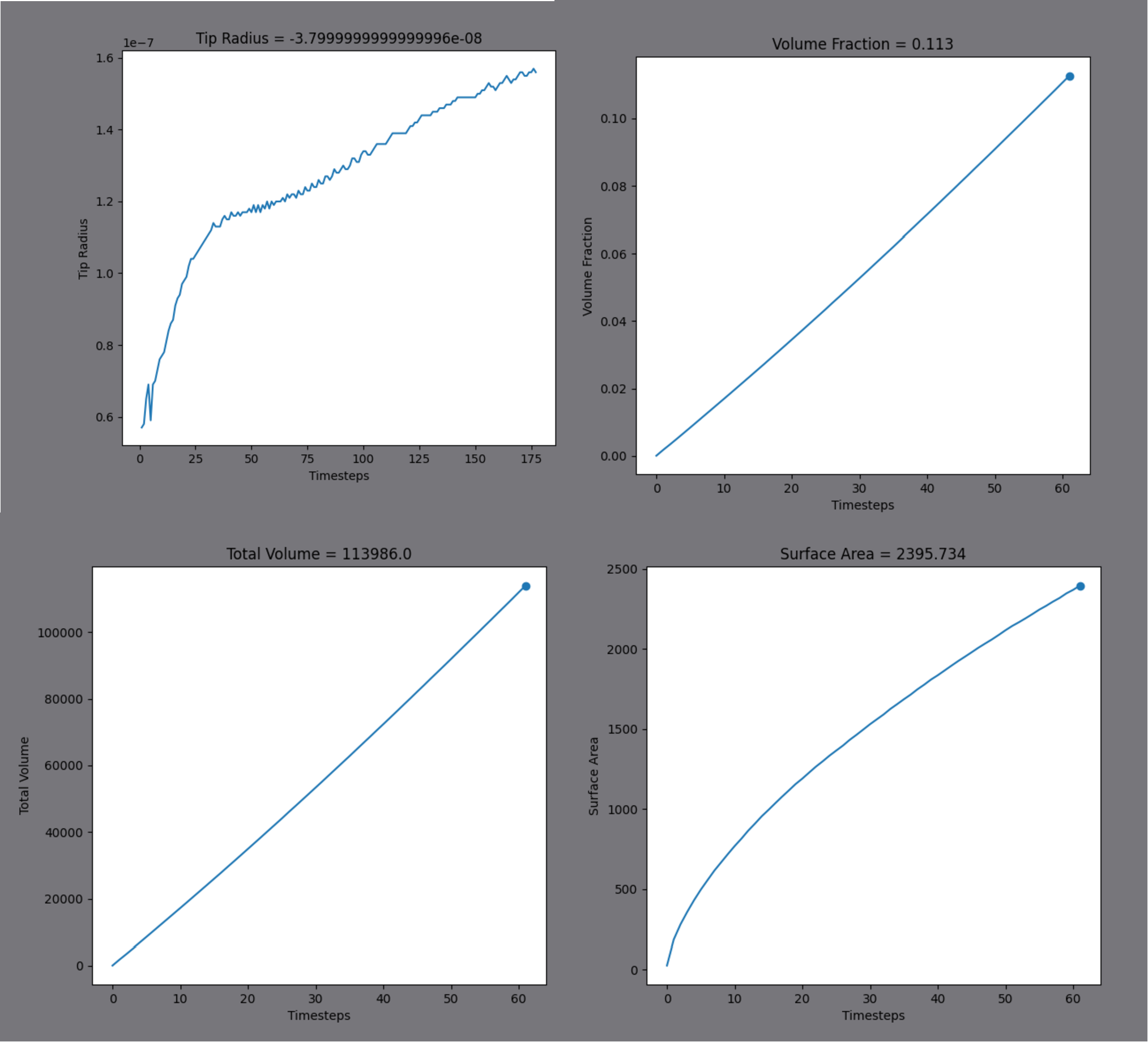}
    \caption{Evolution of radius of AlZn dendrite tip during solidification obtained for grand-potential MPI solver with Function\_F $=$ 2 at undercooling temperature, $\Delta T$ of 4K.}
    \label{pptools-line}
\end{figure}

The results obtained from the solver modules can be processed further using a set of post-processing tools developed with Python. The users can access the tools using the graphical user interface. 
In the following, we present certain snippets of the available tools and their applications. The set of post-processing tools is shown in the left panel of Fig.~\ref{pptools-dendrite}, along with the composition of Al obtained from the MPI solver with Function\_F $=$ 2 at 4K undercooling in the AlZn alloy. The data of the dendrite tip velocity presented in Fig.~\ref{fig:velocity} is obtained using the tool ``Front velocity" for the single dendrite simulation results shown in Fig.~\ref{fig:interface}. The dendrite tip has been located by determining the maximum distance of the solid-liquid interface from the origin (that is, the center of the dendrite). Thus, the dendrite tip velocity is calculated from the change in the tip's distance from the origin at each time step. The evolution of dendrite tip radius for the MPI solver with Function\_F $=$ 2 at 4K undercooling case is displayed in Fig.~\ref{pptools-line} and is estimated using the option ``Tip Radius". The points at the interface have been fit into a polynomial curve. Then, the radius of curvature at the dendrite tip is obtained from the second-order derivative of the polynomial, computed at the dendrite tip. In addition, the volume fraction, volume, and surface area (in terms of the number of grid points) of a dendrite can be calculated using the tools `Phase Fraction,' `Volume,' and `Surface Area,' respectively, as shown in Fig.~\ref{pptools-line}. It can be observed that the volume fraction and volume increase linearly while the surface area increases non-linearly.

\begin{figure}[!htbp]
    \centering
    \includegraphics[width=0.8\textwidth]{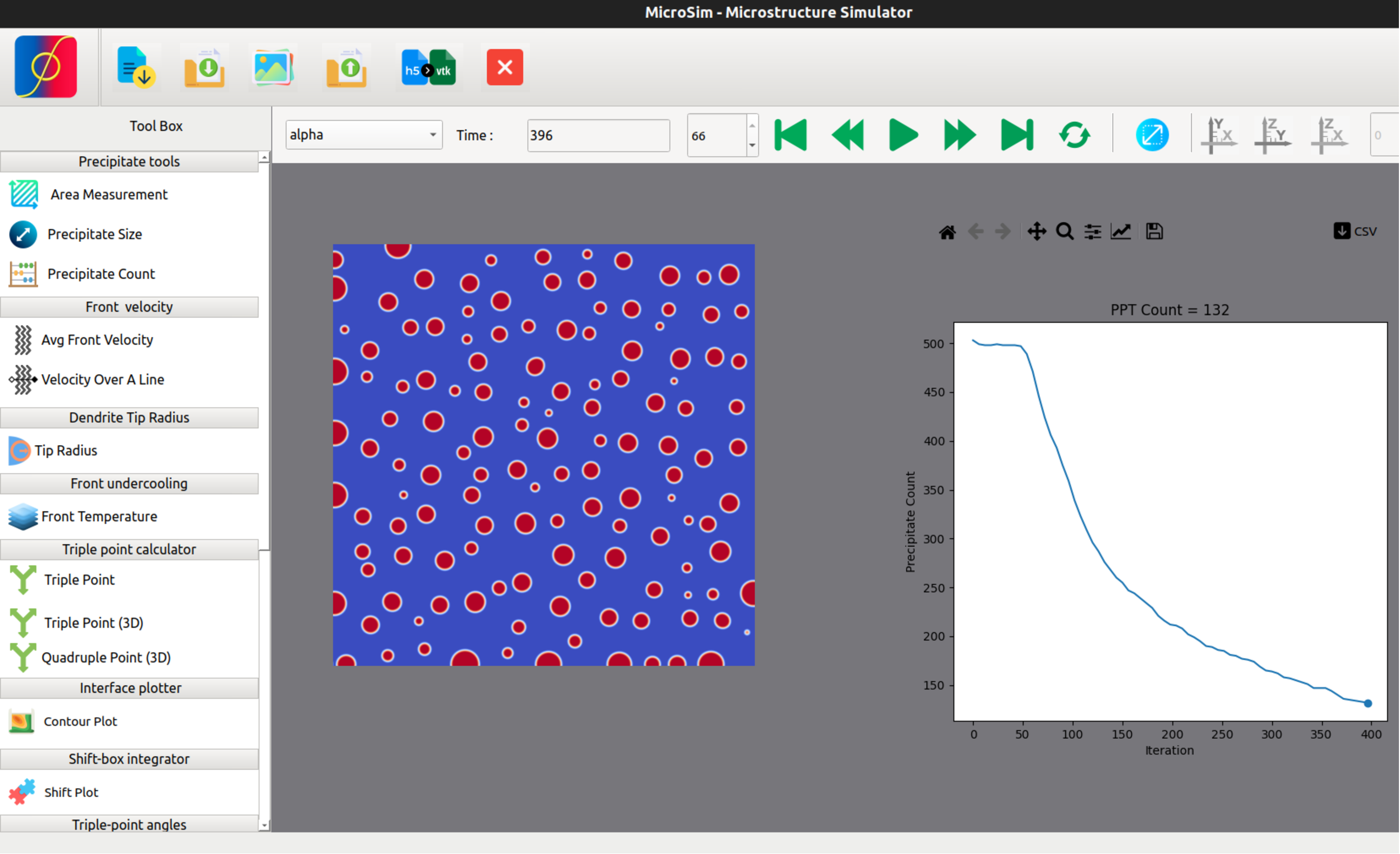}
    \caption{Post-processing tools available at the left panel in the graphical user interface of MICROSIM. The figure depicts a snapshot from a coarsening simulation along with the variation of the number of precipitates on the right.}
    \label{pptools-ppt}
\end{figure}

\begin{figure}[!htbp]
    \centering
    \includegraphics[width=0.8\textwidth]{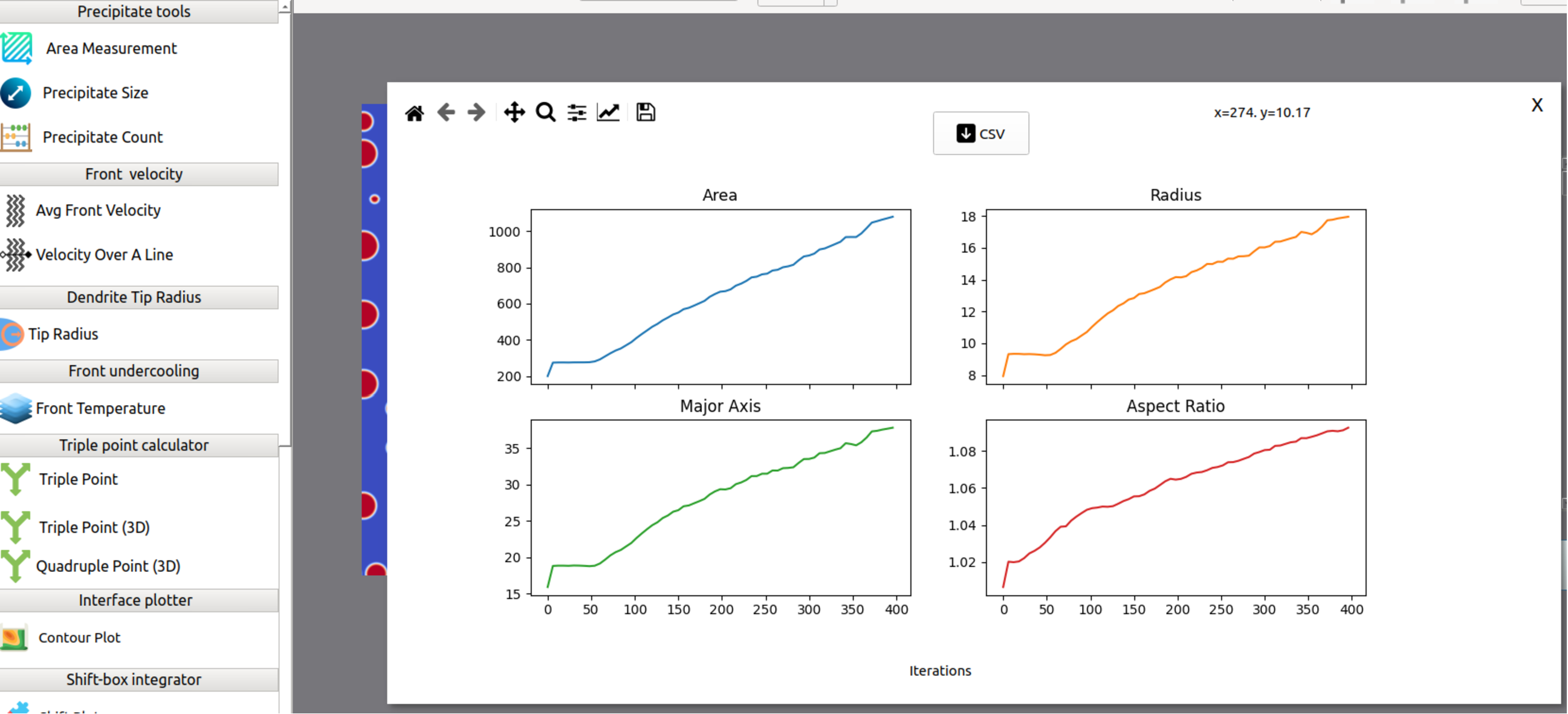}
    \caption{Graphs generated from the precipitate measurement routines in the post-processing toolbox.}
    \label{pptools-ppt2}
\end{figure}

The contours of phase fraction and composition obtained from precipitation growth simulations can be used to estimate the different attributes of the precipitates present in the microstructure. Fig.~\ref{pptools-ppt} displays the evolution of the number of precipitates. The tools `Area Measurement' and `Precipitate Size' allow the users to calculate the area, mean radius, aspect ratio, etc of each precipitate. For instance, a variation of these parameters is presented in Fig.~\ref{pptools-ppt2} for the simulation displayed in Fig.~\ref{pptools-ppt}.

\begin{figure}[!htbp]
    \centering
    \includegraphics[width=0.8\textwidth]{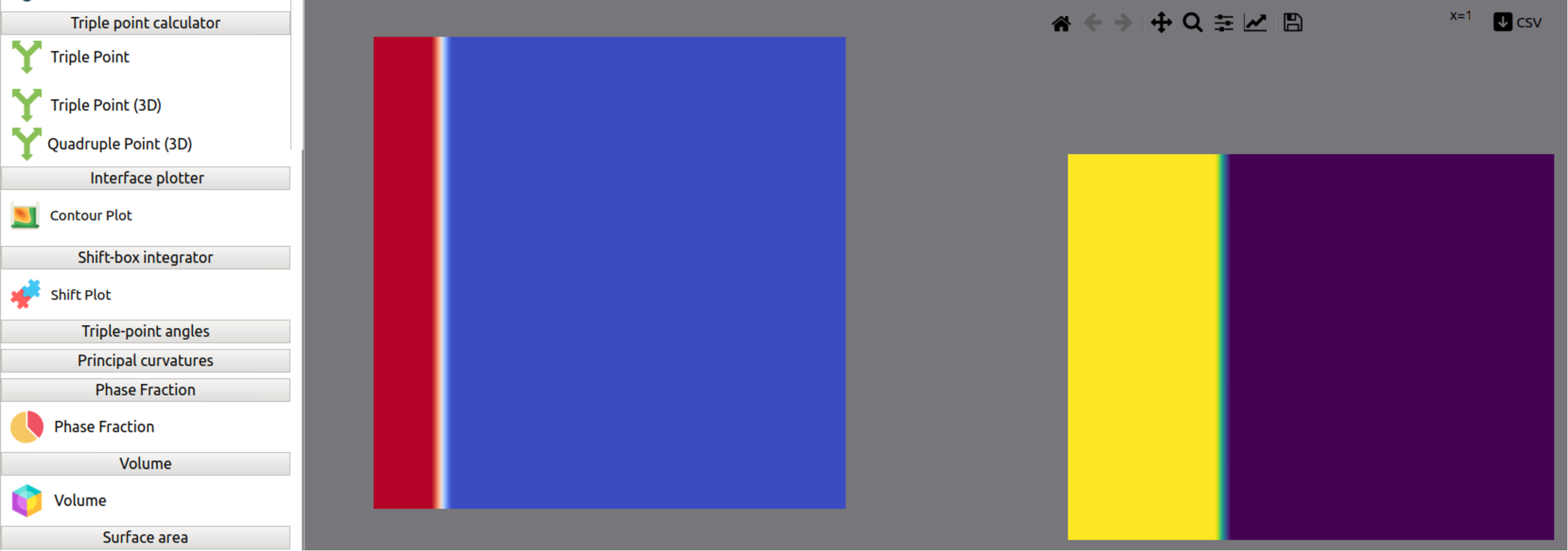}
    \caption{Illustration of the usage of "Shiftplot" that integrates simulation results into one single frame such that the entire microstructural evolution history may be accessed.}
    \label{pptools-shift}
\end{figure}

\begin{figure}[!htbp]
    \centering
    \includegraphics[width=0.8\textwidth]{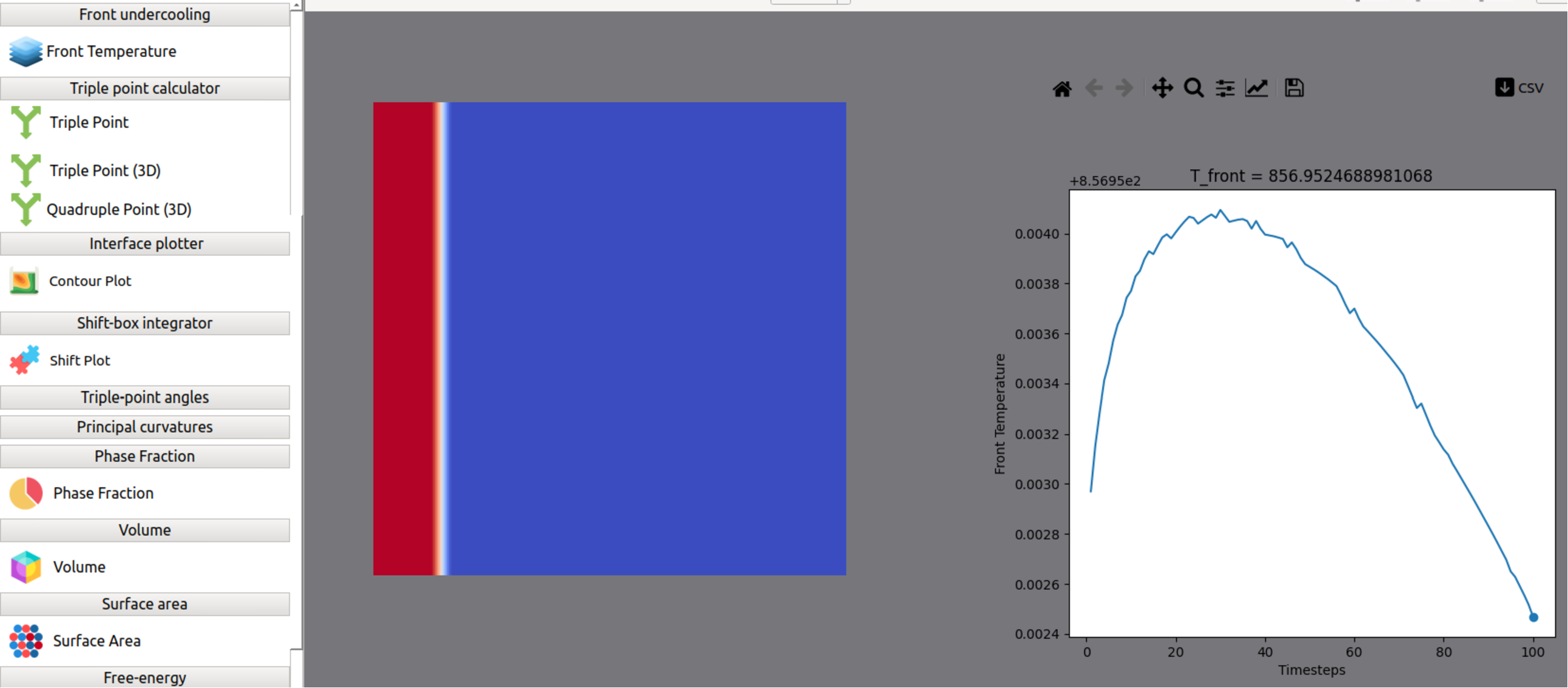}
    \caption{Tool for estimating the front undercooling during directional solidification in a temperature gradient.}
    \label{pptools-temp}
\end{figure}

The directional solidification simulations in the solver module can be executed using the moving box method as described in Subsection~\ref{multiphase}. They can be activated using the `SHIFT' key in the input file. This allows the simulation domain to move with the interface in the positive-x direction with a certain pulling velocity. In this way, part of the solidified region moves outside the simulation domain. The whole domain can be assembled from the result obtained for the partial domains for each time step. To achieve this, the tool `Shift Plot' can be used. As shown in Fig.~\ref{pptools-shift}, the field on the right side is assembled from the result available on the left side for multiple time steps. Further, the temperature evolution at the interface can be derived using the tool `Front Temperature,' as shown in Fig.~\ref{pptools-temp}. 

\begin{figure}[!htbp]
    \centering
    \includegraphics[width=0.8\textwidth]{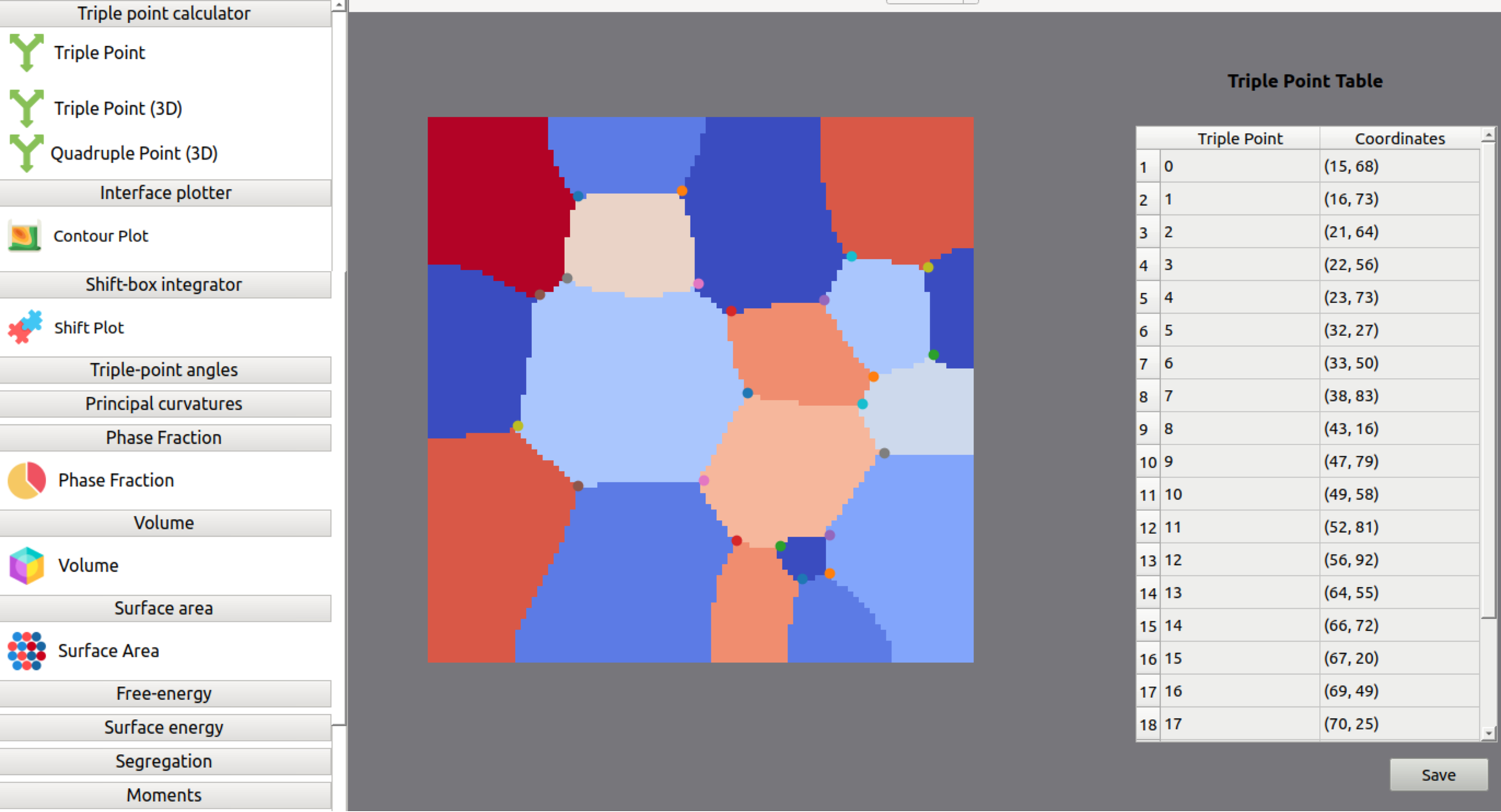}
    \caption{Determination of triple-points in a multi-phase microstructure. The output illustrates the position of the triple-points on the original microstructure and tabulates the co-ordinates in a separate .csv file.}
    \label{pptools-triple}
\end{figure}

The location of triple points in microstructures can be calculated from VTK files obtained from multi-phase simulations. Fig.~\ref{pptools-triple} shows triple point coordinates calculated using the button `Triple Point.' The triple point coordinates can also be exported to .csv files for further processing by the user.

\begin{figure}[!htbp]
    \centering
    \includegraphics[width=0.8\textwidth]{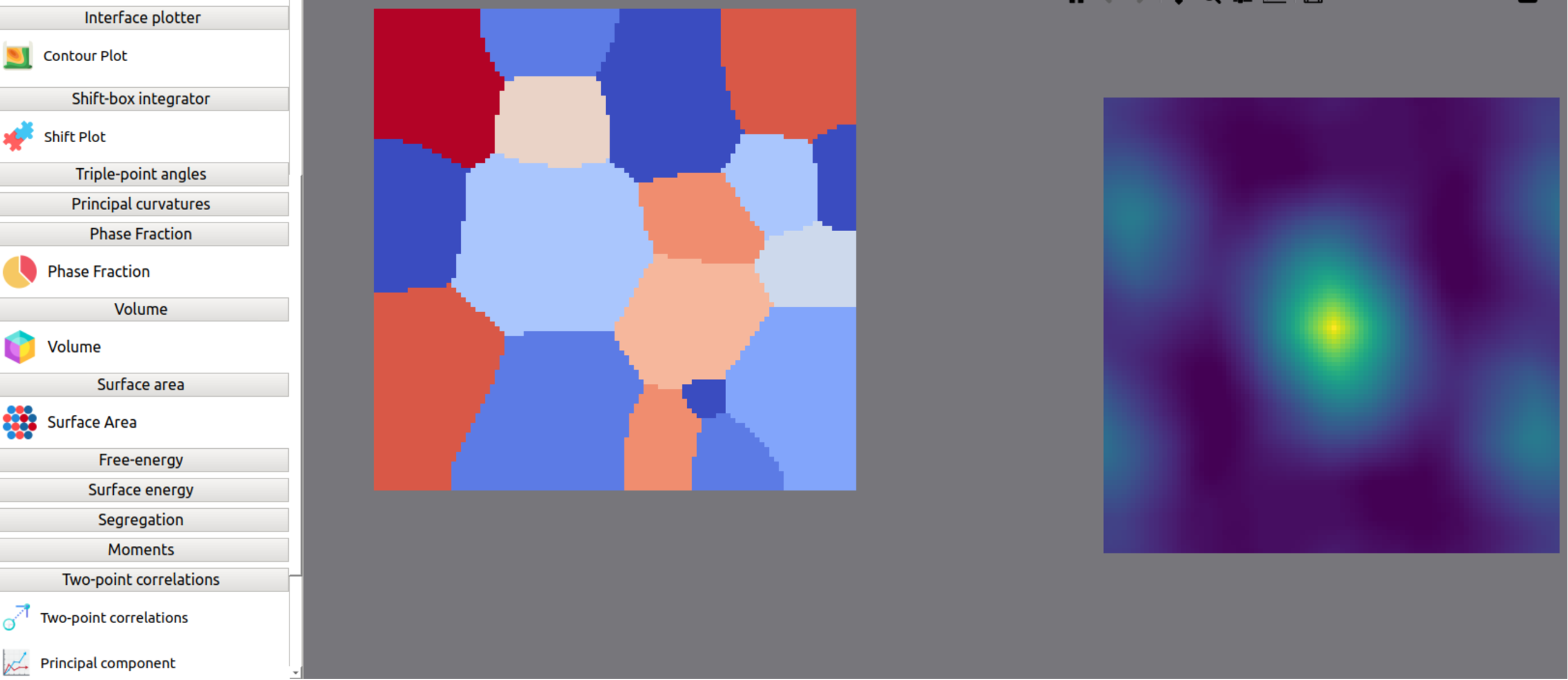}
    \caption{Estimation of two-point correlations in the microstructure on the left using the post-processing toolbox.}
    \label{pptools-2point}
\end{figure}

\begin{figure}[!htbp]
    \centering
    \includegraphics[width=0.8\textwidth]{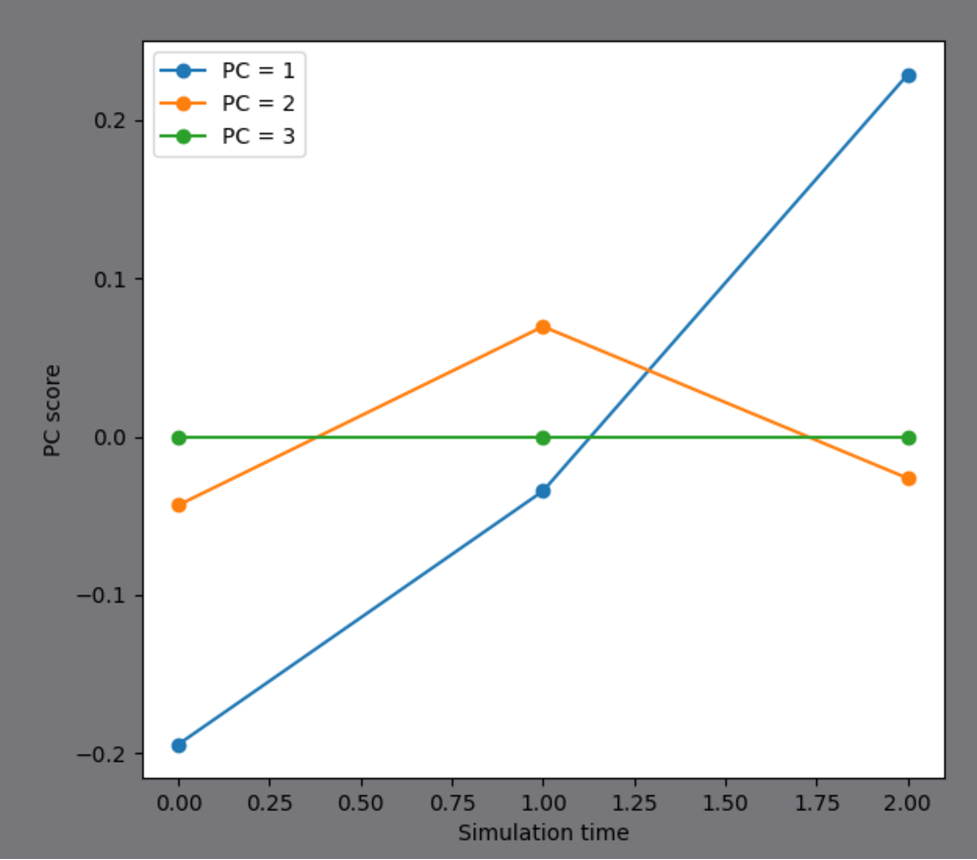}
    \caption{Determination of the principal components from the two-point spatial correlations for the multi-grain microstructure in the preceding figure.}
    \label{pptools-pc}
\end{figure}

Further, one can determine the statistical features of microstructures. Fig.~\ref{pptools-2point} represents a two-point spatial correlation plot obtained from the grain structure depicted in Fig.~\ref{pptools-triple}. It signifies an alternative way of expressing the microstructure and is suitable for deriving feature representation. One way to derive features is by estimating principal components calculated from the two-point spatial correlations. Fig.~\ref{pptools-pc} shows the three major principal components obtained from the two-point spatial correlations. These can be utilized for the formulation of structure-property correlation models.

Finally, there are tools that allow for calculating average velocity, or velocity along a given direction, and the variation of the computed fields along given directions, that become useful for comparison with experimental data. For all the post-processing tools, the results are either available as graphical images that can be directly utilized for publications or reports, or the data can be exported to .csv files for further analysis. This feature provides immense flexibility to the user for not only conducting simulations but also for the interpretation of the results through the extraction of relevant microstructural metrics as a function of time.

\section{Summary}
This paper describes the structure and capabilities of a new open-source phase-field software stack (MICROSIM), a collection of solver modules with the GP and KKS model formulations implemented using different discretization strategies. The combination of different models, discretization, and implementation strategies provides the user immense flexibility regarding the methodology and the accessible hardware (CPU and GPU). The software stack has modules that can be readily adapted for most metallurgical and materials processing industry situations, particularly because of the methods used to couple with thermodynamic databases. With the availability of the software on our GitHub repository (https://github.com/ICME-India/MicroSim) as well as resources and videos on our website (https: //microsim.co.in/), MICROSIM promises to provide new and advanced users with a collection of tools for applications directed both towards academia and industry. Further, the scalability of these solvers ensures efficient usage of the available supercomputing facilities. Finally, the availability of scalable solver modules will enhance the utilization of phase-field models for larger microstructure simulations that open possibilities for deeper scientific insights and impact. 

\section{Acknowledgements}
The authors would like to thank the National Supercomputing Mission project funded by the Department of Science and Technology(DST), India, for the funding for the MICROSIM projects (DST/NSM/R\&D\_HPC \_Applications /2021/03) and (DST/NSM/R\&D\_HPC\_Applications/ Extension/2023/15).
We would also like to thank SERC at IISc for providing the computational resources on the PARAM-PRAVEGA supercomputer that was utilized for all the computations and comparisons between the solver modules. We would also like to thank the Ph.Ds in our groups (Dr. Arka Lahiri, Dr. Sumeet Khanna, Dr. Tushar Jogi, Dr. Gerald Tennyson, Dr. Bhalchandra Bhadak) whose theses formed the bases for the solver modules that were integrated in MICROSIM.

\appendix


\section{Code}

Link to code and test cases: https://github.com/ICME-India/MicroSim

 \bibliographystyle{elsarticle-num} 
 \bibliography{References_bib}

\end{document}